\documentclass[10pt,journal,compsoc]{IEEEtran}
\def\autoref{\ref}
\usepackage{amsmath,amssymb,amsfonts,amsthm}
\usepackage{graphicx}
\usepackage{subfig}
\usepackage{enumerate}
\usepackage{enumitem}
\usepackage{multirow}
\usepackage[table,xcdraw]{xcolor}
\usepackage{xspace}
\usepackage[ruled,vlined]{algorithm2e}
\usepackage{bm}
\usepackage{enumitem}
\newtheorem{remark}{Remark}

\usepackage{color}
\def\cbl{\textcolor{black}}
\def\cred{\textcolor{black}}

\newcommand{\wrt}{\emph{w.r.t.}\xspace}
\newcommand{\ie}{\emph{i.e.,}\xspace}
\newcommand{\aka}{\emph{a.k.a.}\xspace}
\newcommand{\eg}{\emph{e.g.,}\xspace}

\newcommand{\eat}[1]{}
\usepackage{makecell}

\newtheorem{definition}{Definition}
\newtheorem{example}{Example}

\usepackage{graphicx}
\ifCLASSOPTIONcompsoc
  \usepackage[nocompress]{cite}
\else
  \usepackage{cite}
\fi
\ifCLASSINFOpdf
\else
\fi
\begin{document}
%
\title{A Learned Index for Exact Similarity Search in Metric Spaces}
%
%
%
%

\author{
    Yao~Tian,
    Tingyun~Yan,
    Xi~Zhao,
    Kai~Huang,
    and~Xiaofang~Zhou,~\IEEEmembership{Fellow,~IEEE}
\IEEEcompsocitemizethanks{
    \IEEEcompsocthanksitem Y. Tian, X. Zhao, K. Huang and X.F. Zhou are with the Department of Computer Science and Engineering, Hong Kong University of Science and Technology, Clear Water Bay, Kowloon, Hong Kong.\protect\\
    E-mail: ytianbc@cse.ust.hk, \{xizhao, ustkhuang\}@ust.hk, zxf@cse.ust.hk.
    
    \IEEEcompsocthanksitem T.Y. Yan is with the Cyberspace Institute of Advanced Technology, Guangzhou University, Guangzhou, China. \protect\\
    E-mail: tingyun$\_$yan@e.gzhu.edu.cn.

}
\thanks{Manuscript received December 28, 2021; revised 5 April 2022; published online xx xx xx.}
}

\IEEEtitleabstractindextext{%
\begin{abstract}
 Indexing is an effective way to support efficient query processing in large databases. 
 Recently the concept of \textit{learned index}, which replaces or complements traditional index structures with machine learning models, has been actively explored to reduce storage and search costs. However, accurate and efficient similarity query processing in high-dimensional metric spaces remains to be an open challenge. In this paper, we propose a novel indexing approach called LIMS that uses data clustering, pivot-based data transformation techniques and learned indexes to support efficient similarity query processing in metric spaces. In LIMS,
 the underlying data is partitioned into clusters such that each cluster follows a relatively uniform data distribution. Data redistribution is achieved by utilizing a small number of pivots for each cluster. Similar data are mapped into compact regions and the mapped values are totally ordinal. Machine learning models are developed to approximate the position of each data record on disk. Efficient algorithms are designed for processing range queries and nearest neighbor queries based on LIMS, and for index maintenance with dynamic updates. Extensive experiments on real-world and synthetic datasets demonstrate the superiority of LIMS compared with traditional indexes and state-of-the-art learned indexes.

\end{abstract}

}

\maketitle

\IEEEdisplaynontitleabstractindextext

%
\IEEEpeerreviewmaketitle

\IEEEraisesectionheading{\section{Introduction}\label{sec:introduction}}

%
%
%
%
\IEEEPARstart{S}{imilarity} search is one of the fundamental operations in the era of big data. It finds  objects from a large database within a distance threshold to a given query object (called {\em range queries}) or the top-$k$ most similar to the query object (called {\em $k$ nearest neighbor queries}, or $k$NN {\em queries}), based on certain similarity measures or distance functions. For example, in spatial databases, a similarity query can be used to find all restaurants within a given range in terms of \cbl{the} \textit{Euclidean distance}. In image databases, a similarity query can be used to find the top 10 most similar images to a given image in terms of \cbl{the} \textit{Earth mover’s distance}\cite{earth}. To accommodate a wide range of data types and distance functions, we consider similarity search in the context of metric spaces in this paper. A metric space is a generic space that makes no requirement of any particular data representation, but only a distance function that satisfies the four properties, namely non-negativity, identity, symmetry and triangle inequality (Definition \ref{def:metricspace}, Section  \ref{sec:Background}). A number of metric-space indexing methods have been proposed in the literature to accelerate similarity query processing\cite{intrinsic1,metric_survey,metric_survey2,metric_survey3, metric_survey4}. However, these indexing methods that are based on tree-like structures are increasingly challenged by the rapidly growing volume and complexity of data. On the one hand, query processing with such indexes requires traversing many index nodes (\ie nodes on the path from the root node to a leaf node) in the tree structure, which can be time-consuming. On the other hand, tree-like indexes impose non-negligible storage pressure on datasets that store complex and large objects, such as image and audio feature data.  

In recent years, the concept of \textit{learned index} \cite{kraska2018case}  has been developed to provide a new perspective on indexing. By enhancing or even replacing traditional index structures with machine learning models that can reflect the intrinsic patterns of data, a learned index can look up a key quickly and save a lot of memory space required by traditional index structures at the same time. The original idea is limited to the one-dimensional case where data is sorted in an in-memory dense array. Directly adapting this idea for  a multi-dimensional case is unattractive, since  multi-dimensional data has no natural sort order. Several multi-dimensional learned index structures have been proposed to address this issue\cite{wang2019learned,li2020lisa,rsmi,davitkova2020ml,sagedb,flood,tsunami,qdtree} (detailed discussions refer to Section \ref{sec:relatedwork}). Despite the significant success of these learned indexes compared with traditional indexing methods, they still have some limitations.
First, the existing learned index structures do not support similarity search in metric spaces. The  metric space has neither coordinate structure  nor dimension information \cbl{(Remark \ref{rmk:genericmetricspace}, Section  \ref{sec:Background})}, so the numbering rules (\eg z-order\cite{z-curve}) and specific pruning strategies designed for vector spaces are not applicable.  \cbl{The} triangle inequality is the only property we can utilize to reduce the search space. The generality of metric space provides an opportunity to develop unified indexing methods, while it also presents a significant challenge to develop an efficient learned indexing method.
Second, the existing learned multi-dimensional index structures suffer from the phenomenon called {\em curse of dimensionality}. By integrating machine learning models into traditional multi-dimensional indexes, these learned indexes are restricted to certain types of data space partitioning (\eg grid partitioning), which inevitably leads to rapid performance degradation  when the number of dimensions grows. 
Third, the time to train a machine learning model that can well approximate complex data distributions is typically very long, which makes learned indexes  difficult to adapt to frequent insertion/deletion operations and query pattern changes. 
Finally, some existing learned indexes\cite{rsmi} \cbl{can only return approximate query results, \ie there may exist false negatives  in the result set}, because of errors caused by machine learning models.

To address the aforementioned limitations, we develop a novel disk-based learned index structure for metric spaces, called LIMS, to facilitate exact similarity queries (\ie point, range and $k$NN queries).  In contrast to \cbl{the} coordinate-based data partitioning, LIMS adopts a distance-based clustering strategy to group the underlying data into a number of subsets so as to decompose complex and potentially correlated data into clusters with simple and relatively uniform distributions. 
LIMS selects a small set of pivots for each cluster and utilizes the distances to the pivots to perform data redistribution. This reduces the dimensionality of the data to the number of pivots adopted. 
By using a proper pivot-based mapping, LIMS organizes similar objects into compact regions and imposes a total order over the data. Such \cbl{an} organization can significantly reduce  the number of distance computations and page accesses  during query processing. In order to further boost the search performance, LIMS follows the idea of \textit{learned index}, using several simple polynomial regression models to quickly locate data records that might match the query filtering conditions. 
Furthermore, LIMS can be partially reconstructed quickly due to its independent index structure for each cluster, which makes LIMS adaptable to changes.
As we will show later, LIMS significantly outperforms other multi-dimensional learned indexes and traditional indexes in terms of \cbl{the} average query time and the number of page accesses, especially when processing high dimensional data.

The main contributions of this paper include:
\begin{itemize}
    \item We design LIMS, the first learned index structure for metric spaces, to facilitate exact similarity search. 
    
    \item Efficient algorithms for processing point, range and $k$NN queries are proposed, enabling a unified solution for searching complex data in a representation-agnostic way. An update strategy is also proposed for LIMS.
    
    \item To the best of our knowledge, no experiment evaluation between different learned indexes has been performed. In this paper, we compare four multi-dimensional learned indexes.
    Extensive experiments on real-world and synthetic data demonstrate the superiority of LIMS.
\end{itemize}
  
The rest of the paper is organized as follows. Section \ref{sec:relatedwork}  reviews related work. Section \ref{sec:Background} introduces the basic concepts and formulates the research problem. Section \ref{sec:LIMS} describes the details of LIMS. LIMS-based similarity query algorithms are discussed in Section \ref{sec:queryprocessor}. Section \ref{sec:exp} reports the experimental results. Section \ref{sec:conclusion} concludes the paper.

\section{Related Work}
\label{sec:relatedwork}
We focus on reviewing learned multi-dimensional indexes here. Good surveys of various traditional metric-space indexing methods can be found in \cite{intrinsic1,metric_survey,metric_survey2,metric_survey3, metric_survey4}.

The idea of \textit{learned index} is that indexes can be regarded as models which take a key as the input and output the position of the corresponding record. If such a ``black-box" model can be learned from data, a query can be processed by a function invocation in $O(1)$ time instead of traversing a tree structure in $O(\log_{}n)$ time. RMI is the first to explore how to enhance or replace classic index structures with machine learning models\cite{kraska2018case}. It assumes that data is sorted and kept in an in-memory dense array. In light of this,  a machine learning model essentially is to learn a cumulative distribution function (CDF). RMI consists of a hierarchy of models, where internal nodes in the hierarchy are the models responsible for predicting the child model to use, and a leaf model predicts the position of record. Since RMI utilizes the distribution of data and requires no comparison in each node,  it provides significant benefits in terms of storage consumption and query processing time. For the sake of quickly correcting errors caused by machine learning models and supporting range queries,  RMI is limited to index key-sorted datasets, which makes a direct application of RMI to multi-dimensional data infeasible because there is no obvious ordering of points. Even if these points are embedded into an ordered space, guaranteeing the correctness and efficiency of a query (\eg range query and $k$NN query) remains a  challenging task.

ZM index is the first effort to apply the idea of \textit{learned index} to multi-dimensional spaces\cite{wang2019learned}. It adopts the z-order space filling curve \cite{z-curve} to establish the ordering relationship for all points and then invokes RMI to support point and range queries. The correctness is guaranteed by a nice geometric property of the z-order curve, \ie monotonic ordering. However, ZM index needs to check many irrelevant points during the refinement phase, which would get worse for high dimensional spaces. It does not support $k$NN queries and index updates. 

Recursive spatial model index (RSMI) builds on  RMI and ZM\cite{rsmi}. It develops a recursive partitioning strategy to partition the original space, and then groups data according to predictions. This results in a learned point grouping, which is different from RMI that fixes the data layout first and trains a model to estimate positions. For each partition, RSMI  first maps points into \cbl{the} rank space and then invokes ZM to support point, range and $k$NN queries. However,  the correctness of range and $k$NN queries can not be guaranteed. In addition,  since RSMI is still based on space filling curves, the good performance of RSMI is confined to low dimension spaces.

LISA\cite{li2020lisa}, a learned index structure for spatial data can effectively reduces the number of false positives compared with ZM,  by 1) partitioning the original space into grid cells based on \cbl{the} data distribution; 2) ordering data with a partially monotonic mapping function and rearranging data layout according to mapped values; 3) decomposing a large query range into multiple small ones. LISA has a dynamic data layout as RSMI, but the correctness of range and $k$NN queries can be guaranteed by the monotonicity of the models. However, its advantage in low scan overhead  comes with the costly checking procedure and high index construction time. And the grid-based partitioning  strategy makes it unsuitable for high dimensional spaces. Besides, LISA-based $k$NN query processing suffers from many repeated page accesses due to doing range queries with  increasing radius from the scratch.

Similar to LISA, Flood \cite{flood} also partitions data space into grid cells along dimensions such that for each dimension, the number of points in each partition is approximately the same. Flood assumes a known query workload and  utilizes sample queries to learn an optimal combination of indexing dimensions and the number of partitions. Once these are learned, Flood maintains a table to record the position of the first point in each cell. At query time, Flood invokes RMI for each dimension to identify the cells intersecting the query and looks up the cell table to locate the corresponding records. However, it cannot efficiently adapt to correlated data distribution and skewed query workloads. Tsunami \cite{tsunami} extends Flood by utilizing query skew to partition data space into some regions, and then further dividing each region based on  data correlations. However, simply choosing a subset of dimensions could degrade the performance as dimensionality increases.  These studies are not discussed further as we do not assume a known query workload.  

A multi-dimensional learned (ML) index \cite{davitkova2020ml} combines the idea of iDistance \cite{idistance} and RMI. It first partitions data into clusters, and then identifies the cluster center as the reference point. After all data points are represented in a one-dimensional space based on the distances to the reference point, RMI can be applied. Different from iDistance, ML uses a scaling value rather than a constant to stretch the data range. However, points along a fixed radius  have the same value after the transformation, leading to many irrelevant points to be checked.  ML does not support data updates. \cbl{Note that ML cannot be directly applied in metric spaces since reference points
selection is realized by the KMeans algorithm\cite{cluster2}. The reference point that is the mean of points in the clusters may not be in the dataset. It is not always possible to create ``artificial” objects in metric datasets\cite{omni}.}

Different from finding a sort order over multi-dimensional data and then learning the CDF, a reinforcement learning based R-tree for spatial data (RLR-Tree) \cite{rlr} uses machine learning techniques to improve on the classic R-tree index.  Instead of relying on hand-crafted heuristic rules, RLR-Tree models two basic operations in R-tree, \ie choosing a subtree for insertion and splitting a node, as \textit{Markov decision process} \cite{markov}, so reinforcement learning models can be applied. Because it does not need to modify the basic structure of the R-tree and query processing algorithms, it is easier to be deployed in the current databases systems than the learned indexes.  However, due to the curse of dimensionality, the minimum bounding rectangle (MBR) for a leaf node (even in an optimal R-tree) can be nearly as large as the entire data space, such that the R-tree becomes ineffective. 
Similar to RLR-Tree, Qd-Tree \cite{qdtree} uses \cbl{the} reinforcement learning to optimize the data partitioning strategy of kd-tree based on a given query workload, and suffers from the same problem. These studies are not discussed further since they are out of our scope.

\begin{table}[t]
    \renewcommand\arraystretch{1.5}
	\caption{List of key notations}
	\vspace{-2ex}
	\label{tab:Commonly}
	\footnotesize
	\begin{center}
	\begin{tabular}{|c|c|}
		\hline   
		\textbf{Notation} & \textbf{Description}\\ 
		\hline 
		$P$ & The dataset\\
		\hline 
		$\mathbb{U}, dist, d$ & The data space, distance metric,  dimensionality\\
		\hline 
		$p$, $q$ & A data  point, a query point\\
		\hline 
		$r$ &  Query radius \\
		\hline 
		$k$ &  The number of nearest neighbors \\
		\hline 
		$K$ &  The number of clusters\\
		\hline 
		$m$ &  The number of pivots\\
		\hline 
		$N$ &  The number of super rings\\
		\hline 
		$C_i$ &  The $i$th cluster\\
		\hline 
		$O_j^{(i)}$ &  The $j$th pivots in $i$th cluster\\
		\hline 
		$ dist\_max_j^{(i)}$ &  \makecell{The distance of the furthest object in $i$th cluster from \\the $j$th  pivot}\\
		\hline 
		$ dist\_min_j^{(i)}$ &  \makecell{The distance of the nearest object in $i$th cluster from \\the $j$th pivot}\\
		\hline 
	\end{tabular}
	\end{center}
	\vspace{-2ex}
\end{table}

\vspace{-0.1ex}
\section{Background} \label{sec:Background}
In this section, we first introduce some basic concepts and then the formal definition of  learned index for exact similarity search in metric spaces is presented. Table \ref{tab:Commonly} lists the key notations and acronyms used in this paper.

\begin{definition}[Metric space]
	\label{def:metricspace}
	A metric space is a pair $ (\mathbb{U}, dist)$, where $\mathbb{U}$ is a set of objects and $dist: \mathbb{U} \times \mathbb{U} \rightarrow [0, \infty)$ is a function so that $\forall p_1, p_2, p_3 \in \mathbb{U}$, the following holds:
	\begin{itemize}[leftmargin = 9pt]
		\item non-negativity:  $dist(p_1, p_2) \geq 0 ;$
		\item identity: $dist(p_1, p_2) = 0$  iff $p_1 = p_2 ;$
		\item symmetry: $dist(p_1, p_2) = dist(p_2, p_1) ;$
		\item triangle inequality: $dist(p_1, p_3) \leq  dist(p_1, p_2)+dist(p_2, p_3)$.
	\end{itemize}
	
\end{definition}

\begin{remark}
    \label{rmk:genericmetricspace}
	Metric space is generic because it only requires the distance function satisfying the above properties. \cbl{Vector space $\mathbb{R}^d$ with the Euclidean distance is a special metric space, where additional properties, \eg the dimension and coordinates, are specified, which can be used to accelerate the search. }
\end{remark}

In this paper, we consider three types of exact similarity queries in metric spaces: \cbl{the} range query, point query, and $k$NN query.

\begin{definition}[Range query]
	Given a set $P \subseteq \mathbb{U}$, a query object $q \in \mathbb{U}$, and a query radius $r \geq 0$, a range query returns all objects in $P$ within \cbl{the} distance  $r$ of $q$, \ie  $range(q, r) = \{p \in P| dist(p, q) \leq r\}$.  
\end{definition}
\begin{remark}
    Point query is a special case of range query with $r = 0$
and an arbitrary metric. In this case, we say $p=q$.
\end{remark}

\begin{definition}[$k$NN query]
	Given a set $P \subseteq \mathbb{U}$, a query object $q \in \mathbb{U}$, and a positive integer $k$, a $k$NN query returns a set of $k$ objects, denoted as $k$\textnormal{NN}$(q, k)$, such that $\forall p \in k$\textnormal{NN}$(q, k), p^\prime \in  P \setminus k$\textnormal{NN}$(q, k), dist(q, p) \leq dist(q, p^\prime).$
\end{definition}
\begin{example}
	Consider a word dataset $P = \{``fame", ``gain", $ $``aim",$ $ ``ACM"\}$ associated with \cbl{the} edit distance \cite{edit}. \cbl{A} range query $range(``game", 2)$ returns all words in $P$ within \cbl{the} edit distance 2 to $``game"$, \ie $\{``fame", ``gain"\}$. \cbl{The} $1$NN query $k$NN$(``game", 1)$ returns the nearest neighbor of $``game"$, \ie $\{``fame"\}$.
\end{example}


\begin{figure*}[htbp]
	\centering
	\includegraphics[width= \linewidth,height=5cm]{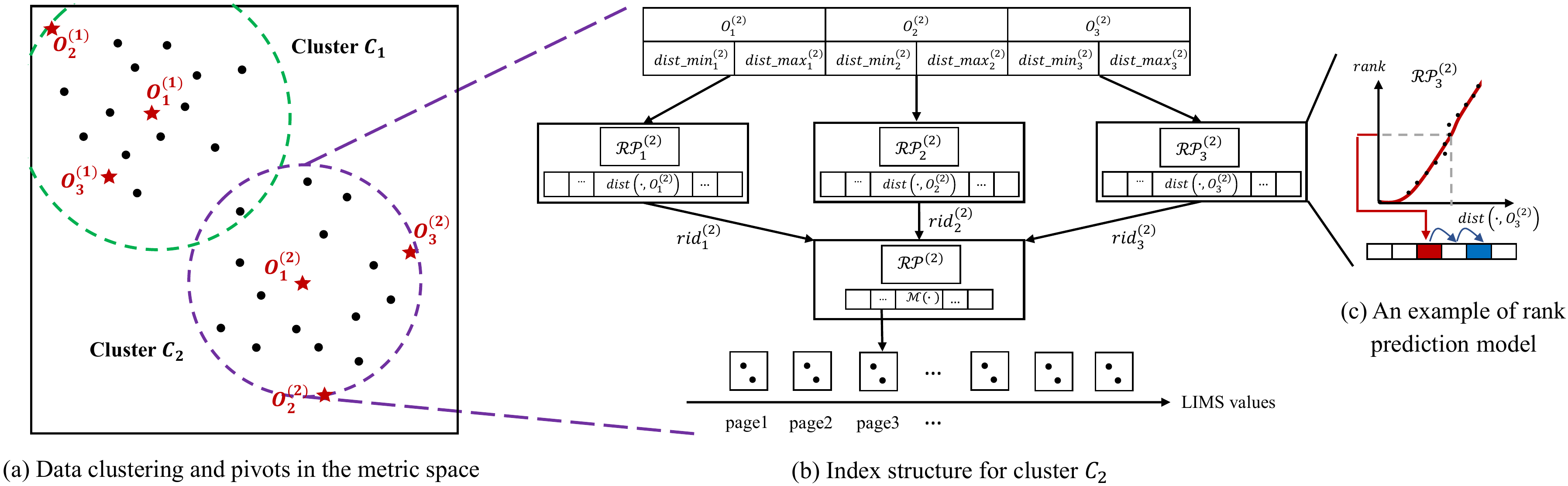}
	\vspace{-3ex}
	\caption{LIMS index structure} 
 	\vspace{-1ex}
	\label{fig:structure}
\end{figure*}
{
	\noindent \textsc{Problem Statement}. 	Let $ (\mathbb{U}, dist)$ be a metric space and $P = \{p_1, p_2,\ldots, p_n\} \subseteq \mathbb{U}$ be a set of objects.  The\textbf{ learned index for exact similarity search in metric spaces} is to learn an index structure for $P$ so that point query, range query and $k$NN query can be processed  accurately and efficiently. In addition, the index structure is supposed to support insertion and deletion operations.
	
	\label{def:observe1}
}

\section{LIMS} \label{sec:LIMS}

In this section, we first give an overview of the index structure of LIMS and then present everything needed to build LIMS. LIMS-based query processing will be discussed in Section \ref{sec:queryprocessor}.

\vspace{-1ex}
\subsection{Overview}
LIMS consists of three parts: \cbl{the} data clustering and pivot selection, \cbl{the} pivot-based mapping function and associated binary relationship, as well as rank prediction models. Fig. \ref{fig:structure} gives an overview of LIMS index structure in a metric space associated with  the Euclidean distance, although other metric spaces also apply for LIMS. 
LIMS first partitions the underlying data into a set of clusters, \cbl{\eg 2 clusters in Fig. \ref{fig:structure}(a),} so that each of them follows a relatively uniform data distribution, and then a set of data-dependent pivots for each cluster are picked, \eg \cbl{3 pivots $O_1^{(1)}, O_2^{(1)}$ and $O_3^{(1)}$} for cluster $C_1$ \cbl{(detailed
discussions refer to Section \ref{subsec: cluster}).} Then, LIMS maintains a learned index for each cluster separately. \cbl{Since the index structure is same for each cluster, we take cluster $C_2$ for example. } LIMS computes the distances from each object in the cluster to the well-chosen pivots, \eg $dist(\cdot, O_3^{(2)})$ in Fig. \ref{fig:structure}(b). \cbl{The} maximum and minimum distances from each pivot to the corresponding objects, \eg $dist\_max_3^{(2)}$ and $dist\_min_3^{(2)}$  are stored so as to support efficient queries. \cbl{Since objects are sorted by distance values,} LIMS can learn a series of rank prediction models, \eg $\mathcal{RP}_3^{(2)}$ in Fig. \ref{fig:structure}(c), \cbl{for quick computation of the rank of an object} given its distance to the pivot. After that, a well-defined pivot-based mapping function $\mathcal{M}$  is called to transform each object into an ordered set (Definition \ref{pivot-based mapping}, \ref{def:relation}, Section  \ref{sect:index}). We call elements in this set \textit{LIMS values}. Finally, we \cbl{physically maintain all data objects sequentially on disk in ascending order of their LIMS values} and the relationship between LIMS values and the addresses of data objects in  disk pages can be learned by another rank prediction model, \eg $\mathcal{RP}^{(2)}$ (detailed
discussions refer to Section \ref{sect:index}).

In what follows, we first focus on the specific learned index structure for each cluster, and then turn back to clustering and pivot selection methods. In other words, we assume that the data space has been partitioned, and the pivots in each cluster have been determined.

\subsection{Index Structure}
\label{sect:index}
Suppose that $K$ clusters, say $\{C_1, C_2, \ldots, C_K\}$, and $m$ pivots for each cluster, say $\{O_1^{(i)}, O_2^{(i)}, \ldots, O_m^{(i)}\}$, $i= 1, \ldots, K$, have been determined. Then,  for each cluster $C_i, i = 1, \ldots, K$ and pivot $O_j^{(i)}, j = 1, \ldots, m$, all data objects \cbl{(unique identifiers)} are sorted in ascending order of their distances to the pivot. \cbl{Based on $m$ sorted lists in  $C_i$, LIMS learns $m$} rank prediction models. For model reuse, we define it formally as follows:

\begin{definition}[Rank]
	Let $A$ be a finite multiset drawn from an ordered set $B$. For any element $x \in A$, we define the rank of $x$ as the number of elements smaller than $x$, \ie
	\begin{equation}
		rank(x) = |\{x^\prime\in A| x^\prime < x\}|.
	\end{equation}
	\begin{example}
		Let $A = \{1.5, 1.5, 1.8, 1.8, 2.0\}$ be a multiset of \cbl{distance values to a given pivot sorted in ascending order.} Then, $rank(1.5) = 0, rank(1.8) = 2, rank(2.0) = 4$.
	\end{example}
\end{definition}
\begin{definition}[Rank Prediction Model]
	Let $A$ be a finite multiset drawn from an ordered set $B$. Rank prediction model $\mathcal{RP}: B \rightarrow [0, +\infty)$ is a function learned from $\{(x, rank(x))\}_{x \in A}$, so that it can predict the rank for any element $x \in B$, \ie
	\begin{equation}
		rank(x) \approx \mathcal{RP}(x).
	\end{equation}
\end{definition}
\begin{remark}
	In the strict sense, there is no definition of rank for element $x \in B \backslash A$. What we want to express here is the number of elements in $A$ smaller than $x$. Without confusion, we still use rank for simplicity.
\end{remark}

\cbl{A series of rank prediction models $\mathcal{RP}_j^{(i)}$ (\aka one-dimensional learned indexes) can be trained as follows:} let $D_j^{(i)} = \{dist(p, O_j^{(i)})\}_{p \in C_i}$, then the training set is $\tilde{D}_j^{(i)} =  \{(x, rank(x))\}_{x \in D_j^{(i)}}$. Let $\mathcal{RP}_j^{(i)}$ be a polynomial function of $x$, and \cbl{the loss function $\mathcal{L}$  be the \textit{squared error} as follows:}
\begin{equation}
	\label{eq:loss1}
	\mathcal{L} = \sum_{\tilde{D}_j^{(i)}} \left(\mathcal{RP}_j^{(i)}\left(x\right) - rank\left(x\right)\right)^2.
\end{equation}
\cbl{Then, the rank prediction models can be determined  by minimizing $\mathcal{L}$ loss using the \textit{gradient descent}. }   For example, if the degree of polynomial is 2, then $\mathcal{RP}_j^{(i)}(x)= ax^2 + bx + c$, where $a, b, c$ are parameters to be learned. 
After $\mathcal{RP}_j^{(i)}$ is trained, we can get \cbl{the} approximate rank for any object $p \in \mathbb{U}$ with \cbl{the} distance in $D_j^{(i)}$ and the error can be easily corrected by exponential search in $O(\log_{}err)$ time, where $err$ is the difference between the estimated rank and the correct rank. For object $p \in \mathbb{U}$ with \cbl{the} distance not in $D_j^{(i)}$, we can also use $\mathcal{RP}_j^{(i)}$ and exponential search to find its corresponding rank, \ie the rank of the first element larger than $dist(p, O_j^{(i)})$ in $D_j^{(i)}$ with the same time complexity.

\begin{figure}[htbp]  
	\centering    
	\subfloat [$m = 1, N = 3$]
	{\label{fig:ringa}
		\begin{minipage}[t]{0.4\linewidth}
			\centering          
			\includegraphics[width=\linewidth]{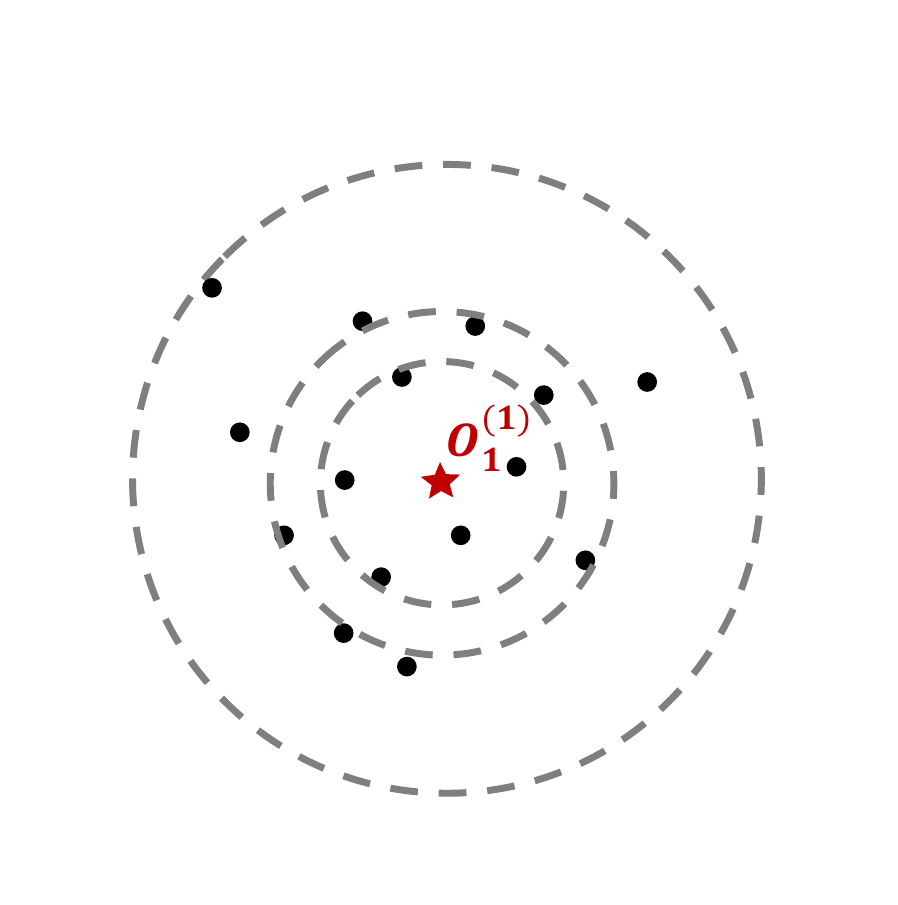}   
		\end{minipage}%
	}
	\subfloat [$m = 2, N = 3$]
	{\label{fig:ringb}
		\begin{minipage}[t]{0.4\linewidth}
			\centering      
			\includegraphics[width=\linewidth]{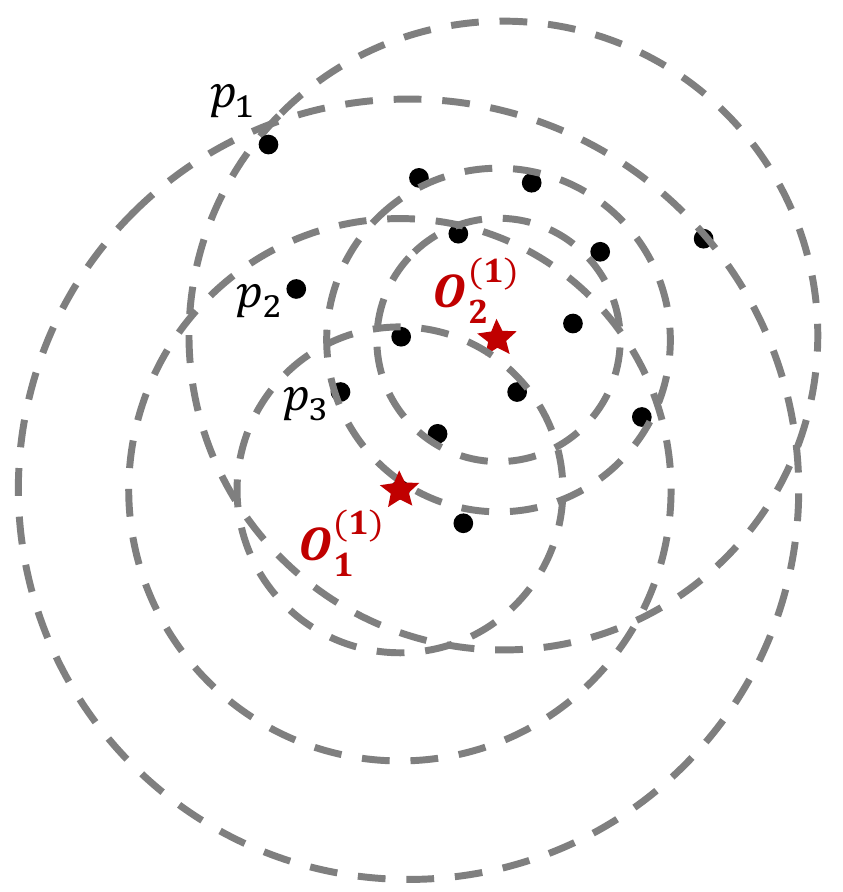}   
		\end{minipage}
	}%
	\caption{Examples of data partitioning}
	\label{fig:rings}  
\end{figure}

\cbl{In order to accelerate query processing, to be introduced in Section \ref{sec:queryprocessor}, LIMS divides ranks of data objects into $N$ equal parts, \ie  makes the data in the cluster covered by $N$ super rings as evenly as possible.}
\cbl{Fig. \ref{fig:rings} gives two examples in a metric space with the Euclidean distance. In Fig. \ref{fig:rings}(a), the number of pivots and rings are specified as 1 and 3, respectively, so we partition all data objects in the cluster into 3 rings \wrt the pivot $O_1^{(1)}$ \cred{such that each ring includes about 5 data objects}. In Fig. \ref{fig:rings}(b), the number of pivots and rings are specified as 2 and 3, respectively, so we partition all data objects into 6 rings, where 3  \wrt the pivot $O_1^{(1)}$ and 3 \wrt $O_2^{(1)}$.} \cred{ In
this way, LIMS effectively avoids that lots of objects whose
ring IDs defined in Equation (\ref{eq:ringID}) are same, while a few objects have different ring IDs.}
\vspace{-0.5cm}

The ring ID, \ie which ring the data object is located, denoted as  $rid_j^{(i)}$, can be computed by:
\begin{equation}
\label{eq:ringID}
	rid_j^{(i)}(p) =\left \lfloor \frac{rank\left(dist\left(p, O_j^{(i)}\right)\right)}{ \left \lceil |D_j^{(i)}| / N \right \rceil }\right\rfloor.
\end{equation}
When the above steps are completed, each data object is equipped with $m$ ring IDs.
Then, LIMS designs a novel pivot-based mapping function $\mathcal{M}$ to transform all data objects in the metric space into ordered sets with an associated binary relation $\leq$. The formal definitions are as follows.

\begin{definition}[Pivot-based mapping function]
\label{pivot-based mapping}
	Given a cluster $C_i$, $i = 1, \ldots, K$, and its corresponding pivots $O_j^{(i)}$, $j = 1, \ldots, m$, let $\{rid_1^{(i)}, \ldots, rid_m^{(i)}\}$ be a set of $m$ ring ID functions. Then, for any data object $p \in \mathbb{U}$, we define a pivot-based mapping function $\mathcal{M}$ as follows:
	\begin{equation}
		\mathcal{M}(p) = rid_1^{(i)}(p)\oplus rid_2^{(i)}(p) \oplus \cdots \oplus rid_m^{(i)}(p).
	\end{equation}
	We call $\mathcal{M}(p)$ LIMS value of $p$.
	
\end{definition}

\begin{example}
\label{exp:rid}
	\cbl{Consider the cluster in Fig. \ref{fig:rings}(b). The ring IDs of $p_1, p_2$ and $p_3$ are $rid_1({p_1}) = 2$ and $rid_2({p_1}) = 2$, $rid_1({p_2}) = 1$ and $rid_2({p_2}) = 2$, $rid_1({p_3}) = 0$ and $rid_2({p_3}) = 1$, respectively. The ring IDs of other data objects can be computed similarly. According to the definition of pivot-based mapping function, $\mathcal{M}({p_1}) = 2\oplus 2, \mathcal{M}({p_2}) = 1\oplus 2$ and $ \mathcal{M}({p_3}) = 0\oplus 1$.}
\end{example}
In order to build a learned index structure on LIMS values, we need to impose a binary relationship $\leq$ between LIMS values as follows:

\begin{definition}[Binary Relation $\leq$]
	\label{def:relation}
	Let $S$ be a multiset of LIMS values in cluster $C_i$ with $m$ pivots, $i = 1, \ldots, K$. Then, $S$ can be ordered as follows:
	\begin{equation}
		rid_1^{(i)}(p) \oplus \cdots \oplus rid_m^{(i)}({p}) \leq rid_1^{(i)}({p}^\prime)\oplus \cdots \oplus rid_m^{(i)}({p}^\prime)
	\end{equation}
	if and only if condition $(1)$ or $(2)$ is satisfied.
	\begin{enumerate}
		\item $\forall j \in \{1,\ldots,m\}$
		\begin{equation}
			rid_j^{(i)}(p) = rid_j^{(i)}(p^\prime);
		\end{equation}
		\item 
		$\exists k \in \{1,\ldots,m\} ~ \text{such that} $
		\begin{equation}
			rid_j^{(i)}(p) = rid_j^{(i)}(p^\prime) ~\text{for} ~j<k 
			~\text{and} ~rid_k^{(i)}(p) < rid_k^{(i)}(p^\prime),
		\end{equation}
	\end{enumerate}
	where $``="$ and $``<"$ in conditions is the order of natural numbers.
\end{definition}
It's straightforward to prove that the binary relation $\leq$ in Definition \ref{def:relation} is well-defined, \ie it is reflexive, antisymmetric and transitive \cite{order}, so $(S, \leq)$ is an ordered set. In our implementation, we use the concatenation of ring IDs as LIMS value, which satisfies conditions in Definition \ref{def:relation} obviously.
\begin{example}
	\cbl{Reconsider Example \ref{exp:rid}, LIMS values of $p_1, p_2$ and $p_3$ are sorted as follows: $\mathcal{M}({p_3})< \mathcal{M}({p_2})< \mathcal{M}({p_1})$.}
\end{example}
Now that all objects in the metric space are transformed into \cbl{the} corresponding ordered sets, we can sort them sequentially in ascending order of LIMS values, and store data in a number of disk pages with each page fully utilized. To quickly locate the addresses of data objects, LIMS learns a rank prediction model. Specifically, let $D^{(i)} = \{\mathcal{M}(p)\}_{p \in C_i}$ be a multiset of LIMS values and $\tilde{D}^{(i)}= \{(x, rank(x))\}_{x \in D^{(i)}}$, then a rank prediction model $\mathcal{RP}^{(i)}$ can be learned based on $\tilde{D}^{(i)}$.  Similar to the loss function in Equation (\ref{eq:loss1}), we still try to minimize the squared error.
After $\mathcal{RP}^{(i)}$ is trained, we can get \cbl{the} approximate rank (address) for any object $p \in \mathbb{U}$ with \cbl{the} LIMS value $\mathcal{M}(p) \in D^{(i)}$. The error can be easily corrected by exponential search that stops when the first occurrence of $\mathcal{M}(p)$  is found. For object $p \in \mathbb{U}$ with \cbl{the} LIMS value not in $D^{(i)}$, we can also use $\mathcal{RP}^{(i)}$ and exponential search to find its corresponding rank, \ie the position where the first occurrence of the element larger than $\mathcal{M}(p)$ in $D^{(i)}$.

\subsection{Data Clustering and Pivot Selection}
\label{subsec: cluster}
\cbl{As mentioned in \cite{kraska2018case}, 
one challenge of replacing traditional tree-like indexes with learned indexes is that it is difficult to approximate complex data distributions with a single model. If we construct a learned index on the whole dataset directly, the function to be learned would be very steep in regions where the data objects are dense, while very gentle for sparse regions. Such a complicated relationship can be fit by a neural network, but it will incur expensive query costs in practice. Based on the observation that real-life data are usually clustered and correlated \cite{local},  LIMS groups the underlying data into a number of clusters and maintains a learned index for each cluster instead. } This strategy gives two advantages: first, the data distribution of each cluster becomes simpler, which simplifies the model to be learned (\eg
a polynomial function); second, simple models have lower query and (re-)construction costs.  In this paper, we simply adopt the k-center algorithm \cite{kcenter}, a simple yet effective algorithm that guarantees to return a 2-approximate optimal centroid set. We can follow the same steps to build LIMS on top of other clustering algorithms such as the kMeans\cite{cluster2}, which can potentially further improve our approach. \cbl{Different from a general clustering problem where the number of clusters can be flexible,  the number of clusters used for indexing affects the index structure and search performance.} Therefore, we propose a statistic to determine the number of clusters, to be discussed in detail in Section \ref{sec:lastpiece}. For now, we assume that the number of clusters has been determined.

\cbl{Once the clusters are obtained, LIMS picks a few data objects, \cred{named pivots \cite{pivots}}, for each cluster and computes distances from each data object in the cluster to pivots. 
This strategy gives three advantages: first, we can use the triangle inequality on these pre-computed distances to prune the search space; second, learned indexes can be built naturally on these 1-dimensional distance values; third, re-distributing data with reference to well-chosen pivots may effectively ease the curse of dimensionality} because metric search performance depends critically on the intrinsic dimension, a property relying on the data distribution itself, as opposed to the dimension where the data is represented\cite{deflating, intrinsic1,intrinsic2,omni}. For example, the intrinsic dimension of a plane is two no matter if it is embedded in a high dimensional space.
While LIMS is not dependent on the underlying pivot selection  method, the number  and locations of pivots have an influence on the retrieval performance. The more high-quality pivots there are, the more information they provide and the higher pruning power for query processing. 
However, the time taken for checking pruning conditions also increases (For LIMS, it mainly refers to the cost of generating search intervals, to be discussed in detail in Section \ref{sec:queryprocessor}). Extensive methods for pre-defining an optimal set of pivots have been proposed. A good survey can be found in \cite{ZhuYifang}. In our implementation, we adopt the farthest-first-traversal (FFT) algorithm\cite{kcenter} because of its linear time and space complexity. 

\section{LIMS-based Query Processing} \label{sec:queryprocessor}
In this section, we proceed to present query processing algorithms for point, range and $k$NN queries using LIMS. \cbl{Section \ref{updates}} explains  dynamic updates. \cbl{Section \ref{sec:lastpiece}} discusses the choice of $K$. Without loss of generality, we assume there is no two objects in $P$ are totally same.

\vspace{-1ex}
\subsection{Range Query}
\label{sec:rangequery}

\begin{algorithm}[t]
	\caption{Range Query}
	\footnotesize
	\label{alg:range}
	\LinesNumbered 
	\KwIn{ $ {q}$: a query object; $r$: a query radius}
	\KwOut{$\mathcal{S}$: objects in $P$ satisfying  $dist(p,q) \leq r$ }
	Let $flag[K]$ be an array of all $TRUE$, $rid\_min[K][m], rid\_max[K][m]$ be arrays of all $0$\;
	\label{fun1}\For {each cluster $C_i$} 
	{
		\For{ each pivot ${O}_j^{(i)}$}
		{
			\If {$dist({O}_j^{(i)}, {q}) > dist\_max_j^{(i)} + r$ OR $dist({O}_j^{(i)}, {q}) < dist\_min_j^{(i)} - r$} 
			{\label{condition} 
				$flag[i] = FALSE$\; 
				$break$\;
			}
		}
	} \label{fun11}
	
	\For{each TRUE cluster $C_i$ }
	{\label{fun2}
		\For{each pivot ${O}_j^{(i)}$}
		{

			$r_{min} \leftarrow \max\{ dist({O}_j^{(i)}, {q}) - r, dist\_min_j^{(i)}\} $\;
			$r_{max} \leftarrow \min\{   dist({O}_j^{(i)}, {q}) + r, dist\_max_j^{(i)}\} $\;
			
			$rank_{min}^\prime, rank_{max}^\prime \leftarrow \mathcal{RP}_j^{(i)}(r_{min}), \mathcal{RP}_j^{(i)}(r_{max})$\;\label{11}
			$rank_{min}\leftarrow ExpSearch(rank_{min}^\prime, r_{min})$\; 
			$rank_{max} \leftarrow ExpSearch(rank_{max}^\prime, r_{max}); $ \label{13} 
			/*assume $rank_{min} < rank_{max}$, otherwise discard $C_i$ */\;
			$rid\_min[i][j] \leftarrow rid_j^{(i)}(rank_{min})$\; \label{15}
			\eIf{$D_j^{(i)}[rank_{max}] = r_{max}$}
			{
				$rid\_max[i][j]\leftarrow rid_j^{(i)}(rank_{max})$\;
			}
			{
				$rid\_max[i][j]\leftarrow rid_j^{(i)}(rank_{max} - 1)$\;
			}

		}
		
	} \label{fun22} 
	generate LIMS-value ranges $\mathcal{R}$ based on $rid\_min$ and $rid\_max$\; \label{fun3} 
	\For{each range $I \in R$}
	{\label{fun4} 
		$lbound^\prime, ubound^\prime \leftarrow \mathcal{RP}^{(i)}(I.left), \mathcal{RP}^{(i)}(I.right)$\;
		$lbound \leftarrow ExpSearch(lbound^\prime, I.left)$\;
		$ubound \leftarrow ExpSearch(ubound^\prime, I.right)$\;
		\eIf{$D^{(i)}[ubound] = I.right$}
		{
			$ubound \leftarrow ExpSearch2(ubound, I.right)$\;
		}
		{
			$ubound \leftarrow ubound - 1$\;
		}
		/* assume $lbound < ubound$, otherwise discard $I$ */\;
		add to $\mathcal{P}$ all unvisited pages from $\lfloor lbound/ \Omega \rfloor$ to $ \lfloor ubound/ \Omega \rfloor$\;
		
	}
	
	add to  $\mathcal{S}$ all objects saved in $\mathcal{P}$ satisfying  $dist(p,q) \leq r$\; \label{refinement} 
	\Return {$\mathcal{S}$} \label{fun44} 

\end{algorithm}

Given a query object $q$, query radius $r$ and dataset $P$,  \cbl{a} range query is to retrieve all objects in $P$ within \cbl{the} distance  $r$ of $q$, \ie $range(q, r) = \{p \in P| dist(p, q) \leq r\}$.  Algorithm \ref{alg:range}  outlines the range query processing, which consists of 4 subprocedures. \textit{TriPrune (Line \ref{fun1}-\ref{fun11}):}  prune irrelevant clusters by triangle inequality; \textit{AreaLocate (Line \ref{fun2}-\ref{fun22}):} determine affected areas of relevant clusters with the help of rank prediction functions $\mathcal{RP}_j^{(i)}$; \textit{IntervalGen (Line \ref{fun3}):} generate search intervals on LIMS values; \textit{PosLocate (Line \ref{fun4}-\ref{fun44}):} locate positions of records in the disk by rank prediction functions $\mathcal{RP}^{(i)}$.

\noindent\textbf{\textit{TriPrune (Line \ref{fun1}-\ref{fun11}).}} 
The algorithm starts by computing the distances between the query $q$ and the pivots, and then utilizes triangle inequality property in the metric space to prune a number of irrelevant clusters and thus accelerate the search. Specifically, according to triangle inequality, an object $p$ in cluster $C_i$ may fall into the query range, it must satisfy the following: $\forall j \in \{1,\ldots,m\}$,
\begin{equation}
	\label{eq:1}
	dist(O_j^{(i)}, q) - r \leq dist(O_j^{(i)}, p) \leq dist(O_j^{(i)}, q) + r.
\end{equation}
Recall that LIMS also maintains both maximum distance $dist\_max_j^{(i)}$ and minimum distance $dist\_min_j^{(i)}$ of the cluster, hence, for any object $p$ in the cluster, we must have:  $\forall j \in \{1,\ldots,m\}$,
\begin{equation}
	\label{eq:2}
	dist\_min_j^{(i)} \leq dist(O_j^{(i)}, p) \leq dist\_max_j^{(i)}.
\end{equation}
Combing Equations (\autoref{eq:1}) and (\autoref{eq:2}), we can derive that a cluster $C_i, i = 1, \ldots, K$ is needed to be further checked if and only if the following condition (Line \ref{condition}) is satisfied:


\begin{equation}
	\label{eq:intersection}
	\begin{split}
	\bigwedge_{j = 1}^m ~~&[dist({O}_j^{(i)}, {q}) \leq dist\_max_j^{(i)} + r \\
	&\wedge~ dist({O}_j^{(i)}, {q}) \geq dist\_min_j^{(i)} - r].
	\end{split}
\end{equation}

\noindent\textbf{\textit{AreaLocate (Line \ref{fun2}-\ref{fun22}).}}
For a cluster $C_i$ that needs to be searched, LIMS first determines the affected areas in the metric space according to the following equations:
\begin{align}
	{r_{min}}_j^{(i)} &= \max\{dist({O}_j^{(i)}, {q}) - r, dist\_min_j^{(i)}\};\label{eq:rmin}\\
	{r_{max}}_j^{(i)} &= \min\{dist({O}_j^{(i)}, {q}) + r, dist\_max_j^{(i)}\} \label{eq:rmax},
\end{align}
where $j = 1,\ldots, m$. Then, LIMS invokes corresponding rank prediction models $\mathcal{RP}_j^{(i)}$ to predict the min and max ranks of affected areas and the error is fixed via exponential search (Line \ref{11}-\ref{13}). Min ring ID can be easily calculated by calling $rid_j^{(i)}$ function, while the max ring ID is divided to 2 cases in order to narrow down search range as much as possible (Line \ref{15}-\ref{fun22}).

\noindent\textbf{\textit{IntervalGen (Line \ref{fun3}).}}
Instead of doing intersection of several candidate sets in the metric space via costly distance computations, LIMS reduces the search space by doing intersection of LIMS-value intervals directly. Specifically, for a relevant cluster $C_i, i = 1, \ldots, K$, let $L_j^{(i)} = \{rid\_min[i][j], $ $rid\_min[i][j] + 1, \ldots, rid\_max[i][j]$\}, $j = 1,\ldots, m -1$ and $L_m^{(i)} = \{rid\_min[i][m],$ $ rid\_max[i][m] \}$. Depth-first search (DFS) is run on a directed acyclic graph (DAG) composed of vertexes $ \cup_{j = 1}^m L_j^{(i)}$ and fully connected edges from $L_j^{(i)}$ to $L_{j + 1}^{(i)}$, $j = 1, \ldots, m -1$, to find all paths from $L_1^{(i)}$ to $L_m^{(i)}$. These paths form a total of $\prod_{j = 1}^{m - 1}|L_j^{(i)}|$ LIMS-value search ranges. It is through \textit{IntervalGen} that the number of data objects to be accessed is significantly reduced, while the pruning cost remains low. Here is an example of this step.

\begin{figure}[t]
		\centering
		\includegraphics[width=0.5\linewidth]{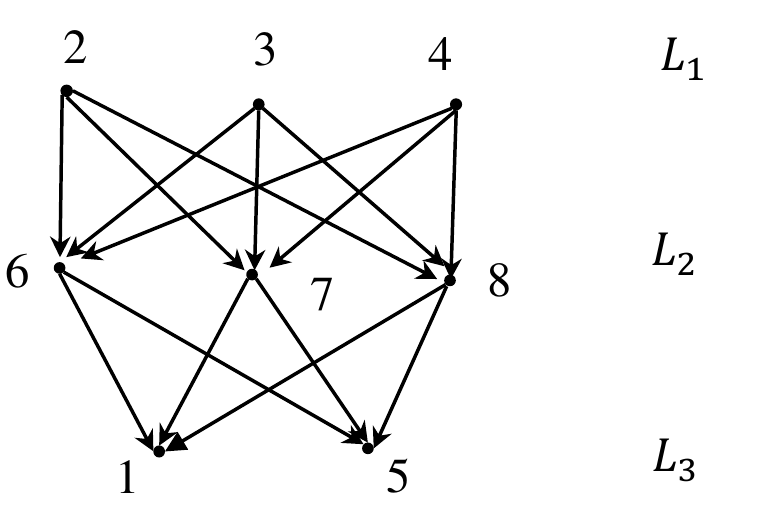}
			\vspace{-1ex}
 \caption{An illustration of finding  LIMS-value search ranges } 
 \vspace{-3ex}
		\label{fig:dfs}
	
	\end{figure}

\begin{example}
	Consider a cluster with $m = 3$ pivots and $N = 10$ rings. Suppose it is relevant to a range query such that the minimum and maximum ring IDs of affected region w.r.t. the 1st, 2nd and 3rd pivots are $rid\_min_1 = 2$, $rid\_max_1 = 4$, $rid\_min_2 = 6$, $rid\_max_2 = 8$, $rid\_min_3 = 1$, $rid\_max_3 = 5$, respectively, i.e., $L_1 = \{2, 3, 4\}, L_2 = \{6, 7, 8\}$ and $L_3 = \{1, 5\}$. Then, LIMS-value search ranges can be computed by running DFS on the DAG shown in Figure \ref{fig:dfs}. The final search ranges are the union of $[2\oplus6\oplus1, 2\oplus6\oplus5]$, $[2\oplus7\oplus1, 2\oplus7\oplus5]$, $[2\oplus8\oplus1, 2\oplus8\oplus5]$, $[3\oplus6\oplus1, 3\oplus6\oplus5]$, $[3\oplus7\oplus1, 3\oplus7\oplus5]$, $[3\oplus8\oplus1, 3\oplus8\oplus5]$, $[4\oplus6\oplus1, 4\oplus6\oplus5]$, $[4\oplus7\oplus1, 4\oplus7\oplus5]$ and $[4\oplus8\oplus1, 4\oplus8\oplus5]$.

\end{example}


\noindent\textbf{\textit{PosLocate (Line \ref{fun4}-\ref{fun44}).}}
For each search range, the position of lower bound in the disk can be easily located via rank prediction model $\mathcal{RP}^{(i)}$ and exponential search. $\Omega$ is the maximum number of objects each page can hold. The upper bound is divided into 2 cases to make sure the results correct. For example, it's possible that different objects have the same LMIS value, so we find the last occurrence of the LIMS value via another exponential search, denoted as $ExpSearch2$, to guarantee no objects missed (\ie no false negatives).  Finally, all retrieved objects are further refined in the refinement step (Line \ref{refinement}), where exact distance computations are performed.
	
\noindent \cbl{
\textbf{Correctness.} To prove that Algorithm  \ref{alg:range}  offers exact answers for a range query, we need to show that 1) all data objects in the result set satisfy $dist(p,q) \leq r$, \ie no false positives; and 2) no objects satisfying $dist(p,q) \leq r$ are missed \ie no false negatives. Obviously, no false positives can be returned because a final refinement step is applied, where the exact distance computations are performed to guarantee all data objects in the result set satisfy $dist(p,q) \leq r$. We prove there are no false negatives by contradiction. Assume that there exists an object $p \in C_i$ that satisfies $dist(p,q) \leq r$ but is not returned. According to the triangle inequality, we know that $\forall j \in \{1, \ldots, m\}$, $dist(O_j^{(i)}, q) \leq dist(O_j^{(i)}, p) + dist(p, q) \leq dist\_max_j^{(i)} + r.$
Similarly, we can derive $dist({O}_j^{(i)}, {q}) \geq dist\_min_j^{(i)} - r$. It indicates that the cluster $C_i$ (and thus $p$) will not be pruned in \textit{TriPrune} step (Equation (\ref{eq:intersection})). Since $dist(O_j^{(i)}, p) \leq  dist(O_j^{(i)}, q) + r$ and $dist(O_j^{(i)}, p) \leq dist\_max_j^{(i)}$, we know that $dist(O_j^{(i)}, p) \leq \min\{dist({O}_j^{(i)}, {q}) + r, dist\_max_j^{(i)}\} = {r_{max}}_j^{(i)}$ (Equation (\ref{eq:rmax})). Similarly, we can derive $dist(O_j^{(i)}, p) \geq {r_{min}}_j^{(i)}$ (Equation (\ref{eq:rmin})). Although the rank prediction models $\mathcal{RP}_j^{(i)}$ have an error, the error is fixed by exponential search. Thus, we have $rid\_min[i][j] \leq rid_j^{(i)}(p) \leq rid\_max[i][j]$ after \textit{AreaLocate} step. Denote by $l_j^{(i)}$ the value in $L_j^{(i)}$ satisfying  $l_j^{(i)} = rid_j^{(i)}(p)$, then $\mathcal{M}(p)$ can be written as $l_1^{(i)}\oplus \cdots \oplus l_m^{(i)}$. According to the procedure of generating LIMS-value search ranges in \textit{IntervalGen} step, there must exist a range $[l_1^{(i)}\oplus \cdots \oplus l_{m-1}^{(i)} \oplus rid\_min[i][m], l_1^{(i)}\oplus \cdots \oplus l_{m-1}^{(i)} \oplus rid\_max[i][m]]$. Based on the binary relation in Definition \ref{def:relation},  we know that $\mathcal{M}(p)$ falls in the above search range, and thus in the union of all search ranges after \textit{IntervalGen} step. The rank prediction model $\mathcal{RP}^{(i)}$ in \textit{PosLocate} step may have an error, but the error is fixed by exponential search. Hence, exact distance computations between all objects in the search ranges and the query object are performed. $p$ will be returned, which leads to a contraction. Therefore,  Algorithm  \ref{alg:range} can answer the range query correctly. 
}

\begin{figure*}[htbp]
	\centering
	\includegraphics[width= \linewidth]{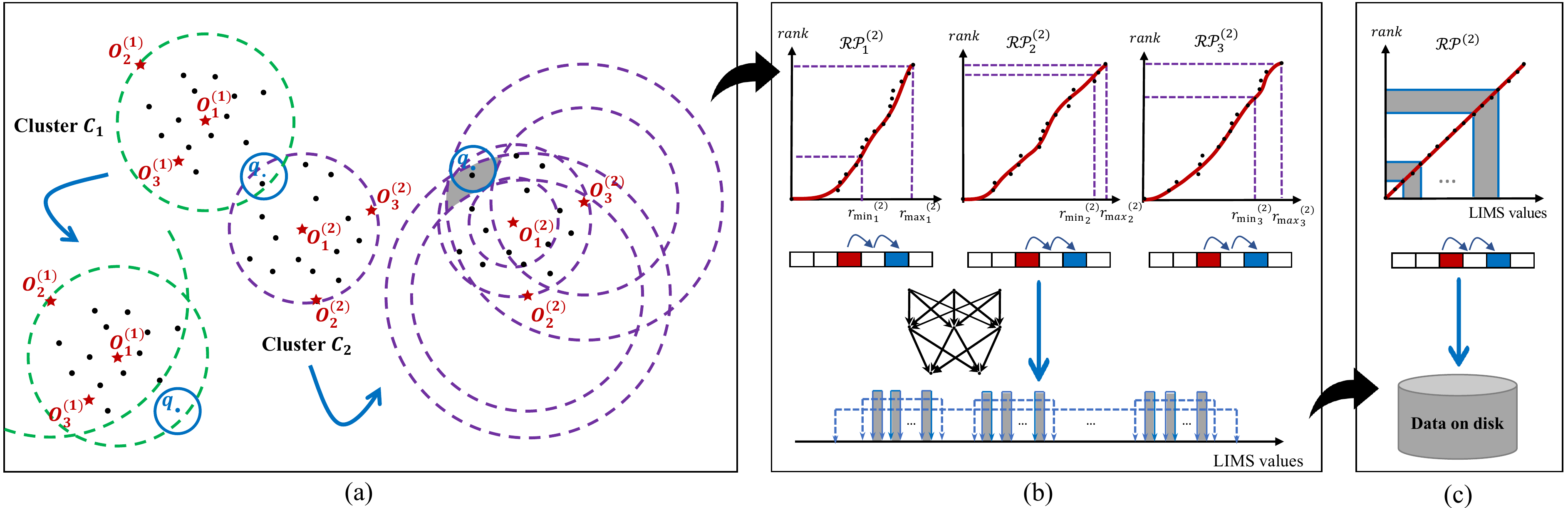}
		\vspace{-2ex}
	 \caption{An example of range query based on LIMS}
	 \vspace{-1ex}
	\label{fig:1}
\end{figure*}
\noindent \textbf{Query Cost.} \textit{TriPrune} takes $O(mKD)$ time, where $D$ represents the cost of distance computation. The cost of \textit{AreaLocate} depends on  rank prediction models. We use $O(\mathcal{RP})$ and $O(\log_{}{err})$ to denote the prediction cost of $\mathcal{RP}_j^{(i)}$ or $\mathcal{RP}^{(i)}$, and the cost of fixing error incurred by models via exponential search, respectively. Hence, we need $O(mK(\mathcal{RP} + \log_{}{err}))$ time to locate affected areas of relevant clusters. The cost of \textit{IntervalGen} is from running DFS on DAG, which takes $O(|V|+|E|)$ time, where $|V|$ is the number of vertexes and $|E|$ is the number of edges. Similar to \textit{AreaLocate} subprocedure, \textit{PosLocate} takes $O(|\mathcal{R}|(\mathcal{RP} + \log_{}{err}))$ time, where $|\mathcal{R}|$ is the number of LIMS-value search ranges. Generally, $|\mathcal{R}| > |E| > |V|$. In addition, we need to access disk pages in $\mathcal{P}$ to refine and retrieve finial result, which takes $O(|\mathcal{P}|\Omega D)$ time. Therefore, the overall query time is $O((mK + |\mathcal{R}|)(\mathcal{RP} + \log_{}{err})+(mK + |\mathcal{P}|\Omega)D)$.

\begin{example}
Figure \autoref{fig:1} is an example of LIMS-based range query. It is straightforward to see that cluster $C_1$ does not satisfy \cbl{$dist(O_2^{(1)}, q) \leq dist\_max_2^{(1)}+ r$}, and thus can be discarded from further processing directly, while cluster $C_2$ cannot be discarded. As for $C_2$, LIMS first inputs the min and max boundaries (\ie purple dashed lines in Figure \ref{fig:1}(a)) of affected areas into corresponding rank prediction models $\mathcal{RP}_j^{(2)}$ and ring ID functions $rid_j^{(2)}$, $j = 1, 2 ,3$. Then, LIMS runs DFS on DAG to transform the intersection of several candidate sets in the metric space (the grey region in Figure \ref{fig:1}(a)) to the intersection of intervals (the grey region in Figure \ref{fig:1}(b)), to significantly reduce the number of distance computations and page accesses. Next, positions of objects on disk are estimated by the rank prediction function $\mathcal{RP}^{(2)}$. Finally, candidate objects are retrieved and a refinement step is applied (Figure \ref{fig:1}(c)).
\end{example}

\subsection{\textit{k}NN Query}
In LIMS, \cbl{a $k$NN query is processed by conducting a series of range queries with increasing search radius $r = \Delta r$, $2\cdot\Delta r$, $3\cdot\Delta r, \ldots$ until $k$ nearest neighbors are found, where $\Delta r$ is a given small initial radius.} Algorithm \ref{alg:knn} outlines the LIMS-based $k$NN query. Given a query $q$ and $ k$, a range query with a radius $r = \Delta r$ is issued at the beginning.  To answer this query, LIMS invokes Algorithm \ref{alg:range} (Line \ref{call}). $\mathcal{Q}$ is a max priority queue to record the candidate $k$ nearest neighbors. If the distance from the retrieved object $p$ to the query object $q$ is smaller than the current furthest distance in $\mathcal{Q}$, LIMS will extract the object with the maximum distance value from $\mathcal{Q}$ and insert $p$ into $\mathcal{Q}$ (Line \ref{check1}-\ref{check2}). The search stops if and only if the furthest object in the current priority queue falls within the current query range and further expansion of the query radius does not change the answer set (Line \ref{stop1}-\ref{stop2}). \cbl{Otherwise, it enlarges the query radius by $\Delta r$ (Line \ref{r}) and invokes Algorithm \ref{alg:range} till the termination conditions are satisfied.} LIMS also maintains an array to record whether a page has been processed. If so, it will be skipped during the next range query to avoid repeated accesses (Line \ref{repeat}). \cbl{The number of range query calls depends on $\Delta r$ and the distance between $q$ and the $k$-th NN in the dataset, which can be determined by $\lceil\frac{dist(\mathcal{Q}[1], q)}{\Delta r}\rceil + 1$.}
\begin{algorithm}[t]
    \footnotesize
	\caption{$k$NN Query}
	\label{alg:knn}
	\LinesNumbered 
	\KwIn{ ${q}$: a  query point; $k$: a positive integer; $\Delta r$: a positive number}
	\KwOut{$\mathcal{Q}$: top $k$ nearest neighbors to ${q}$ }
	$r \leftarrow 0,  flag \leftarrow FALSE$,
	$\mathcal{Q}$ is a max priority queue on $k$ objects initialized to $\infty$\; \label{Q}
	\While{$flag = FALSE$}
	{
		$r \leftarrow r + \Delta r$\; \label{r}
		\If{$dist(\mathcal{Q}[1], {q}) < r$} 
		{
			\label{stop1} $flag \leftarrow TRUE$\;
		}
		call $range({q}, r)$ to get pages set $\mathcal{P}$\; \label{call}
		call $check(\mathcal{P})$\;
		\Return{$\mathcal{Q}$}\;\label{stop2}
	}
	
	\textbf{procedure} $check(\mathcal{P})$:\\
	\For{each unvisited page $P \in \mathcal{P}$ \label{repeat}}
	{
		\For{each point ${p} \in P$}
		{            
			\label{check1}\If{ $dist(\mathcal{Q}[1], {q}) > dist({p}, {q})$}
			{
				Extract-Max($\mathcal{Q}$)\;
				Insert($\mathcal{Q}, {p}$)\;
			}\label{check2}
		}
	}
\end{algorithm} 

\cred{
\begin{remark}
\vspace{-5ex}
Intuitively, the initial radius $\Delta r$ affects the number of range query calls, and thus the query efficiency. We observed that a small $\Delta r$ would not degrade the query performance severely. As we know, the extra cost of using range queries with increasing radius to answer a $k$NN query mainly comes form 1) traversing the index multiple times and 2) accessing the same pages multiple times. In LIMS, a query is processed by a function invocation in $O(1)$ time instead of $O(\log_{}n)$ time in tree-like indexes, which makes the cost of multiple traverses negligible. Besides, LIMS would not access visited pages. However, a too large initial radius can degrade the query performance. Therefore, we recommend a small initial search radius, which can be simply estimated based on distances between pairs of data objects sampled from the dataset. 
\end{remark}
}

\vspace{-2ex}
\subsection{Updates}
\label{updates}
\cbl{LIMS allows both insertions and deletions of data.} To support the efficient insertion, LIMS maintains a sorted array for each cluster in ascending order of distance values to the centroid. Given an object $p$ to be inserted, LIMS first runs a point query to find if there is a page containing $p$. If so, the algorithm terminates immediately. Otherwise, LIMS finds the cluster closest to $p$ and inserts $p$ into the sorted array of this cluster. All newly inserted data are arranged in a number of pages with each page fully utilized. During query processing, LIMS uses the triangle inequality and exponential search to retrieve these inserted objects matching query filters. Given an object $p$ to be deleted, LIMS first runs a point query for $p$. If $p$ is found, it is marked as `deleted'. Then, LIMS updates maximum and minimum distances to each pivot of the cluster $p$ belongs to. Due to the space limitation, we omit the pseudocode here.
\cbl{Such an easy update strategy is effective and efficient because} 1) LMIS partitions the data space into many clusters that amortize the additional search time on the sorted arrays; 2) LIMS maintains an index for each cluster separately, \cbl{which allows for partially rebuilding the index, \ie  retraining rank prediction models for certain clusters, especially if deletions invalidate a cluster or insertions result in too much overlapping between 
some clusters. The procedure of retraining rank prediction models is the same as that of training them}; 3)  the short index construction (reconstruction) time of LIMS ensures the feasibility of this strategy in practice (Section \ref{sec:exp}).

\vspace{-1ex}
\subsection{Last Piece}
\label{sec:lastpiece}
The last piece of LIMS is to determine the number of clusters. The optimal number is related not only to \cbl{the} data distribution but also to \cbl{the} query workload. However, \cbl{the} query workload is not available during data clustering and we do not assume a known query workload in this paper. Therefore, an alternative  should be developed to pre-define the number of clustering parameter $K$. 
Recall that the goal of clustering in LIMS is to decompose complex and potentially correlated data into  a few clusters. Ideally, these clusters are independent and each can be accurately fit by a linear function. However, in most real-world use cases, the data do not follow such a perfect pattern. On the one hand, the overlapping between clusters may incur extra pruning and refinement costs. On the other hand, uneven intra-cluster distribution incurs more arithmetic and comparison operations.  In order to pick a $K$ to avoid or reduce such overhead, we introduce \textit{overlap rate} (OR) and \textit{mean absolute error} (MAE) to evaluate the goodness of clustering. OR quantifies the extent of overlapping among clusters. It can be computed as:

\begin{equation}
	OR =\frac{1}{K(K-1)}\sum_{i = 1}^K \sum_{i^\prime \neq i}\frac{r^{(i, i^\prime)}}{dist\_max_1^{(i)}}.
\end{equation}
Without confusion, we use $dist\_max_1^{(i)}$ to represent the distance of the furthest object in $i$th cluster from the centroid, and $r^{(i, i^\prime)}$ is the length of the overlapping area computed as:

\begin{equation}
	\begin{aligned}
		r^{(i, i^\prime)} &= \min\{
		dist(O_1^{(i)}, O_1^{(i^\prime)}) + dist\_max_1^{(i^\prime)}, 
		dist\_max_1^{(i)}
		\} \\
		& - \max\{(dist(O_1^{(i)}, O_1^{(i^\prime)}) - dist\_max_1^{(i^\prime)}, dist\_min_1^{(i)}\}.
	\end{aligned}
\end{equation}
MAE quantifies the quality of the linear regression fit for each cluster.  It can be computed as:

\begin{equation}
	MAE = \frac{1}{m|P|}\sum_{i = 1}^K
	\sum_{j = 1}^m
	\sum_{\tilde{D}_j^{(i)}}|a_j^{(i)}x + b_j^{(i)} - rank(x)|,
\end{equation}
where $|P|$ is the cardinality of dataset,
$\mathcal{RP}_j^{(i)}(x) = a_j^{(i)}x + b_j^{(i)} $ are  linear rank prediction models learned from $\tilde{D}_j^{(i)}$.
\begin{table}[t]
    \renewcommand\arraystretch{1.5}
	\caption{Summary of datasets}
	\vspace{-2ex}
	\label{tab:datasets}
	\footnotesize
	\begin{center}
	\begin{tabular}{|c|c|c|c|c|}
		\hline
		\textbf{Datasets}          &\textbf{Cardinality} & \textbf{Dim.} & \cbl{\textbf{Ins.}} & \textbf{Metric}   \\
		\hline
		\textit{Color}   & 1,281,167   & 32   & \cbl{4.2}          & $L_2$-norm           \\
		\hline
		\textit{Forest} & 565,892     & 6     & \cbl{1.5}         & $L_2$-norm           \\
		\hline
		\textit{GaussMix}         & 5,\underline{10},20,40,60,80M  & 2,4,\underline{8},12,16  & \cbl{6.2}  & $L_2$-norm           \\
		\hline
		\textit{Skewed}            & 10M & 2,4,\underline{8},12,16  & \cbl{5.6}  & $L_1$-norm           \\
		\hline
		\textit{Signature}         & 100K    & 65    & \cbl{36}         & \textit{Edit distance}   \\ 
		\hline
	\end{tabular}
	\end{center}
	\vspace{-5ex}
\end{table}

We model the overhead as $OR + \lambda MAE$, where $\lambda > 0$ is a user-defined weight. Inspired by the elbow method \cite{elbow}, we choose the \textit{elbow} or \textit{knee} of a curve as the clustering number to use, \ie a point where adding more clusters will not give much better modeling of the data. The number estimated by the techniques described above turns out to be very close to the optimal number of clusters observed in practice, as shown in Section \ref{sec:exp}.
\section{Experiments} \label{sec:exp}
In this section, we present the results of an in-depth experimental study on LIMS. We implement LIMS \footnote{https://github.com/learned-index/LIMS} and associated similarity search algorithms in C++. All experiments are conducted on a computer running 64-bit Ubuntu 20.04 with a 2.30 GHz Intel(R) Xeon(R) Gold 5218 CPU, 254 GB RAM, and an 8.2 TB hard disk.

\begin{figure*}[t]  
    \begin{minipage}[c]{\linewidth}
		\centering
		\includegraphics[width=0.7\textwidth]{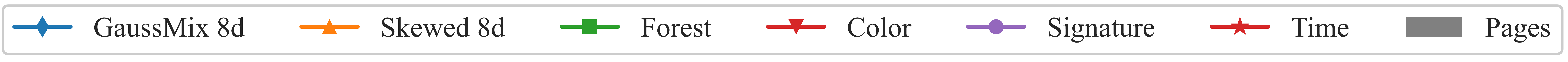}
 		\vspace{-0.5em}
	\end{minipage}
	\centering    
	\subfloat [Estimated effect of $K$]
	{
		\begin{minipage}[t]{0.24\linewidth}
			\centering          
			\includegraphics[width=\linewidth]{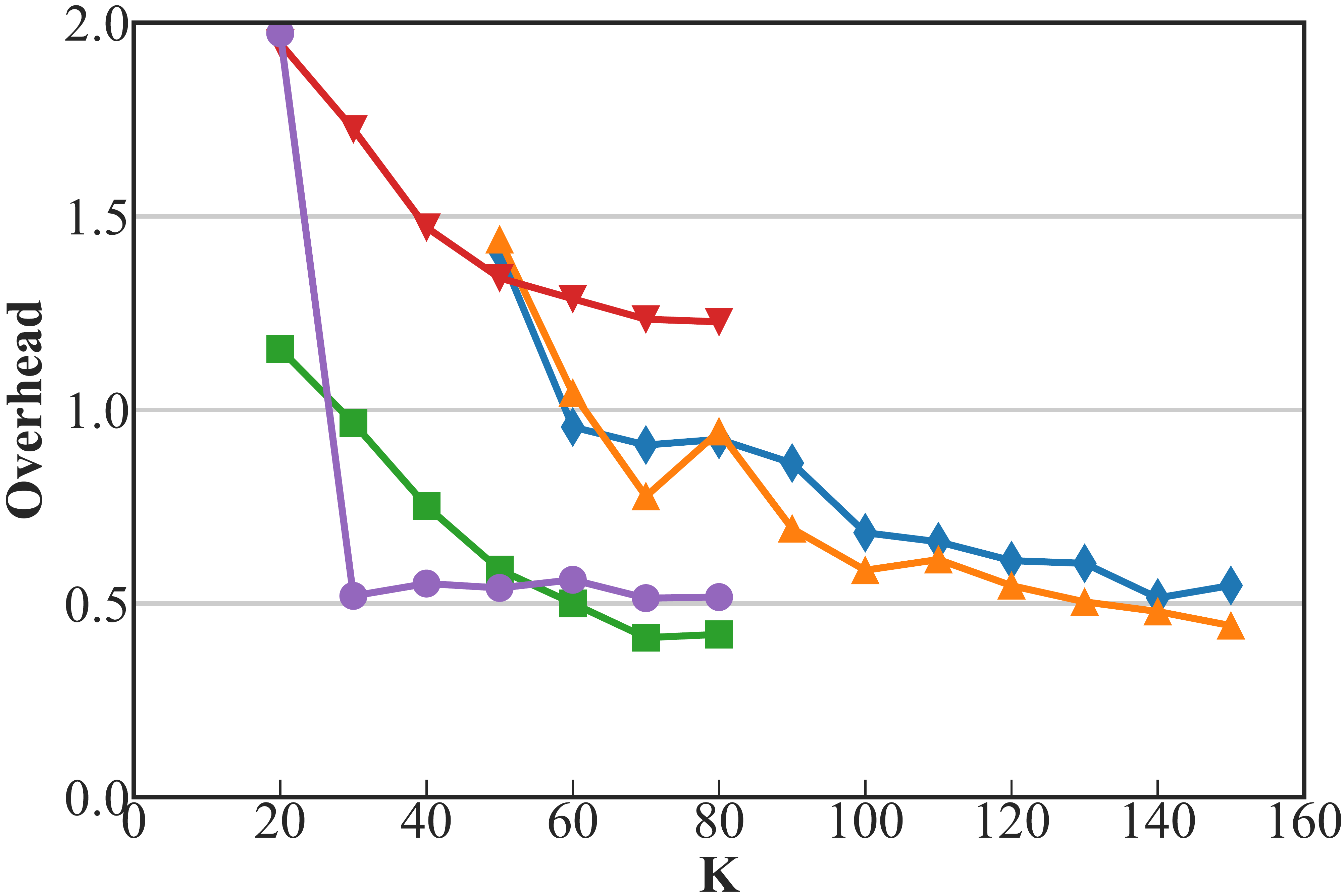}   
		\end{minipage}%
	}
	\subfloat [Effect of $K$ on range query]
	{
		\begin{minipage}[t]{0.25\linewidth}
			\centering      
			\includegraphics[width=\linewidth]{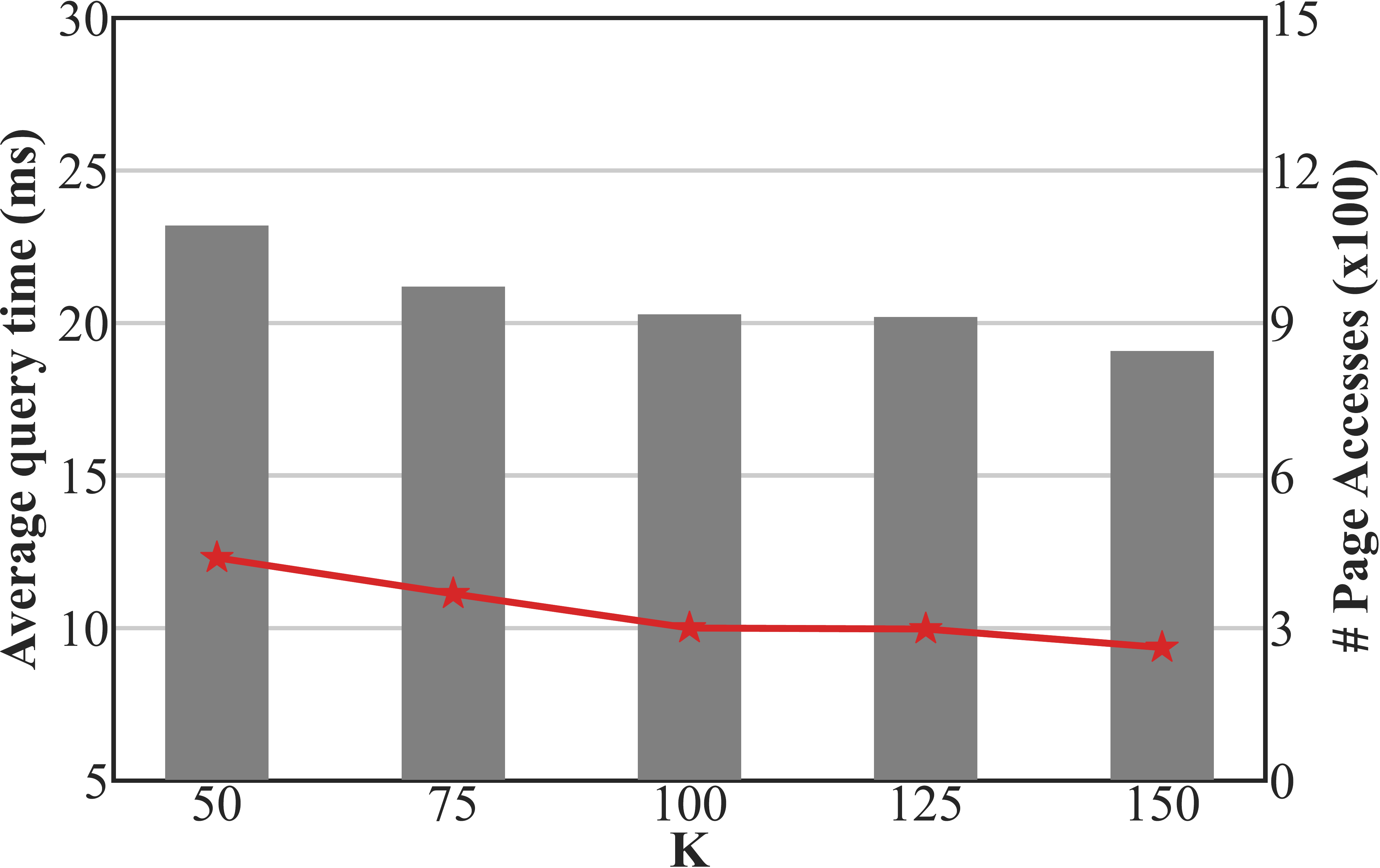}   
		\end{minipage}
	}%
	\subfloat[Effect of $m$ on range query]
	{
		\begin{minipage}[t]{0.25\linewidth}
			\centering          
			\includegraphics[width=\linewidth]{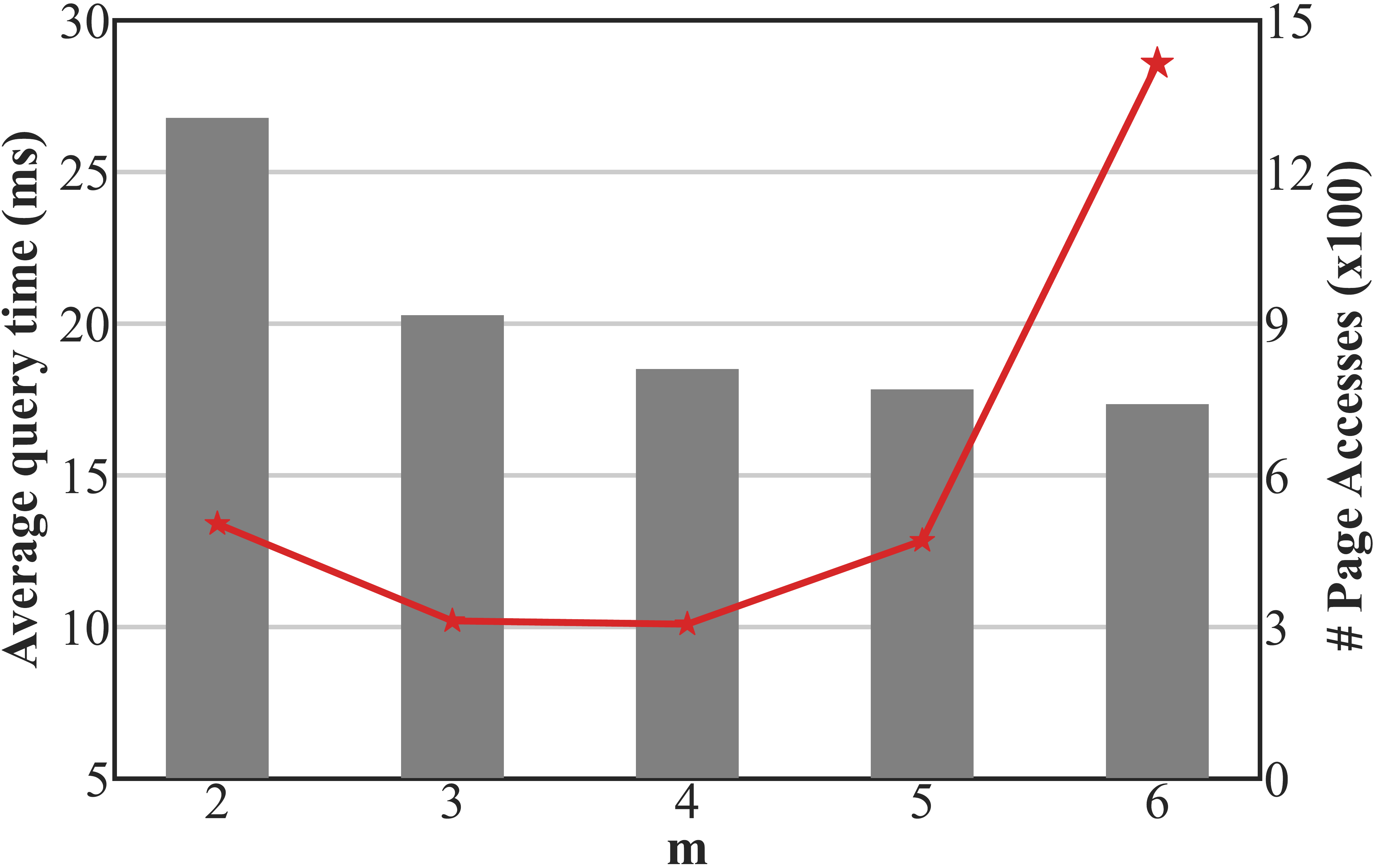}   
		\end{minipage}%
	}
	\subfloat[Effect of $N$ on range query]
	{
		\begin{minipage}[t]{0.25\linewidth}
			\centering          
			\includegraphics[width=\linewidth]{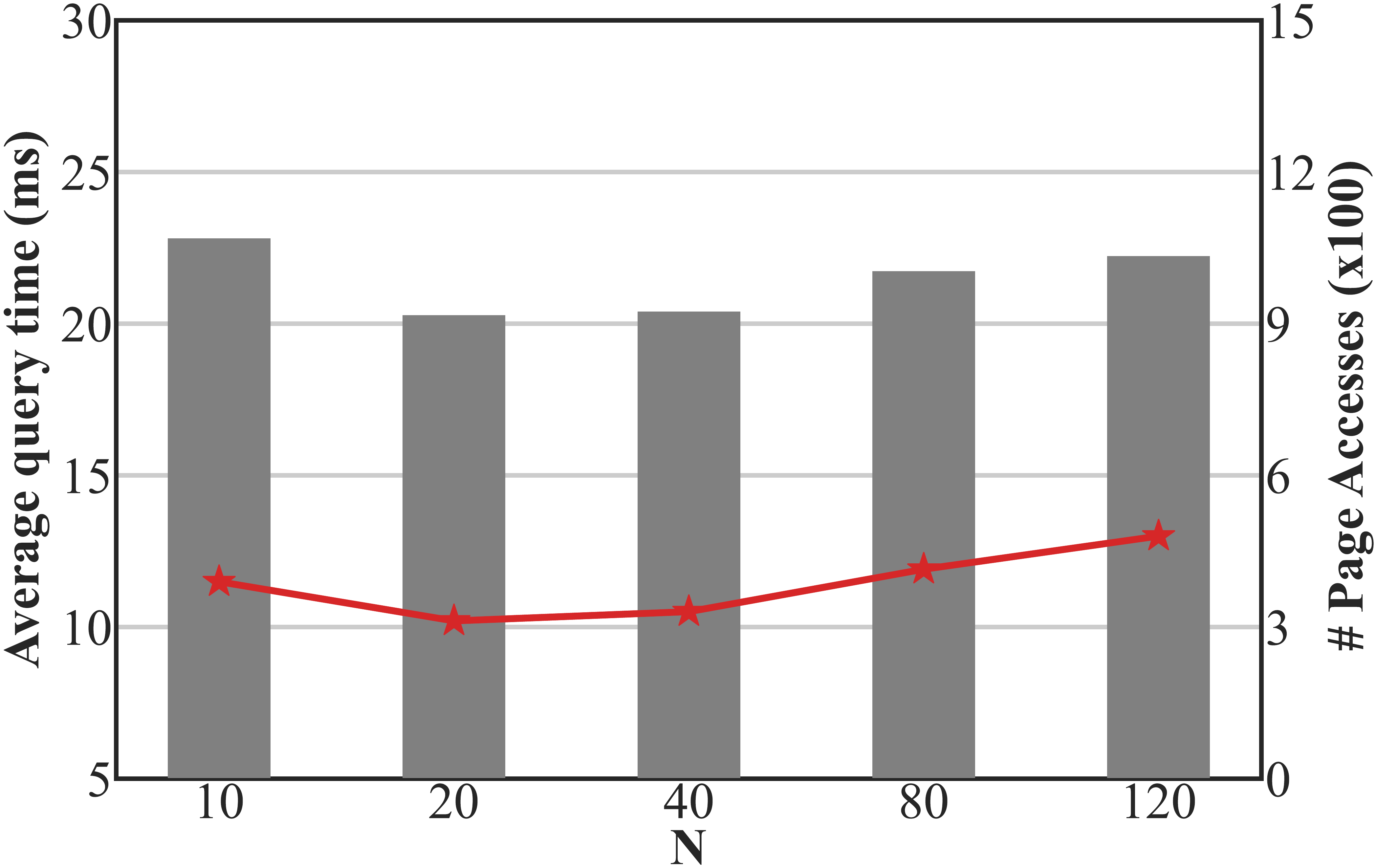}   
		\end{minipage}%
	}
	
	\vspace{-1ex}
	\caption{Effect of parameters} 
	\label{fig:parameters}
	\vspace{-2ex}
\end{figure*}
\subsection{Experimental Settings}
\label{sec:expsettings}
\noindent\textbf{Datasets.}
We employ two real-world datasets, namely,  \textit{Color Histogram}\footnote{https://image-net.org/download-images} and \textit{Forest Cover Type}\footnote{https://www.kaggle.com/c/forest-cover-type-prediction/data} following the experimental settings of ML index\cite{davitkova2020ml}. \textit{Color Histogram} contains 1,281,167 32-dimensional image features extracted from the ImageNet\footnote{ http://image- net.org/download- images} dataset.  \textit{Forest Cover Type} is collected by US Geological Survey and US Forest Service. It includes 565,892 records, each of which has 12 cartographic variables. We extract 6 quantitative variables of them as our data object. Following the experimental settings of iDistance\cite{idistance}, we generate 2, 4, 8, 12, 16-dimensional \textit{GaussMix} datasets. Every dataset contains up to 80 million points (5.36GB in size) sampled from 150 normal distributions with the standard deviation of 0.05 and randomly determined means. Default settings are underlined. Without loss of generality, $L_2$ norm is utilized to the above datasets. Following the experimental settings of RSMI\cite{rsmi}, we create 2, 4, 8, 12, 16-dimensional \textit{Skewed} datasets. They are generated from uniform data by raising the values in each dimension to their powers, \ie a randomly generated data point are converted from $(x_1, x_2, \ldots, x_d)$ to $(x_1, x_2^2, \ldots, x_d^d)$. The size of each dataset is 10 million and $L_1$ norm is employed. Without loss of generality, all the data values of the above datasets are normalized to the range $[0, 1]$.
Following the experimental settings in \cite{signature}, we also generate a \textit{Signature} dataset, where each object is a string with 65 English letters. We first obtain 25 `anchor signatures' whose letters are randomly chosen from the alphabet. Then, each anchor produces a cluster with 4,000 objects, each of which is obtained by randomly changing $x$ positions in the corresponding anchor signature to other random letters, where $x$ is uniformly distributed in the range $[1, 30]$. The \textit{edit distance} is used to compute the distance between two signatures. Table \ref{tab:datasets} summarizes the statistics of the datasets.

\noindent\textbf{Competitors.}
We compare LIMS with  three representative multi-dimensional learned indexes as mentioned in Section \ref{sec:relatedwork}, \ie ZM\cite{wang2019learned}, ML\cite{davitkova2020ml}, LISA\cite{li2020lisa},  and \cbl{three} traditional indexes, \ie $\text{R}^*$-tree\cite{Rstar}, M-tree\cite{mtree} \cbl{and SPB-tree\cite{spb_short}\cite{spb_ext}}. ZM and ML are in-memory indexes, so we adapt them to the disk by storing data in ascending order of their z-order/mapped values in a number of pages with each page fully utilized.  For LISA, no open-source C++ code is available, so we implement it following Python version implementation. For $\text{R}^*$-tree, M-tree and \cbl{SPB-tree}, we use the original implementations.  In  addition,  to  study  the effectiveness  of learning components  in LIMS,  we  design a  method  called  non-learned index for metric spaces (N-LIMS) by replacing rank prediction models in LIMS with the traditional B$^+$-trees\cite{TLX}. All competitors are configured to use a fixed disk page size of 4KB.

\noindent\textbf{Evaluation Metrics.}
Four metrics are used to evaluate the performance of indexes: the average number of page accesses, the average query time, indexing time and index size. We randomly select 200 objects from each dataset and repeat each experiment 20 times to get average results.

\subsection{Effect of Parameters}
\label{sec:effectPara}
We first study the effect of parameters, including the number of clusters $K$, the number of pivots $m$  and the number of rings $N$, to optimize LIMS-based similarity search, as summarized in Fig. \ref{fig:parameters}.  Only one parameter varies whereas the others are fixed to their default values in every experiment. By default, the selectivity of range query, \ie the fraction of objects within the query range from the total number of objects, is set to 0.01\%. The $k$ of $k$NN query is 5. Degrees of $\mathcal{RP}_j^{(i)}$ and $\mathcal{RP}^{(i)}$ are 20 and 1. $m$ and $N$ are 3 and 20.
$K$ is determined according to the method described in Section \ref{sec:lastpiece}. We set $K = 100$ for 10M 8$d$ \textit{GaussMix} and 8$d$ \textit{Skewed}, $K = 50$ for \textit{Color Histogram}, \textit{Forest Cover Type} and \textit{Signature}. 
 

\vspace{-1ex}
\subsubsection{Effect of $K$} 
Fig. \ref{fig:parameters}(a) plots the criterion $OR + \lambda MAE$  versus $K$ on both real and synthetic data, where $K = 20, 30, \ldots, 150$ and $\lambda$ is set to $1/\max\{MAE(K)\}$. We can see that different datasets have different elbow points. In order to show the estimation is close to the actual optimal number of clusters, we also plot the actual average query time and the number of page accesses for range query by varying $K$. Due to the space limitation, we only report the performance on 10M 8$d$ \textit{GaussMix} dataset in Fig. \ref{fig:parameters}(b). It can be observed that query 
time decreases slowly after $K = 100$, which is consistent with the recommended choice of $K$ in Fig. \ref{fig:parameters}(a). Therefore, we set $K = 100$ for this dataset.

\vspace{-1ex}
\subsubsection{Effect of $m$} 
Fig. \ref{fig:parameters}(c) reports the query performance on 10M 8$d$ \textit{GaussMix} dataset by varying the number of pivots $m$. We can see that increasing the number of pivots always reduces (or at least does not increase) the number of page accesses.  This is expected because the intersection of metric regions defined by more pivots is always smaller than (or at least equal to) the intersection of metric regions defined by fewer pivots. As discussed in Section \ref{subsec: cluster}, the more pivots, the stronger pruning ability. This observation can also be proven using Equation (\ref{eq:intersection}). 
However, the average query time  decreases when using up to four pivots, and then increases progressively. That is because the cost for filtering unqualified objects grows as well with more pivots. The best number of pivots for a metric index is a trade-off between filter cost and scanning cost. Therefore, the default value for $m$ is set to 3 unless otherwise stated. 

\vspace{-1ex}
\subsubsection{Effect of $N$} Fig. \ref{fig:parameters}(d) reports the query performance on 10M 8$d$ \textit{GaussMix} dataset by varying the number of rings $N$. For the same reason as above, the average query time presents a down and up trend whereas the lowest value turns out at $N = 20$. 

\vspace{-1ex}
\subsection{Range Query Performance}
\begin{figure*}[htbp]  
    \begin{minipage}[c]{\linewidth}
		\centering
		\includegraphics[width=0.9\textwidth]{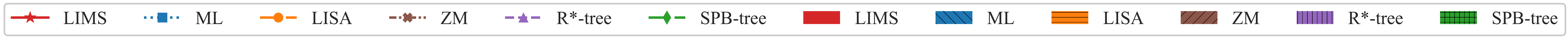}
 		\vspace{-0.5em}
	\end{minipage}
	\centering    
	\subfloat[Query time on \textit{Skewed} ]
	{
		\begin{minipage}[t]{0.24\linewidth}
			\centering          
			\includegraphics[width=\linewidth]{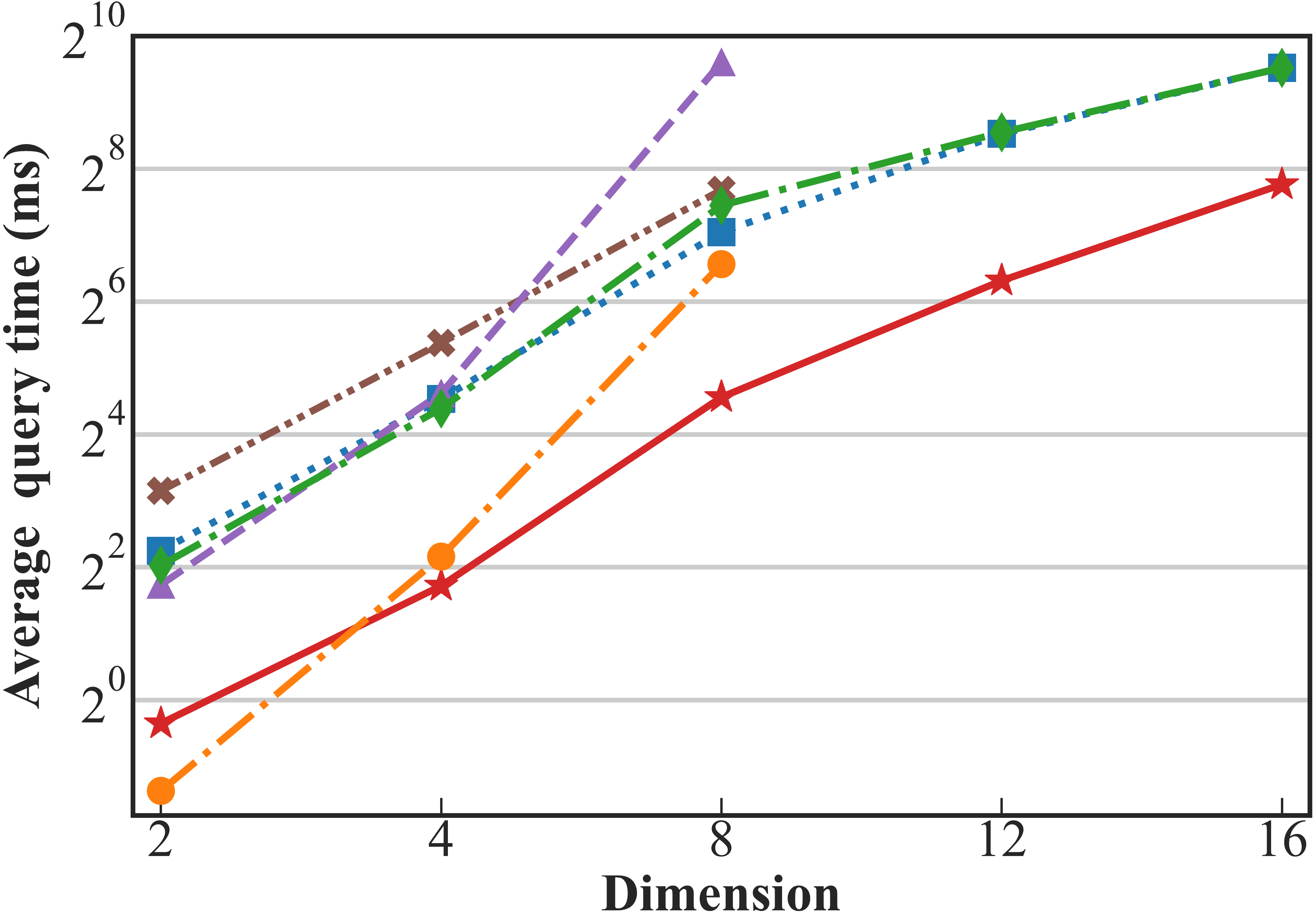}   
		\end{minipage}%
	}
	\subfloat[\# Page accesses on \textit{Skewed}]
	{
		\begin{minipage}[t]{0.24\linewidth}
			\centering      
			\includegraphics[width=\linewidth]{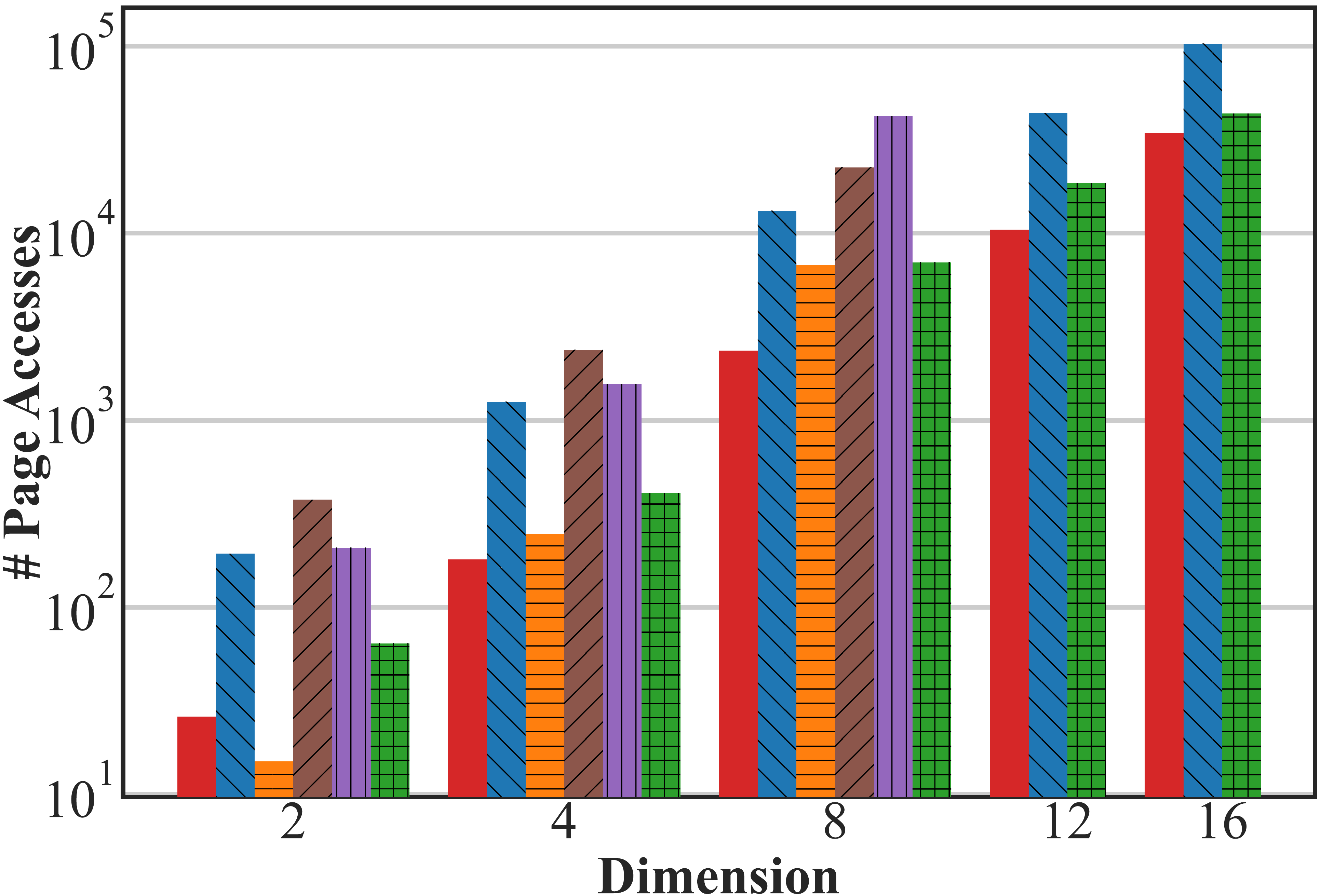}   
		\end{minipage}
	}%
	\subfloat[Query time on \textit{GaussMix}]
	{
		\begin{minipage}[t]{0.24\linewidth}
			\centering      
			\includegraphics[width=\linewidth]{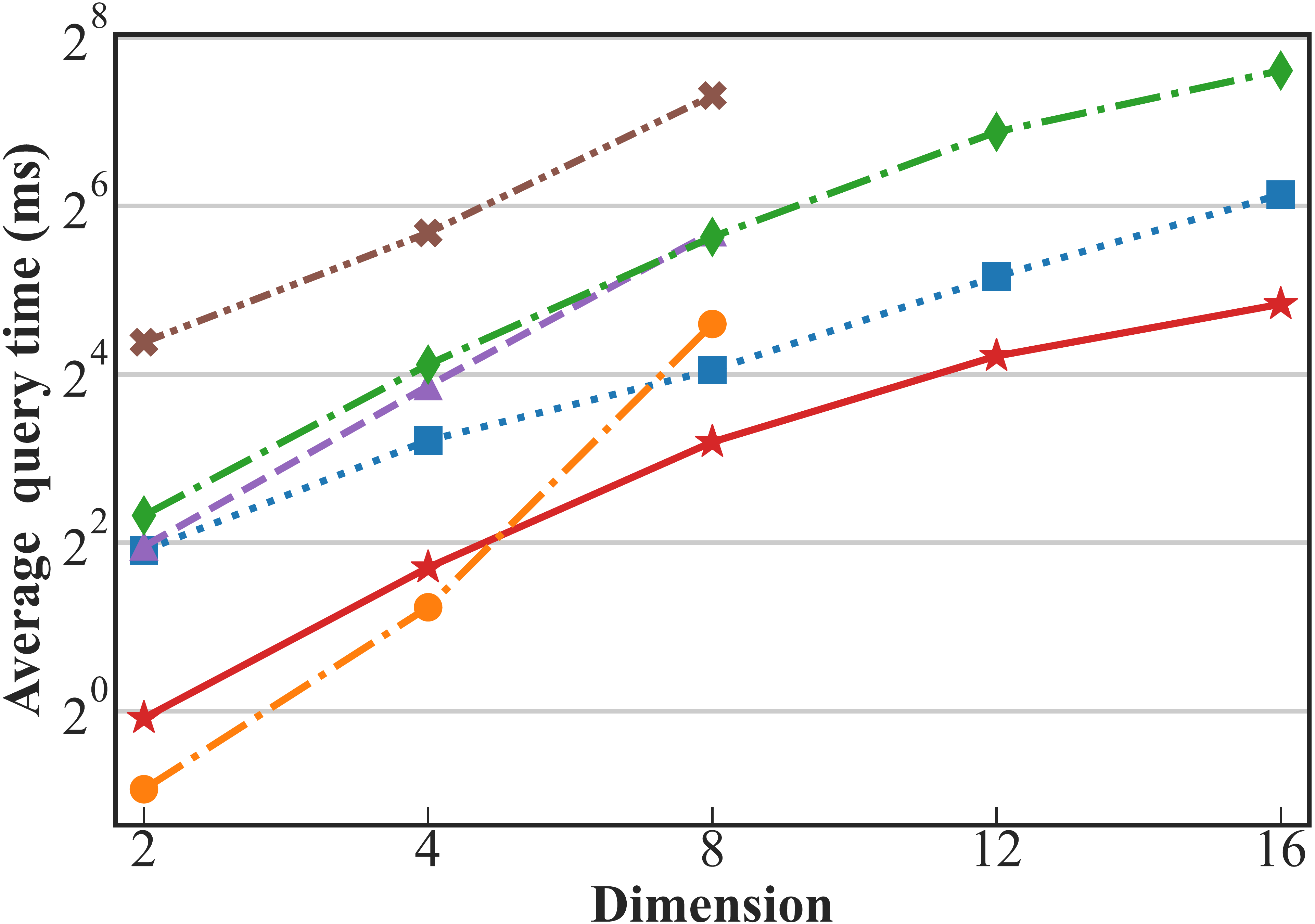}   
		\end{minipage}
	}%
	\subfloat[\# Page accesses on \textit{GaussMix}]
	{
		\begin{minipage}[t]{0.24\linewidth}
			\centering      
			\includegraphics[width=\linewidth]{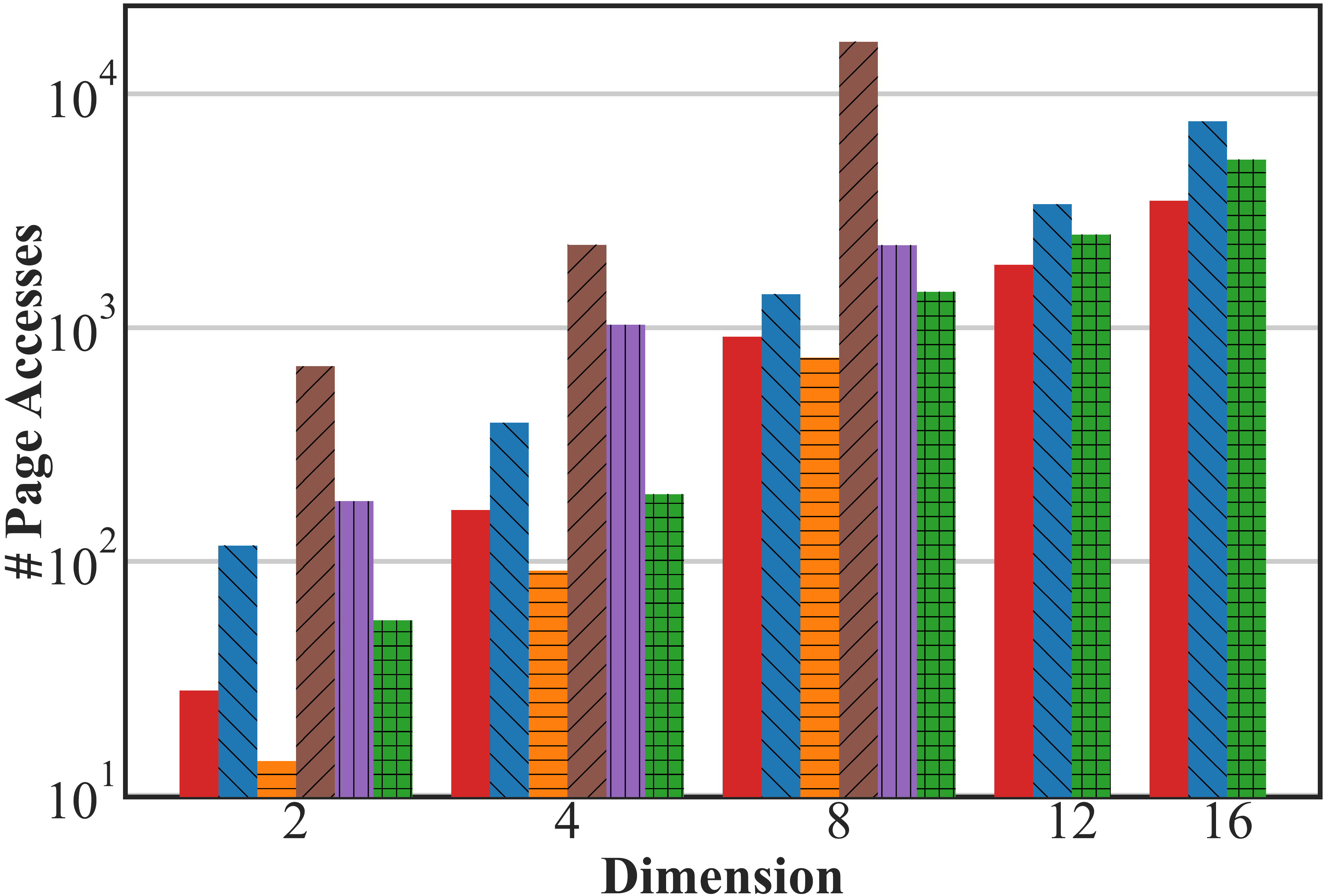}   
		\end{minipage}
	}%
	\vspace{-1ex}%
	\caption{Range query performance with dimensionality}
	\vspace{-2ex}
	\label{fig:range_dim}  
\end{figure*}

\begin{figure*}[t]  
	\centering    
	\subfloat[Query time on \textit{Forest Cover Type} ]
	{
		\begin{minipage}[t]{0.245\linewidth}
			\centering          
			\includegraphics[width=\linewidth]{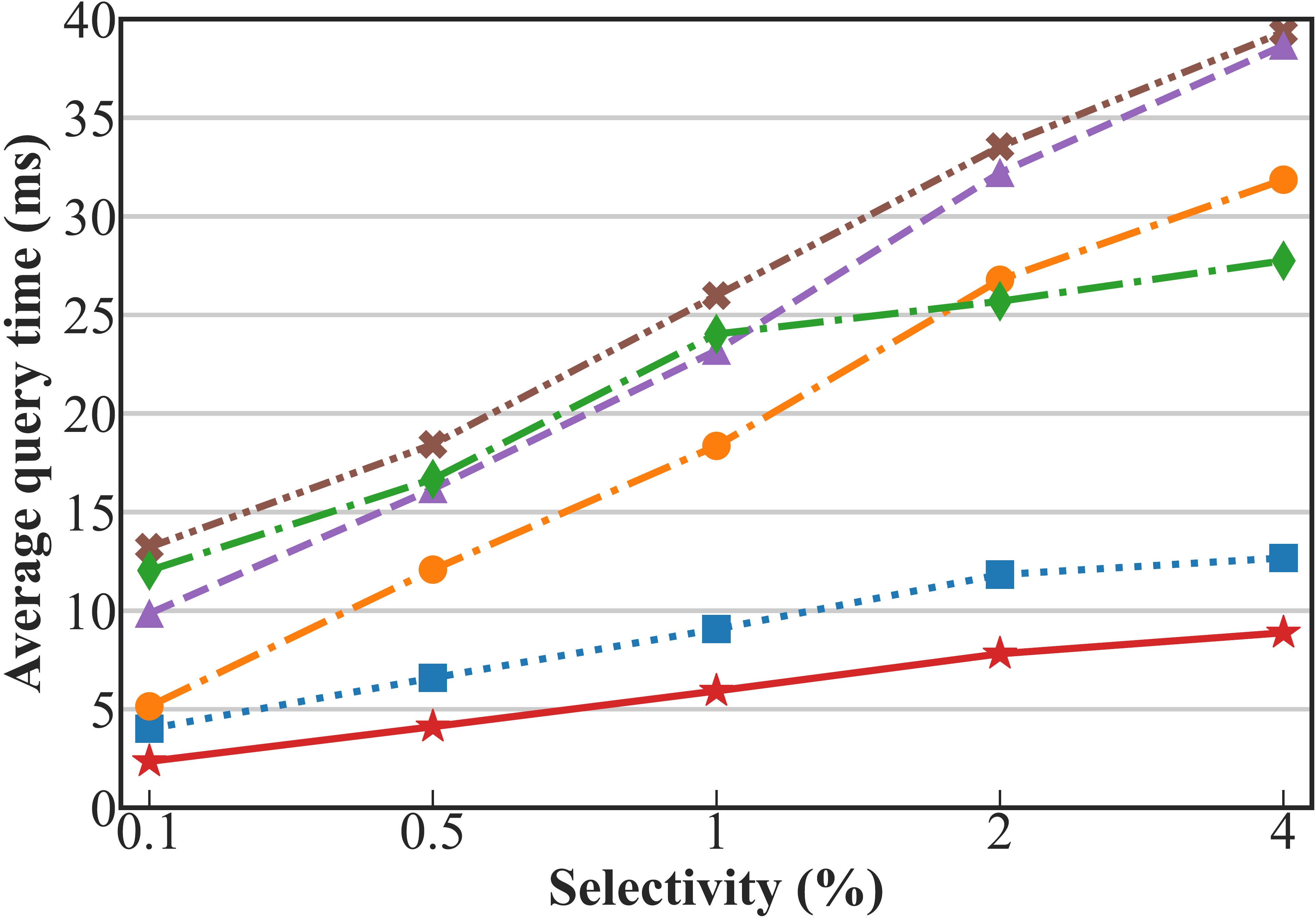}   
		\end{minipage}%
	}
	\subfloat[\# Page accesses on \textit{Forest}]
	{
		\begin{minipage}[t]{0.245\linewidth}
			\centering      
			\includegraphics[width=\linewidth]{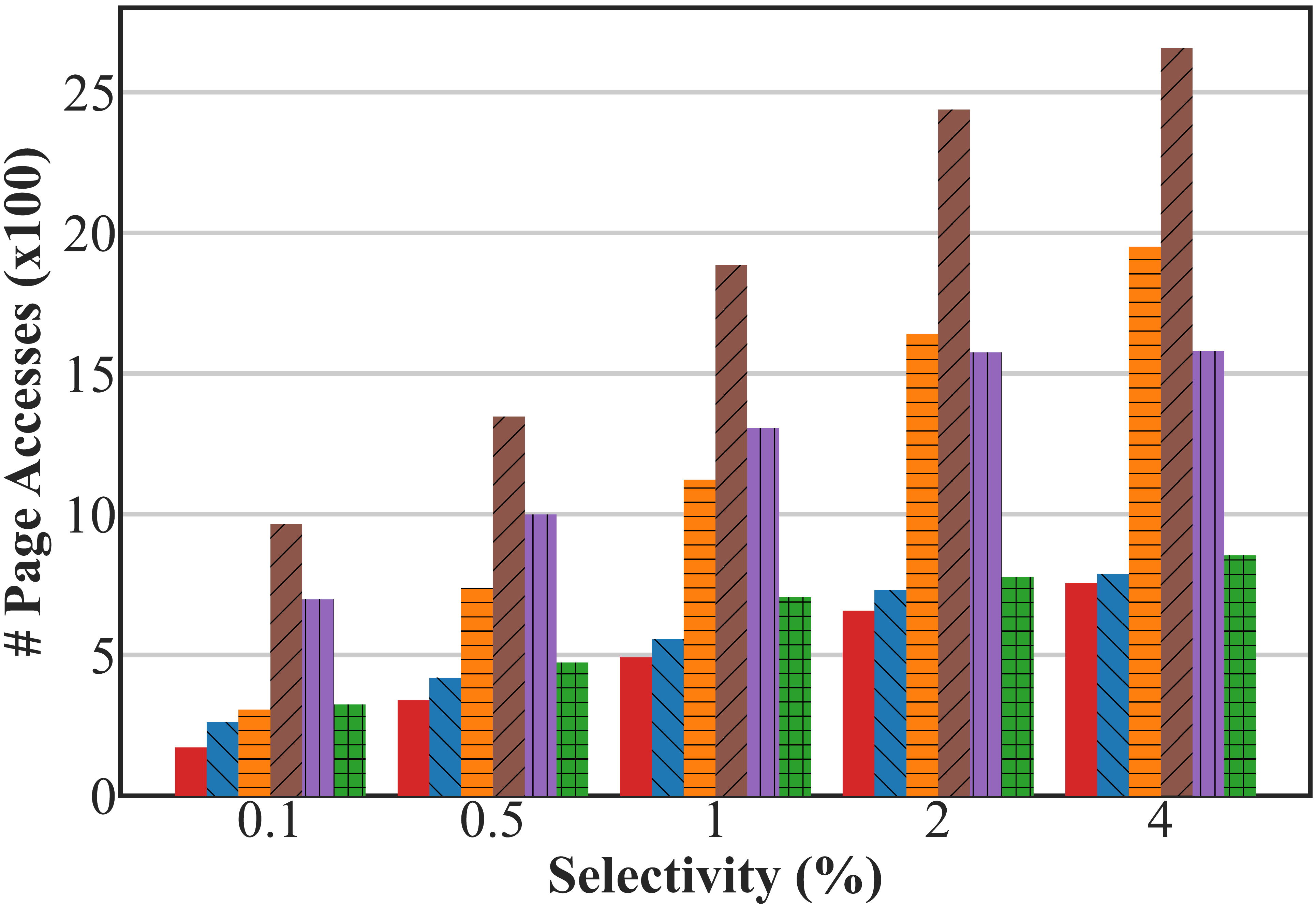}   
		\end{minipage}
	}%
	\subfloat[Query time on \textit{Color}]
	{
		\begin{minipage}[t]{0.245\linewidth}
			\centering      
			\includegraphics[width=\linewidth]{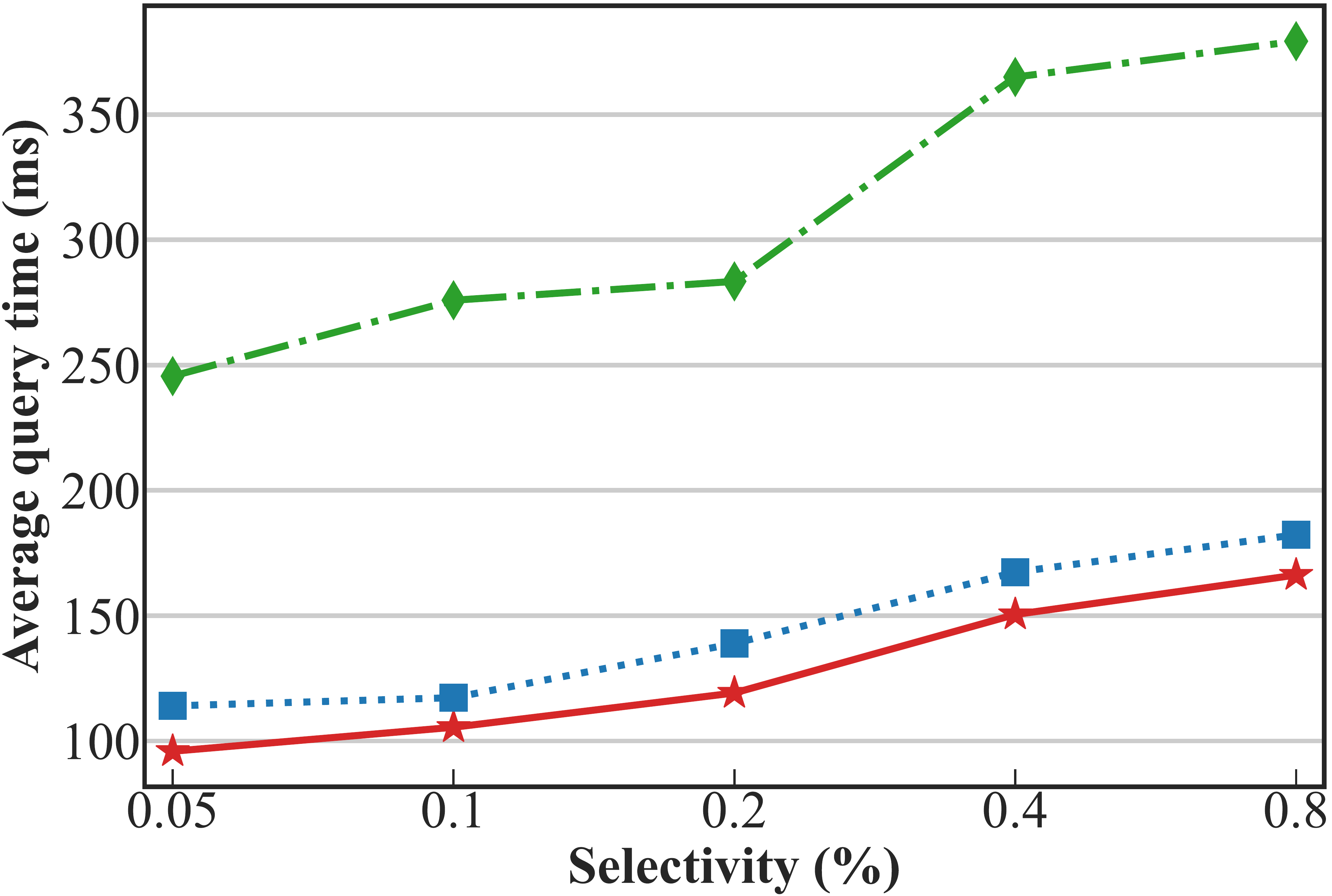}   
		\end{minipage}
	}%
	\subfloat[\# Page accesses on \textit{Color}]
	{
		\begin{minipage}[t]{0.245\linewidth}
			\centering      
			\includegraphics[width=\linewidth]{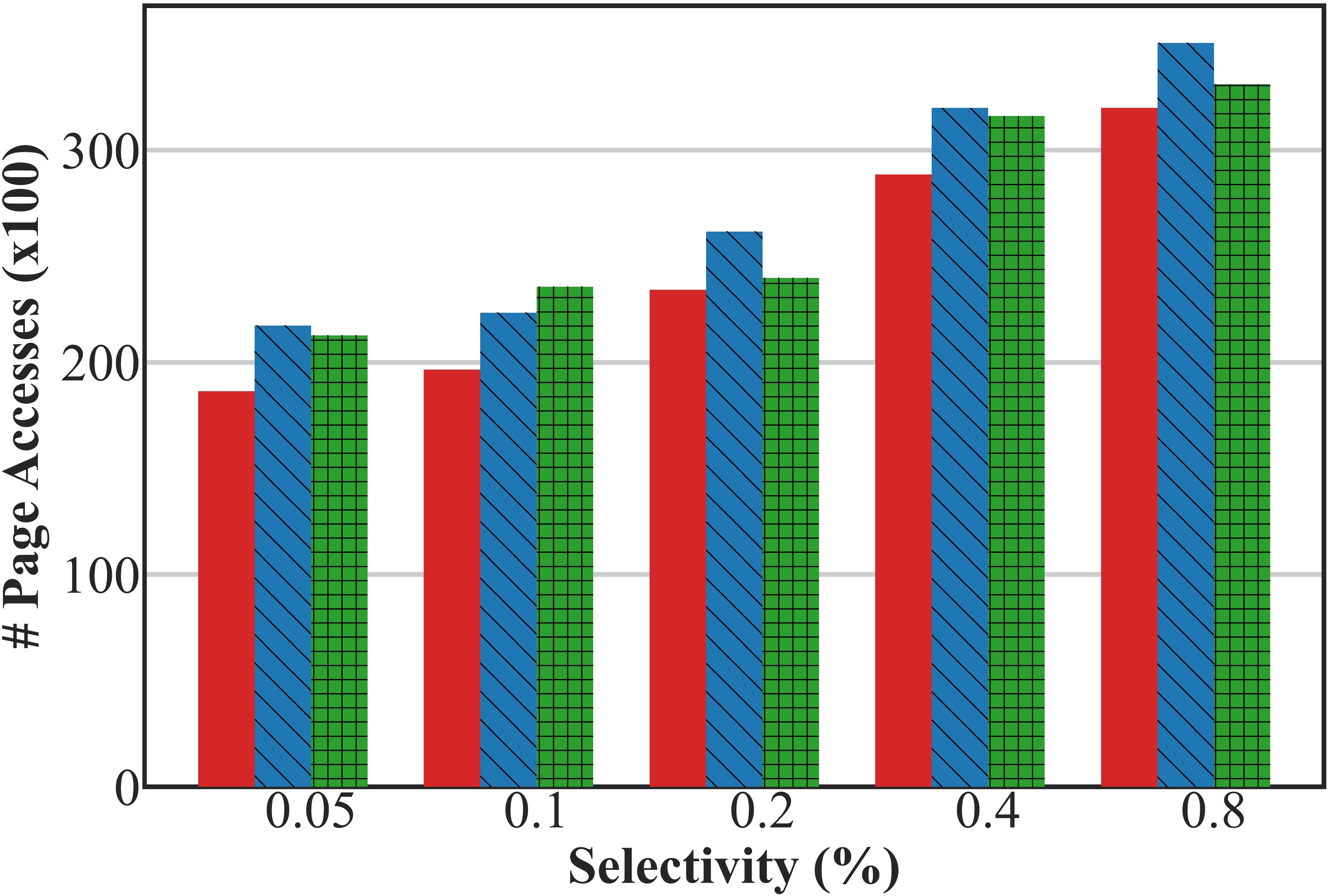}   
		\end{minipage}
	}%
	\vspace{-1ex}
	\caption{Range query performance with selectivity}%
	\vspace{-2ex}
	\label{fig:range_selectivity}  
\end{figure*}

In this subsection, we study the performance of LIMS, ML, LISA, ZM, $\text{R}^*$-tree, M-tree \cbl{and SPB-tree} on range query from different angles, as summarized in Fig. \ref{fig:range_dim}, \ref{fig:range_selectivity} and \ref{fig:range_knn_sg}.

\subsubsection{Performance with dimensionality} The first set of experiments studies the average query time and the number of page accesses under different dimensionalities.  Fig. \ref{fig:range_dim}(a)(b) and Fig. \ref{fig:range_dim}(c)(d) report the results on \textit{Skewed} and \textit{GaussMix} datasets, respectively. From the figures, we have the following observations: 
\cbl{1) The average query time of all methods increases with dimensionality, but LIMS, ML and SPB-tree grow much slower than others, suggesting that data clustering and pivot-based data transformation techniques are effective in alleviating the curse of dimensionality. The coordinate-based methods, \ie LISA, ZM and $\text{R}^*$-tree, degrade rapidly with dimensionality and even do not work, hence we do not report  their results  after  $8d$.}
2) LIMS is slightly slower than LISA when $2d$ and when $4d$ on \textit{GaussMix}, but LIMS offers the best performance  in a higher range of dimensions on both datasets. The reason is that metric-space indexes  can only use the four distance properties to prune \cbl{the} search space, while LISA can easily locate the cells that overlap with the query range by coordinates.  Fewer assumptions about the data result in poorer pruning and slightly larger query costs in the low-dimensional case. However, the filtering cost of LISA increases exponentially with dimensionality and it does not work when the dimension is greater than  $8$.
Note that on $8d$ \textit{GaussMix}, even though LIMS has higher numbers of page accesses than LISA, it is still faster due to the low filter cost. On \textit{Skewed}, LIMS begins to have a lead advantage since $4d$ because LIMS can be applied to a metric space with any distance metrics naturally (\eg $L_1$ norm here) and guarantees a few false positives and fast query response. 3) LIMS is always better than ML in both the query time and page accesses. 
This is intuitive since ML transforms different objects with equi-distance from the pivot into the same 1-dimensional value, while LIMS integrates the pruning abilities from multiple pivots. In addition, with the help of a well-defined pivots-based mapping, LIMS maps nearby data into the compact region, which further reduces the search region, and thus fewer objects to be accessed in the refinement step. 4) LIMS outperforms the learned index, ZM, by over an order of magnitude. This is because LIMS uses clusters and LIMS values to organize objects into compact regions, while ZM uses z-order curve to organize data, which incurs too many false positives, and thus large page accesses and distance computations. 5) LIMS is always better than traditional indexes. \cbl{ The main reason is that  query processing with a traditional index requires traversing many tree nodes multiple times, which is time-consuming.   M-tree is omitted since it is considerably worse than others. 6) 
The performance of all indexes on \textit{GaussMix} is better than that on \textit{Skewed} due to the simple  distribution.}

\subsubsection{Performance with selectivity} 
The second set of experiments studies the average query time and the number of page accesses with different selectivity. Fig. \ref{fig:range_selectivity}(a)(b) show the results on \textit{Forest} dataset  by varying selectivity from  0.1\% to 4\%. Fig. \ref{fig:range_selectivity}(c)(d)  show the results on \textit{Color} dataset by varying the selectivity from  0.05\% to 0.8\%.
$N$ on these small datasets is cut in half. From the figures, we have the following observations: 1) both query time and page accesses among indexes grow with the selectivity since more objects are queried. 2)
LIMS achieves better performance under all selectivities by at least 1.7X and up to 4.4X faster. Consistent high efficiency  of LIMS shows \cbl{the} robustness for range queries in varying settings. 3) On \textit{Color Histogram}, the advantage in fewer page accesses  shows the potential advantages that LIMS can achieve over ML for processing similarity search with costly distance metrics. 

\begin{figure*}[htbp]  
\begin{minipage}[c]{\linewidth}
		\centering
		\includegraphics[width=0.95\textwidth]{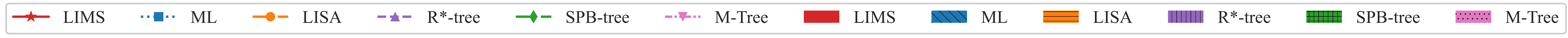}
 		\vspace{-0.5em}
	\end{minipage}
	\centering    
	\subfloat[Time of range queries]
	{
		\begin{minipage}[t]{0.24\linewidth}
			\centering          
			\includegraphics[width=\linewidth]{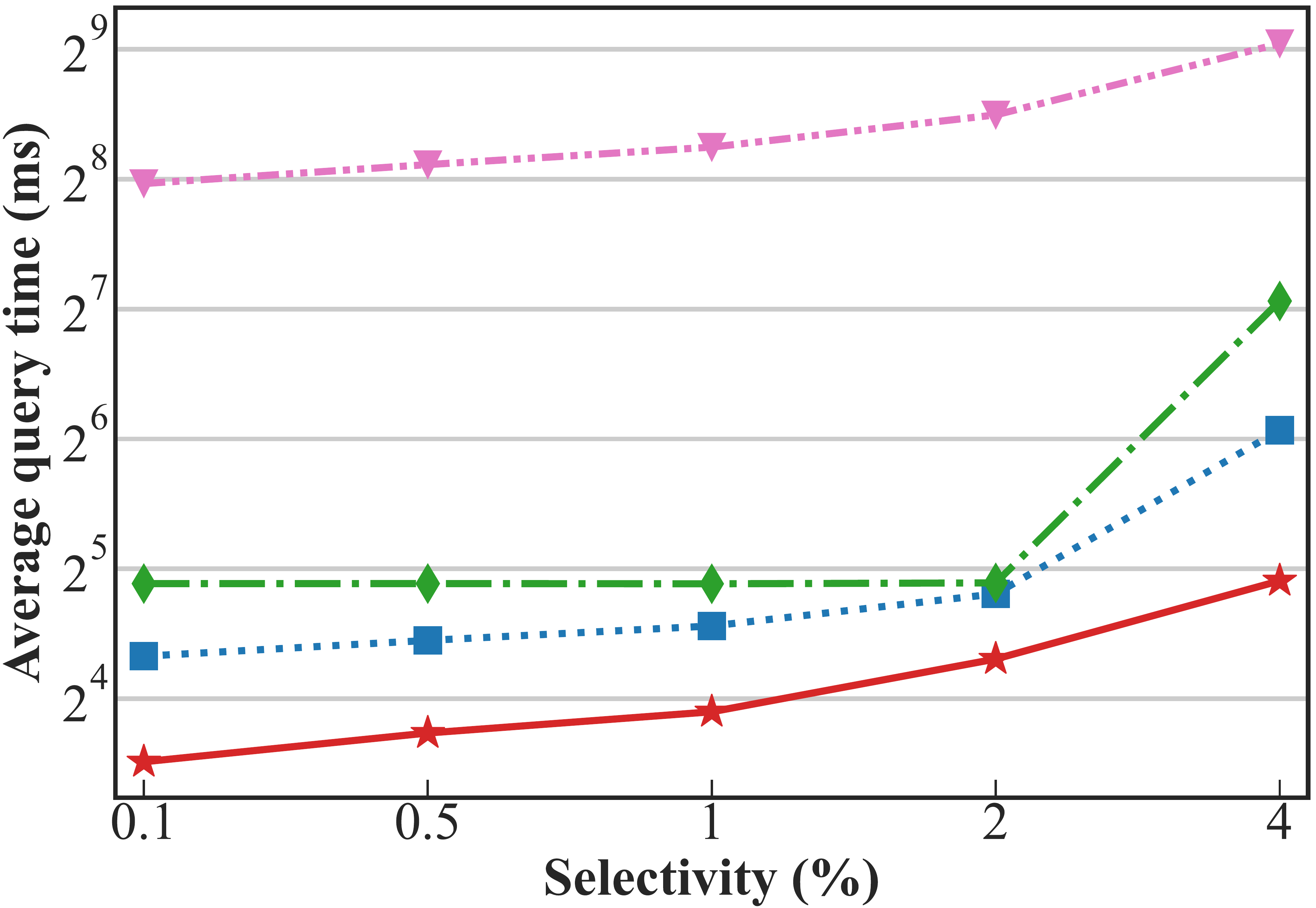}   
		\end{minipage}%
	}
	\subfloat[\# Page accesses of range queries]
	{
		\begin{minipage}[t]{0.24\linewidth}
			\centering      
			\includegraphics[width=\linewidth]{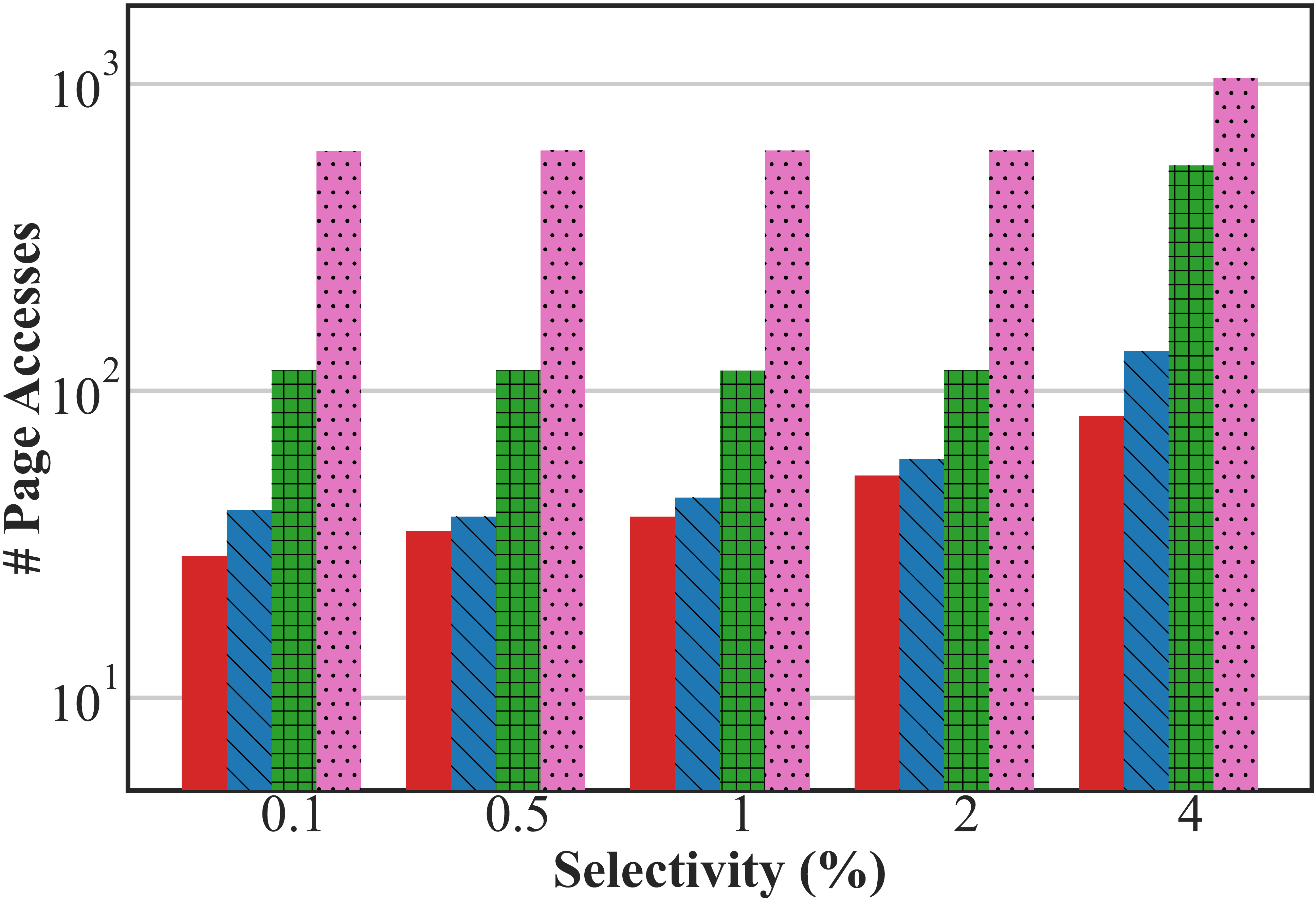}   
		\end{minipage}
	}%
	\subfloat[Time of \textit{k}NN queries]
	{
		\begin{minipage}[t]{0.24\linewidth}
			\centering          
			\includegraphics[width=\linewidth]{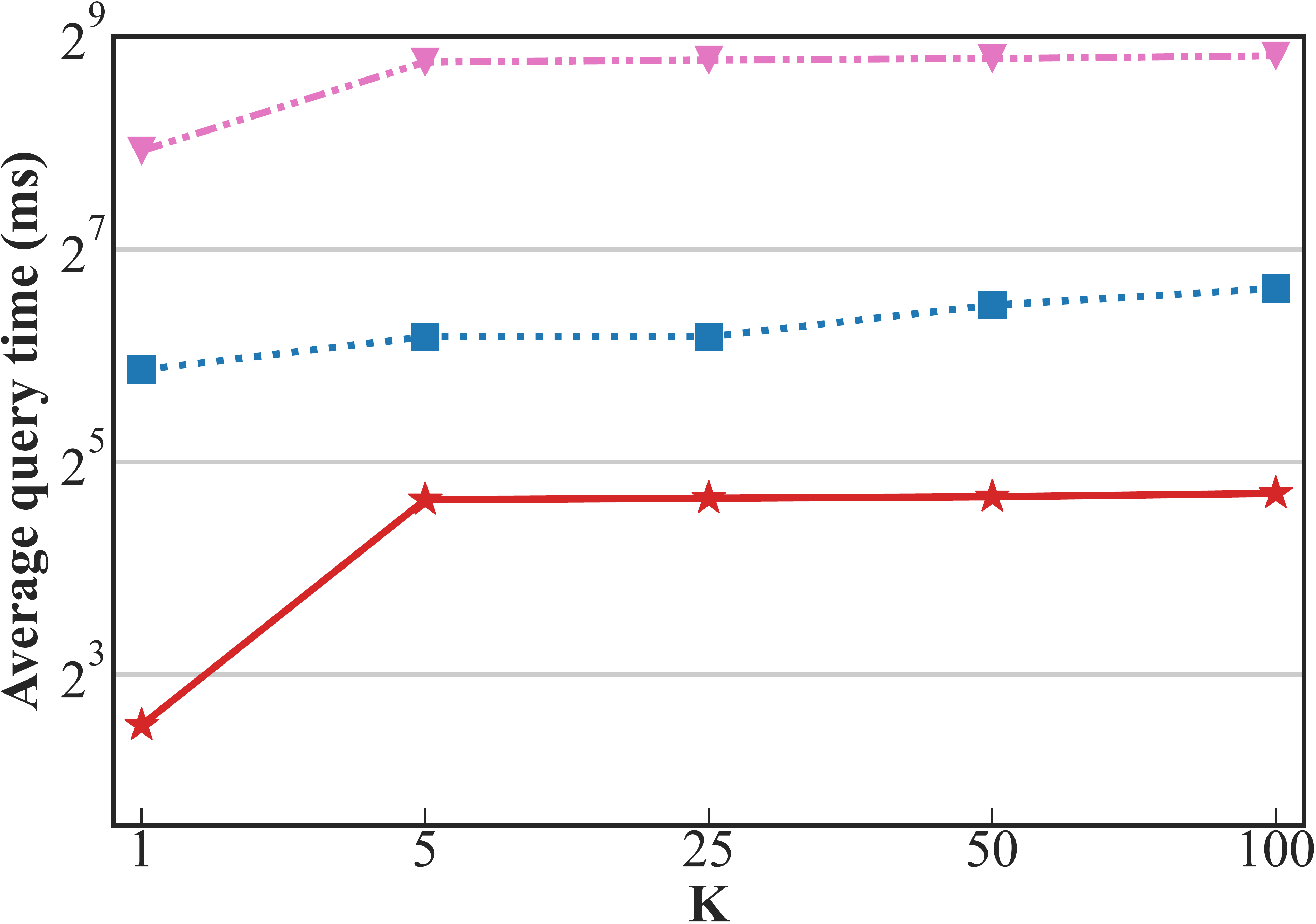}   
		\end{minipage}%
	}
	\subfloat[\# Page accesses of \textit{k}NN queries]
	{
		\begin{minipage}[t]{0.24\linewidth}
			\centering      
			\includegraphics[width=\linewidth]{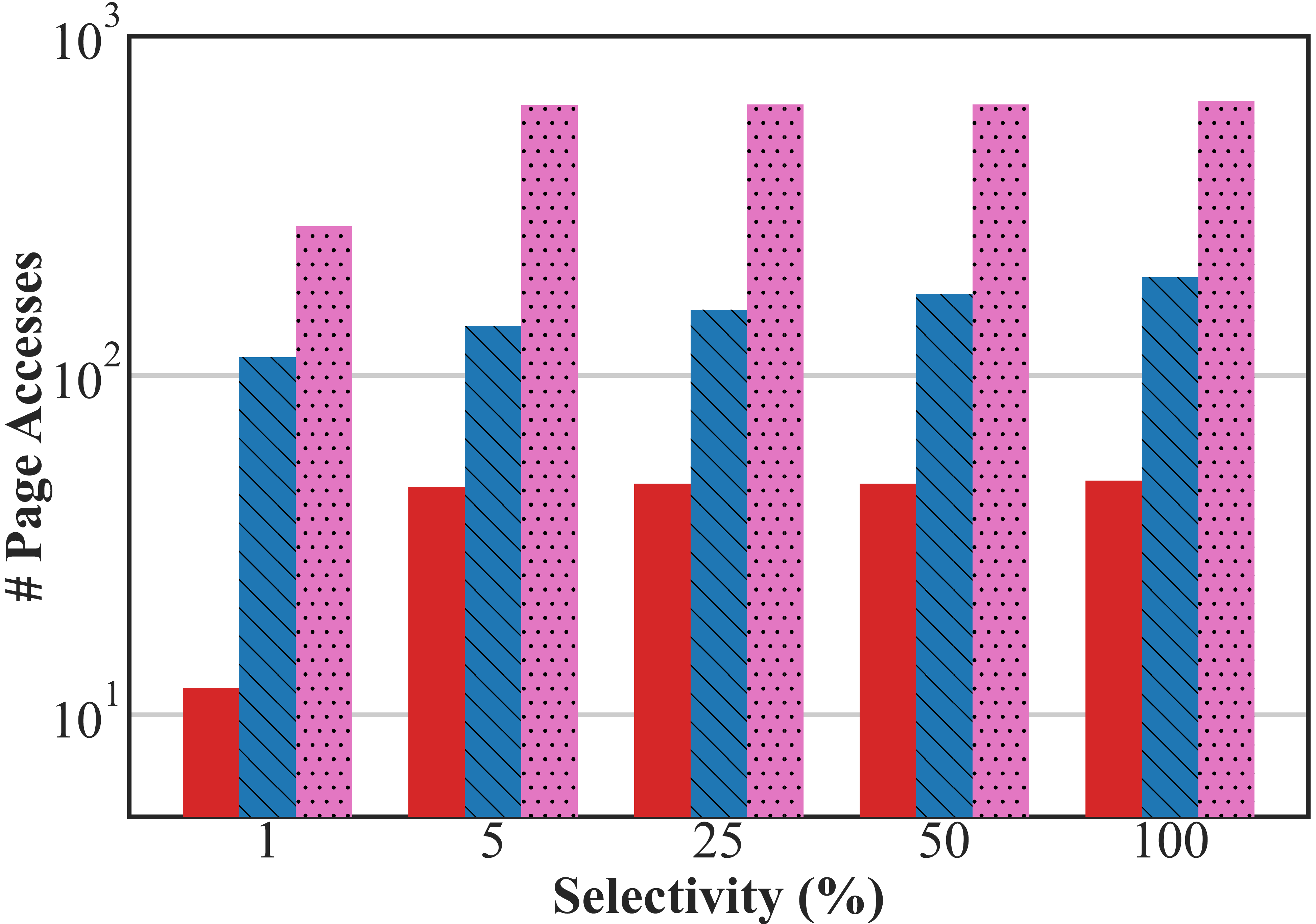}   
		\end{minipage}
	}%
	\vspace{-1ex}
	\caption{Range and \textit{k}NN query performance on \textit{Signature}} 
	\vspace{-2ex}
	\label{fig:range_knn_sg}  
\end{figure*}
\begin{figure*}[htbp]  
	\centering    
	\subfloat[Query time on \textit{Skewed} ]
	{
		\begin{minipage}[t]{0.24\linewidth}
			\centering          
			\includegraphics[width=\linewidth]{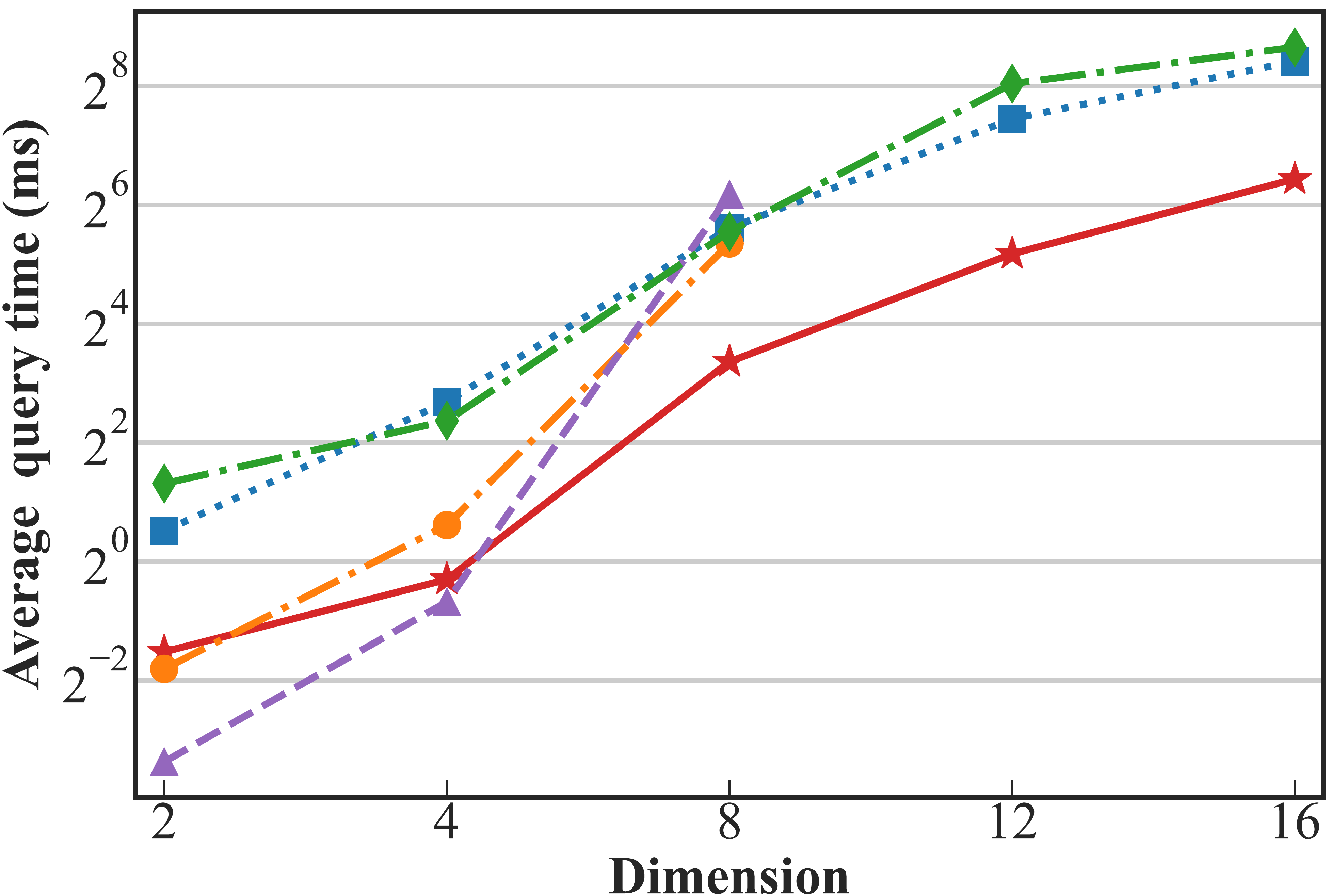}   
		\end{minipage}%
	}
	\subfloat[\# Page accesses on \textit{Skewed}]
	{
		\begin{minipage}[t]{0.24\linewidth}
			\centering      
			\includegraphics[width=\linewidth]{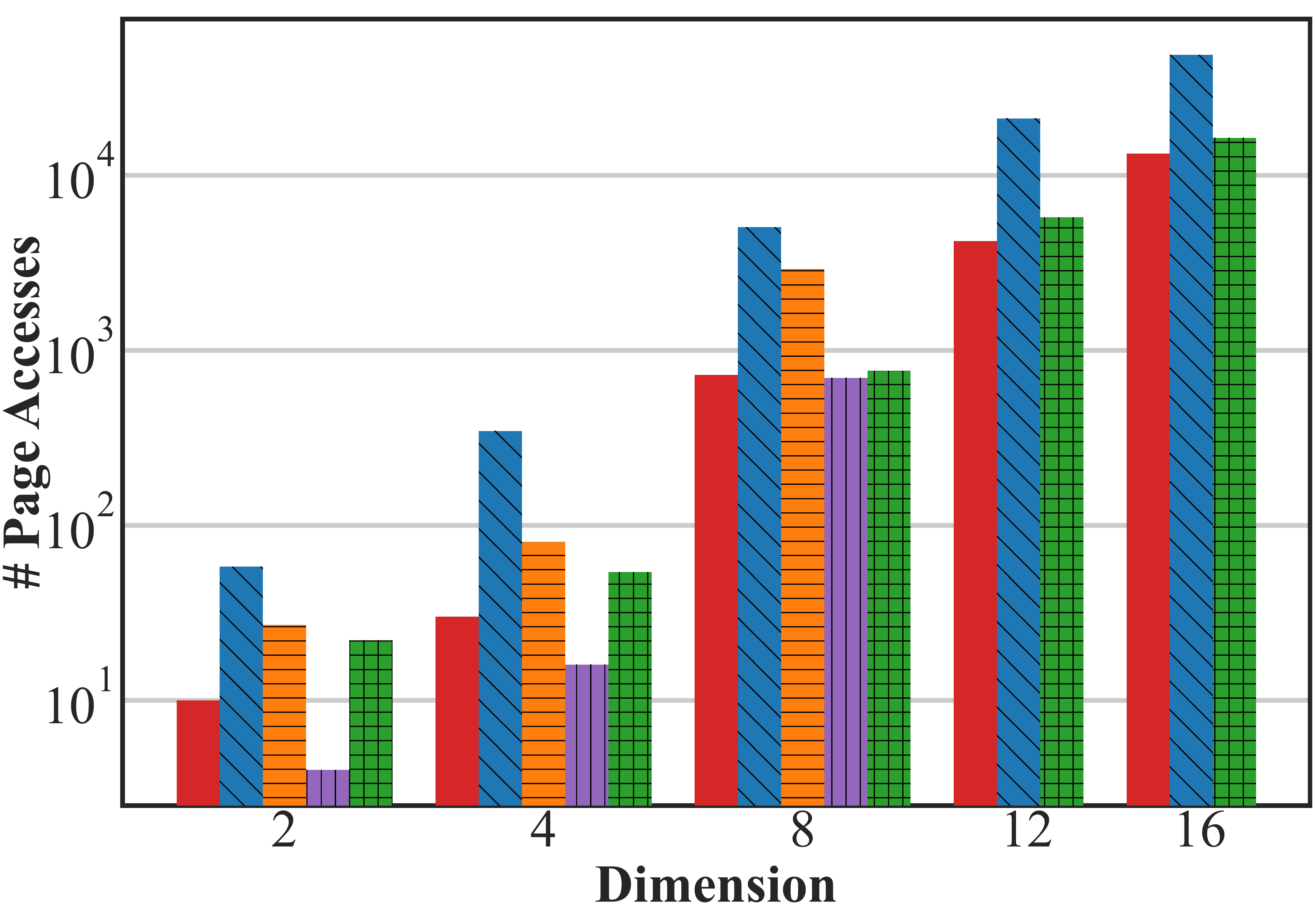}   
		\end{minipage}
	}%
	\subfloat[Query time on \textit{GaussMix}]
	{
		\begin{minipage}[t]{0.24\linewidth}
			\centering      
			\includegraphics[width=\linewidth]{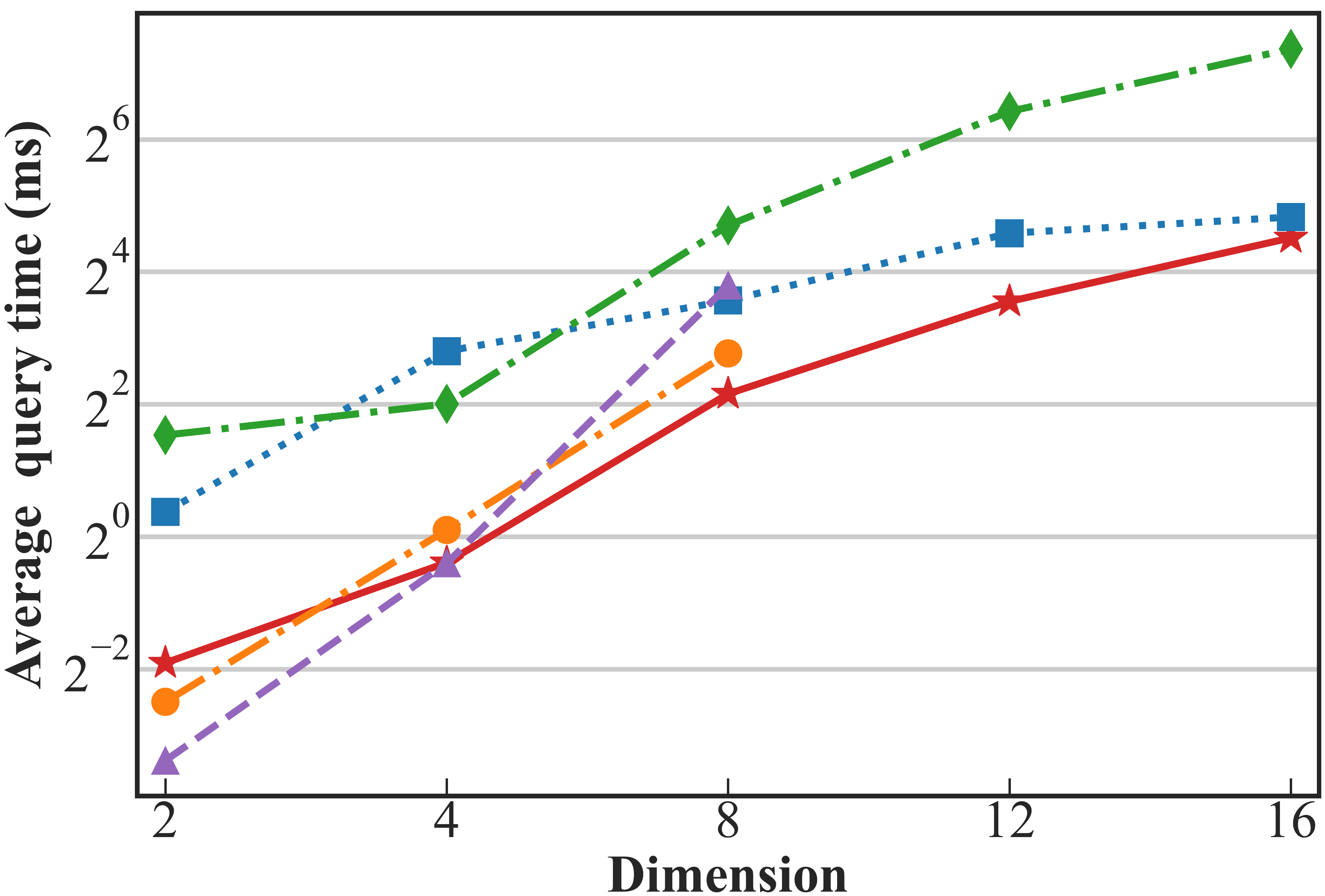}   
		\end{minipage}
	}%
	\subfloat[\# Page accesses on \textit{GaussMix}]
	{
		\begin{minipage}[t]{0.24\linewidth}
			\centering      
			\includegraphics[width=\linewidth]{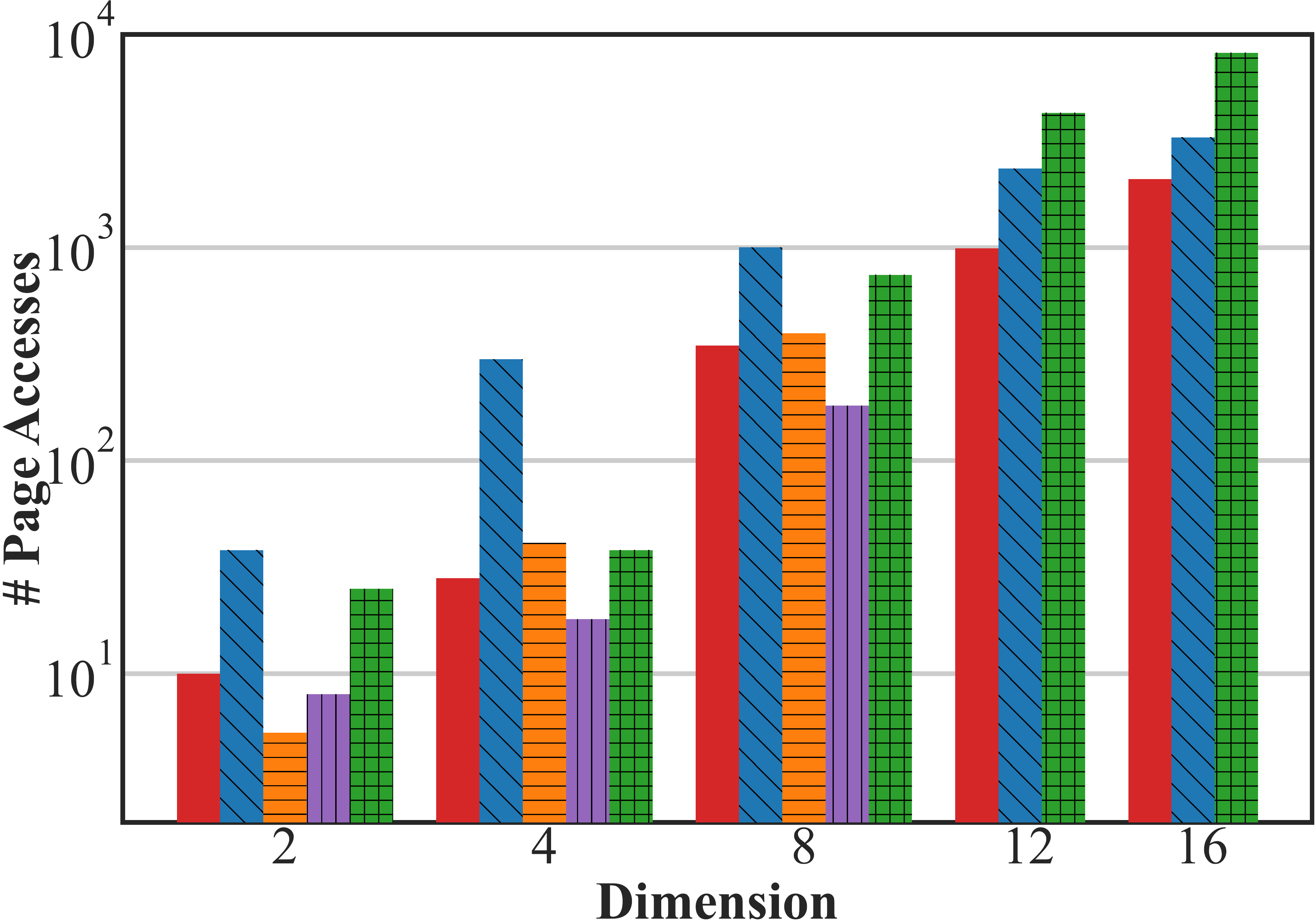}   
		\end{minipage}
	}%
    \vspace{-1ex}
	\caption{\textit{k}NN  query performance with dimensionality}
	\vspace{-2ex}
	\label{fig:knn_dim}  
\end{figure*}
\begin{figure*}[htbp]  
	\centering    
	\subfloat[Query time on \textit{Forest} ]
	{
		\begin{minipage}[t]{0.24\linewidth}
			\centering          
			\includegraphics[width=\linewidth]{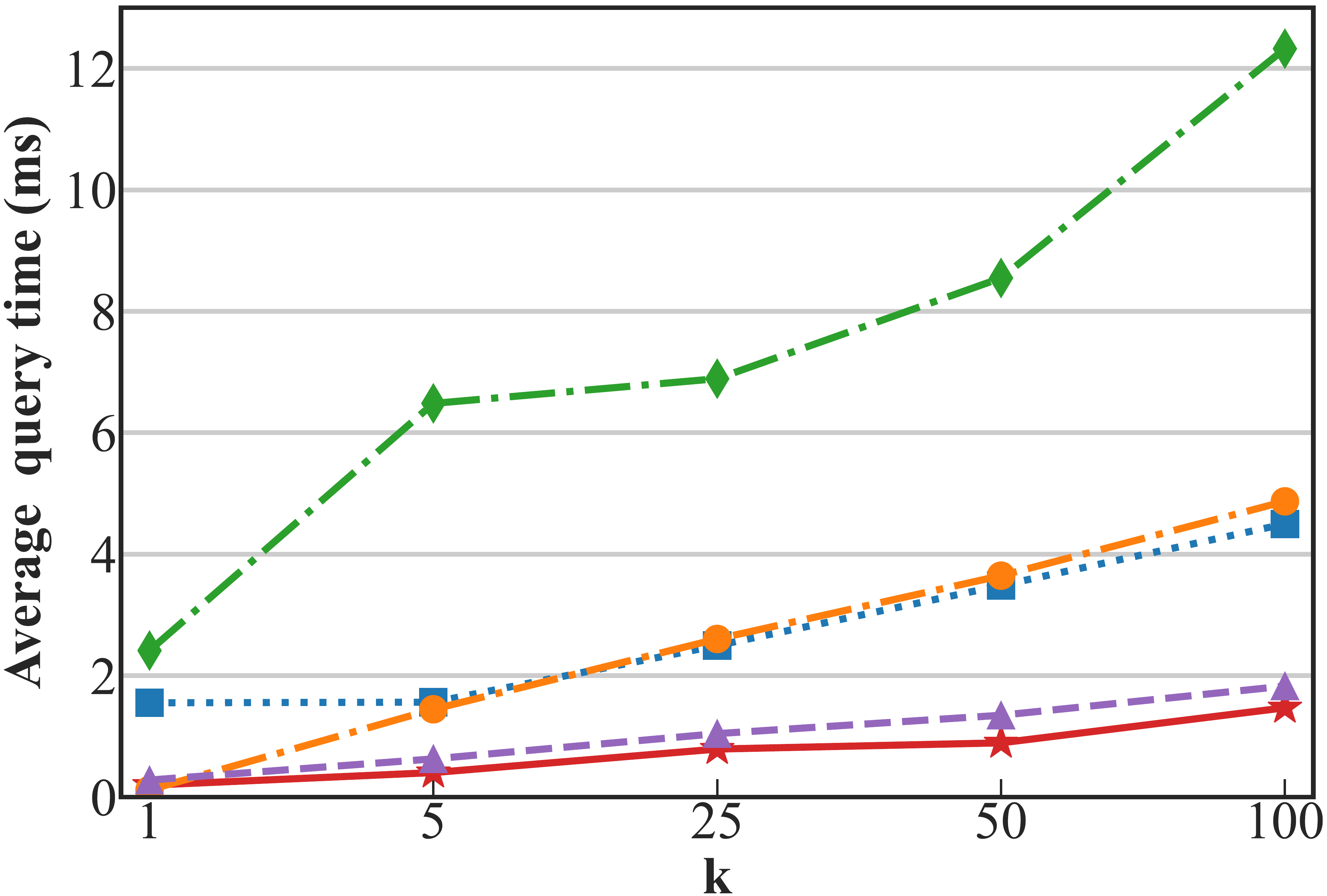}   
		\end{minipage}%
	}
	\subfloat[\# Page accesses on \textit{Forest}]
	{
		\begin{minipage}[t]{0.24\linewidth}
			\centering      
			\includegraphics[width=\linewidth]{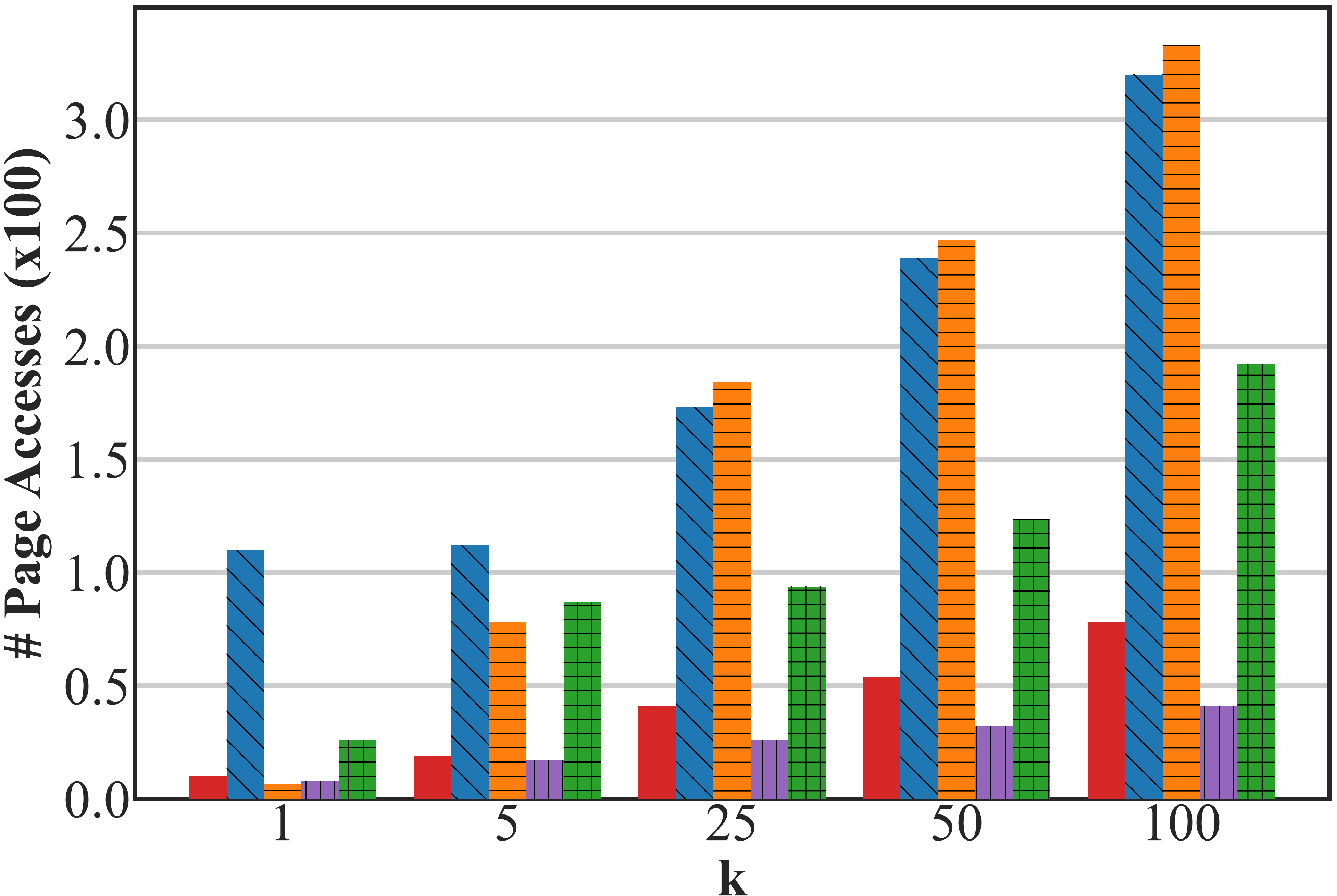}   
		\end{minipage}
	}%
	\subfloat[Query time on \textit{Color}]
	{
		\begin{minipage}[t]{0.24\linewidth}
			\centering      
			\includegraphics[width=\linewidth]{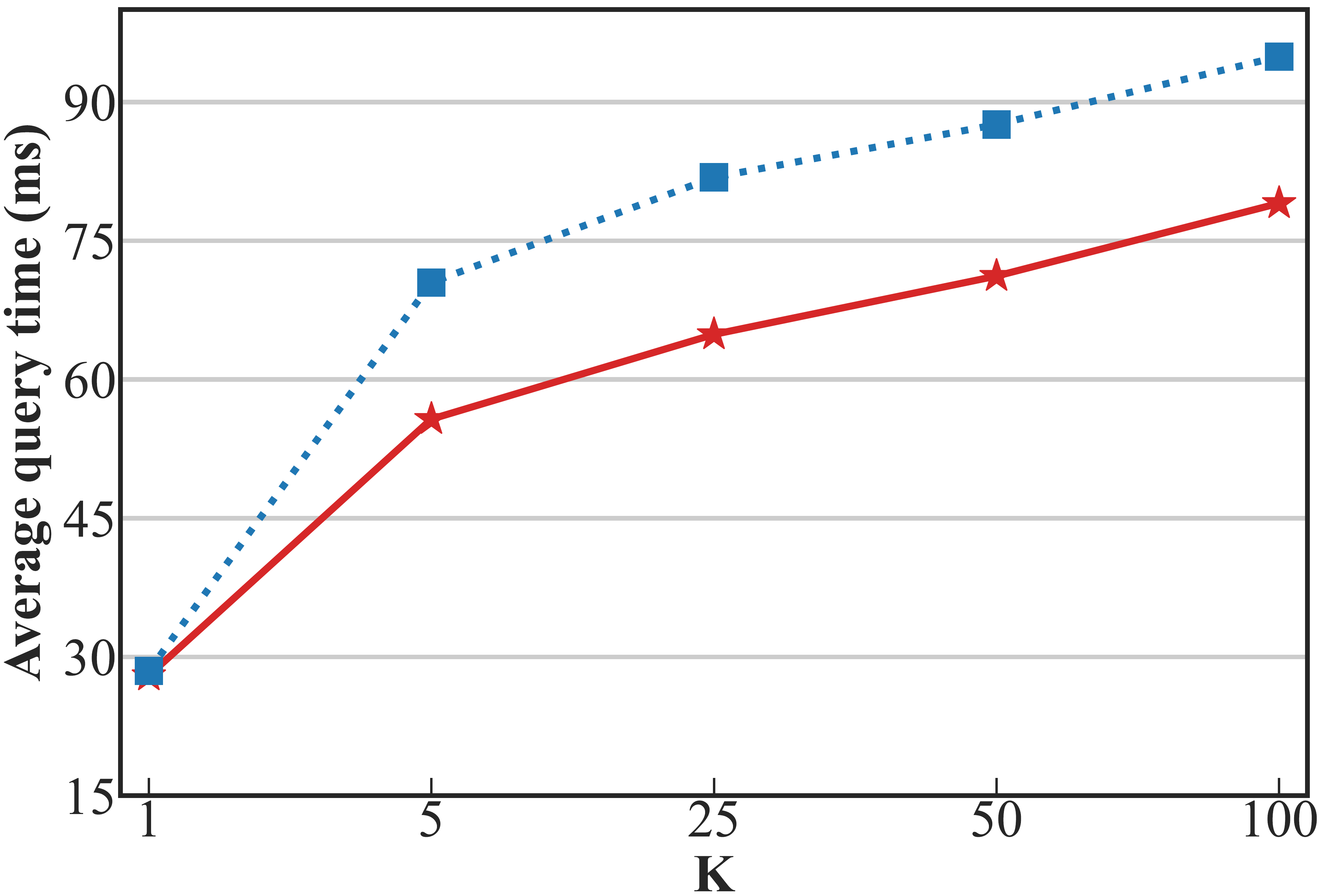}   
		\end{minipage}
	}%
	\subfloat[\# Page accesses on \textit{Color}]
	{
		\begin{minipage}[t]{0.24\linewidth}
			\centering      
			\includegraphics[width=\linewidth]{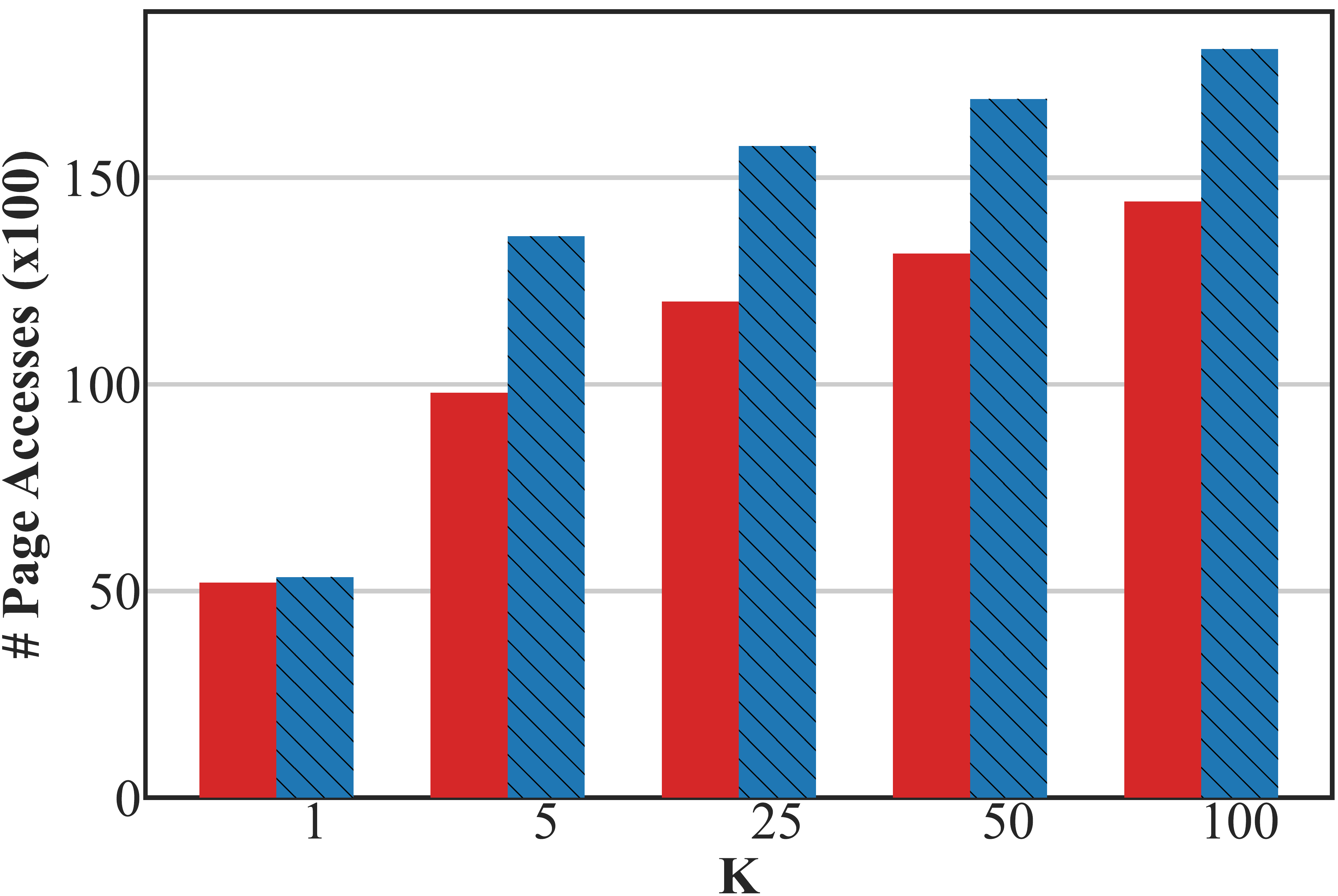}   
		\end{minipage}
	}%
    \vspace{-1ex}
	\caption{\textit{k}NN query performance with \textit{k}}%
	\vspace{-2ex}
	\label{fig:knn_selectivity}  
\end{figure*}

\subsubsection{Performance in metric spaces} The third set of experiments studies the average query time and the number of page accesses in a metric space. We compare LIMS \cbl{with SPB-tree and M-tree. We also extend ML to the metric dataset by replacing the KMeans in ML with the k-center algorithm.} Other competitors are omitted since they are not applicable for metric spaces. Fig. \ref{fig:range_knn_sg}(a)(b) report the results on \textit{Signature} dataset. Clearly, LIMS has a decided advantage over \cbl{all competitors}  under all selectivities, where LIMS is around 20X faster than M-tree and  the page accesses are at least 12X fewer. This is expected because  of expensive and unavoidable costs to traverse the tree structure in traditional index structures, \cbl{and poor pruning powers in ML.}

\subsection{\textit{k}NN Query Performance}
In this subsection, we study the performance of LIMS, ML, LISA, $\text{R}^*$-tree, M-tree \cbl{and SPB-tree} on $k$NN query, as summarized in Fig. \ref{fig:range_knn_sg}, \ref{fig:knn_dim}, and \ref{fig:knn_selectivity}. ZM is excluded because it does not support $k$NN query.

\subsubsection{Performance with dimensionality} 
Fig. \ref{fig:knn_dim}(a)(b) and Fig. \ref{fig:knn_dim}(c)(d) report the average query time and the number of page accesses on \textit{Skewed} and \textit{GaussMix} datasets, respectively.  From the figures, we have the following observations: 1) $\text{R}^*$-tree works best in low dimensions, but it is less effective than LIMS with increased dimensionality. 2) The relative performance of LIMS and ML on $k$NN queries is similar to 
their performance on range queries, since both techniques follow a similar search region expansion paradigm. 3) Even though LISA uses a model learned from the data distribution to estimate the initial radius, it still suffers from too many unnecessary page accesses. The reason is that once the number of retrieved objects is smaller than $k$, LISA will issue a range query with a larger radius from the scratch, leading to the same page accessed repeatedly. When the estimated initial radius is large, it also incurs many page accesses.

\subsubsection{Performance with \bm{$k$}} Fig. \ref{fig:knn_selectivity}(a)(b) and Fig. \ref{fig:knn_selectivity}(c)(d) show the performance when varying the value of $k$  in $\{1, 5, 25, 50, 100 \}$ on \textit{Forest} and \textit{Color} datasets, respectively. LIMS is the fastest except for 1NN on \textit{Forest}, where LISA is slightly faster owing to a proper radius estimation. 
\cred{If setting $\Delta r$ in LIMS to the value recommended by LISA. LIMS can achieve better performance ($0.11ms$ v.s. $0.12ms$). However, the training time of model used to estimate $\Delta r$ in  LISA is long, which is not favorable for frequent insertion/deletion operations and query pattern changes.}
As $k$ grows, LISA becomes not comparable due to the aforementioned reasons. The results of SPB-tree on \textit{Color} is omitted because it is drastically slower. Consistent high efficiency of LIMS indicates that it scales to large $k$ values.

\subsubsection{Performance in metric spaces} 
Fig. \ref{fig:range_knn_sg}(c)(d) present the performance on \textit{Signature} dataset. As expected, LIMS again yields \cbl{the} fastest query time and fewest page accesses.  The curve becomes gentle after $k = 5$ because many signatures share the same edit distance to a given signature.

\begin{figure}[t]  
\begin{minipage}[r]{\linewidth}
		\centering
		\includegraphics[width=0.95\textwidth]{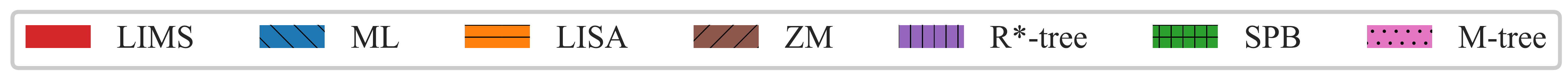}
 		\vspace{-0.5em}
	\end{minipage}
	\centering    
	\subfloat[Indexing time]
	{
		\begin{minipage}[t]{0.48\linewidth}
			\centering          
			\includegraphics[width=\linewidth]{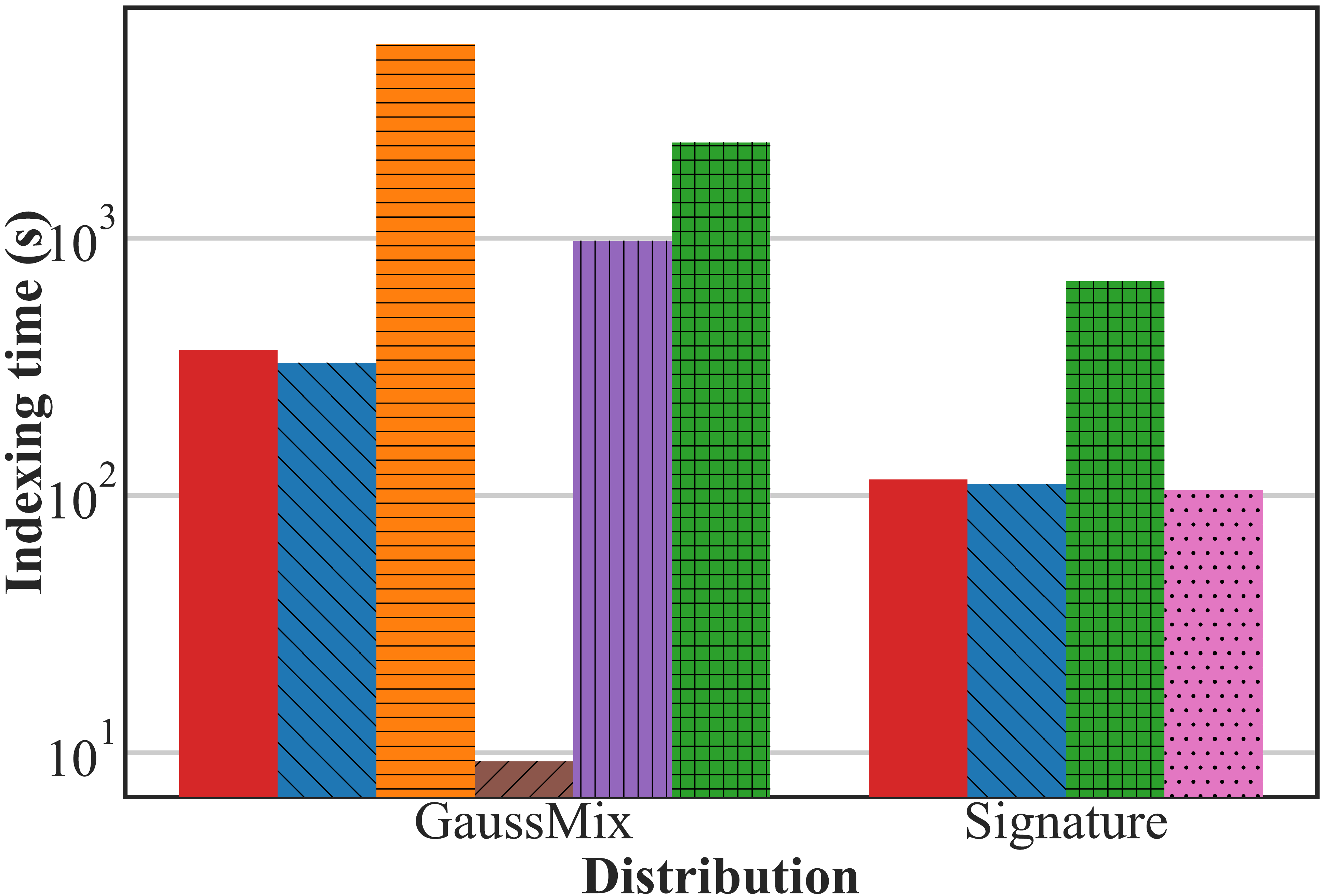}   
		\end{minipage}%
	}
	\subfloat [Index size]
	{
		\begin{minipage}[t]{0.48\linewidth}
			\centering      
			\includegraphics[width=\linewidth]{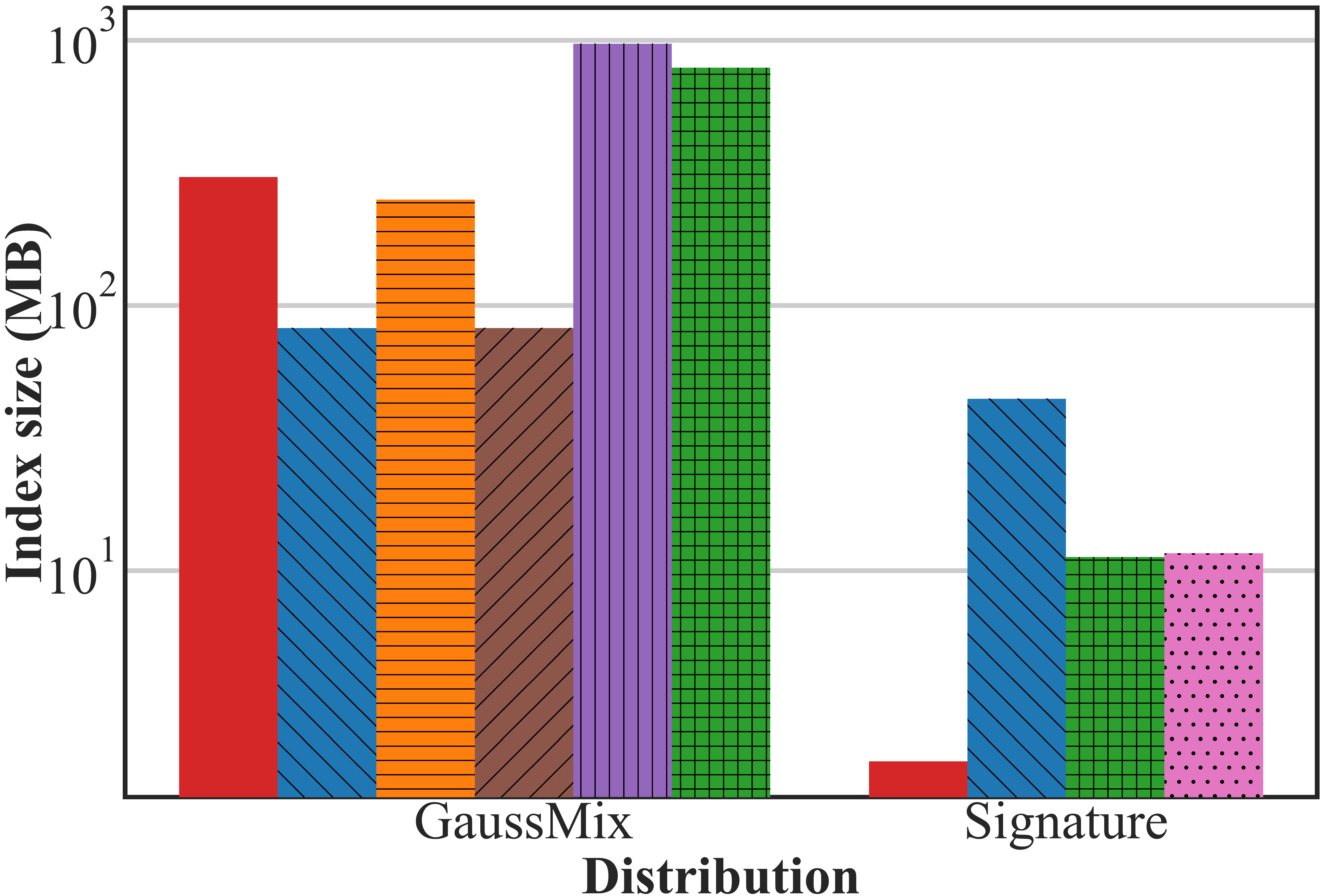}   
		\end{minipage}
	}%
	\vspace{-0.5ex}
	\caption{Indexing time and index size}
	\vspace{-2ex}
	\label{fig:sizetime}
\end{figure}
\vspace{-1ex}
\subsection{Indexing Time and Index Size}
In this subsection, we report the indexing time and index size. Due to the space limitation, we only report the results on $8d$ 10M \textit{GaussMix} and \textit{Signature}. As Fig. \ref{fig:sizetime}(a) shown, LIMS does not suffer from long training time, a common dilemma faced by learned indexes.  On 10M 8$d$ \textit{GaussMix}, LIMS is 15.5X faster than LISA ($368s$ v.s. $1.6 h$), which makes LIMS easy to rebuild and ensures simple update operations effective in practice. \cbl{ As described in Section \ref{updates},  when the query performance
does not degrade severely, we can partially rebuild LIMS
by retraining rank prediction models for some clusters. 
Retraining the index of a cluster only takes an average of $0.5s$.}
LIMS has a longer construction time than ZM because  of expensive distance  computations in metric spaces. However, LIMS has a decided advantage over ZM regardless of the dataset or dimension. Fig. \ref{fig:sizetime}(b) shows the index size. The index size of LIMS is only 1/3 that of  $\text{R}^*$-tree on \textit{GaussMix} and 1/23 that of ML  on \textit{Signature}, respectively, since traditional indexes have to store a large number of internal nodes and ML have to store multiple stages of learned models. LIMS has a slightly larger index size than other learned indexes  on \textit{GaussMix} because  LIMS is a metric-space index that needs to store many pre-computed distances between pivots and data objects. However, it is accepted because a few extra distance values are relatively insignificant compared to the complex and large data, such as images and audio. We expect 
the indexing time and index size to be smaller when using fewer pivots.

\vspace{-1ex}
\begin{figure}[htbp]  
\begin{minipage}[r]{\linewidth}
		\centering
		\includegraphics[width=0.3\textwidth]{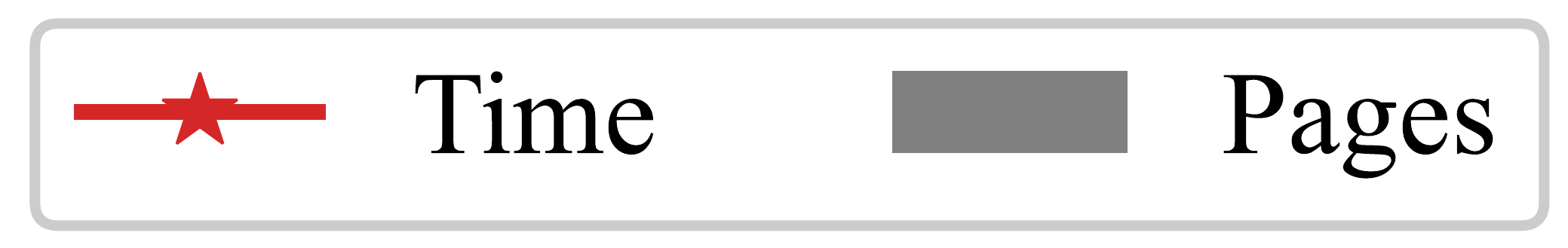}
 		\vspace{-0.5em}
	\end{minipage}
	\subfloat [\textit{GaussMix}]
	{
		\begin{minipage}[t]{0.49\linewidth}
			\centering          
			\includegraphics[width=\linewidth]{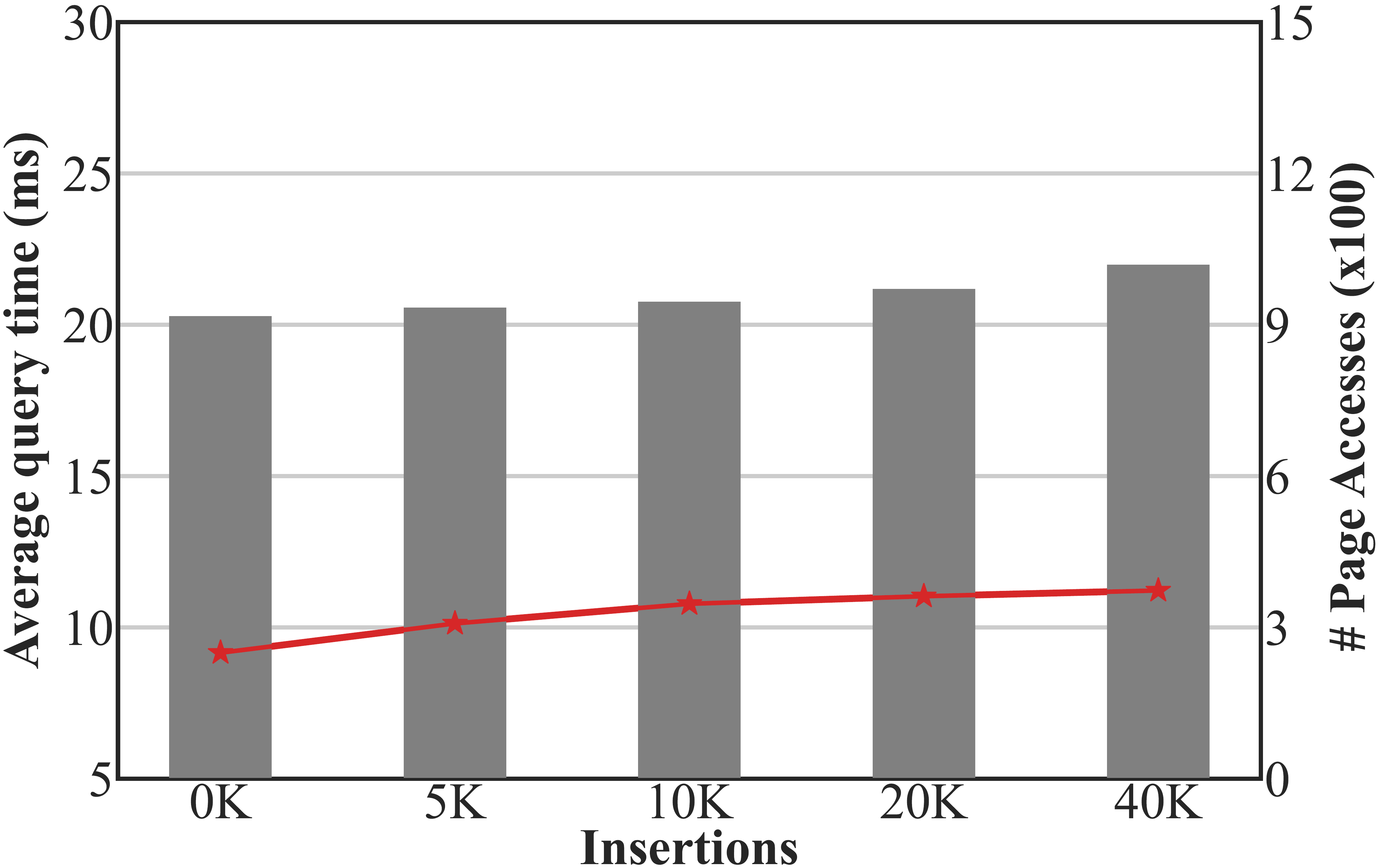}   
		\end{minipage}%
	}
	\subfloat [\textit{Skewed}]
	{
		\begin{minipage}[t]{0.49\linewidth}
			\centering      
			\includegraphics[width=\linewidth]{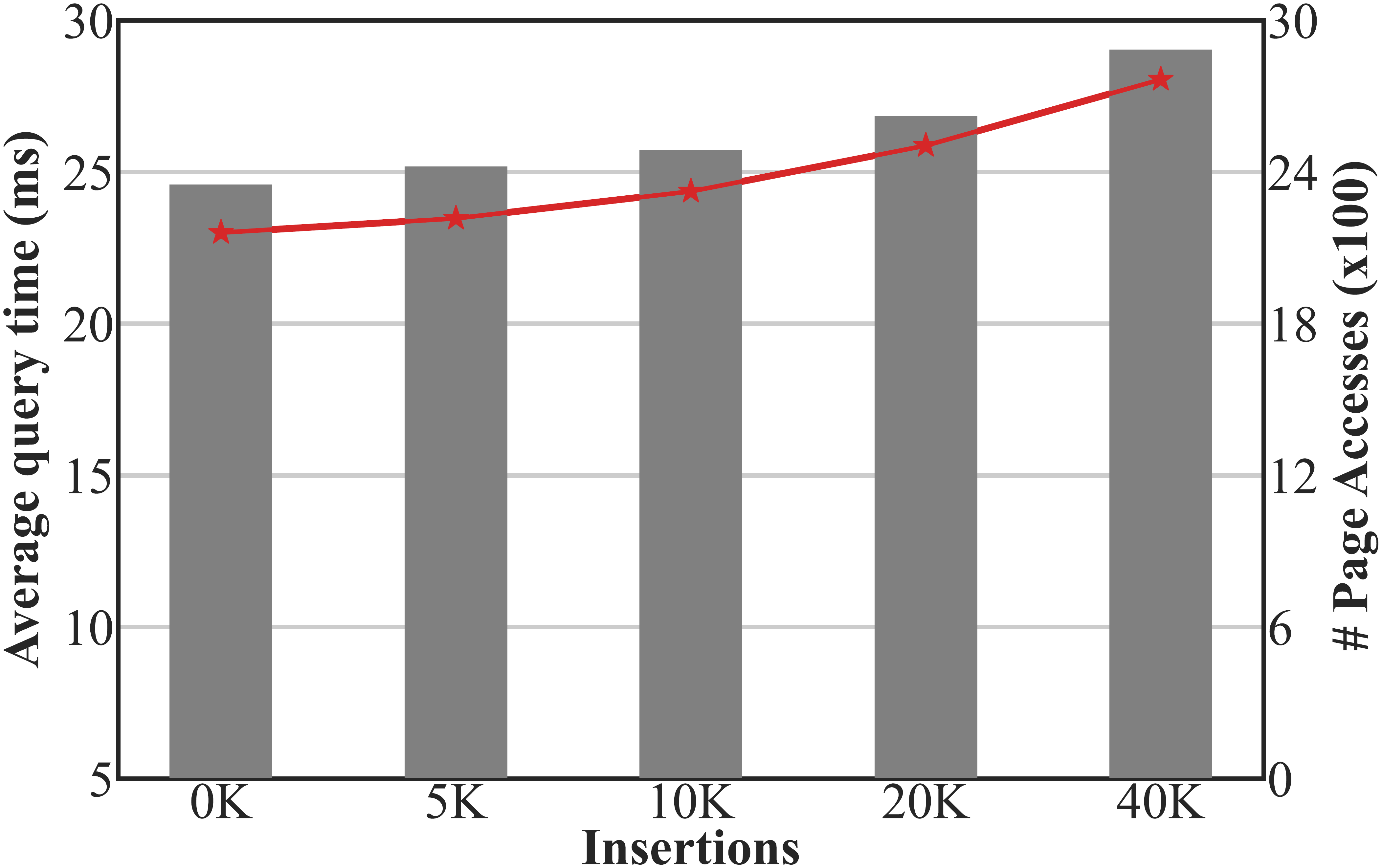}   
		\end{minipage}
	}%
	\vspace{-0.5ex}
	\caption{Range query performance after insertions} 
	\vspace{-2ex}
	\label{fig:last}
\end{figure}
\vspace{-1ex}
\subsection{Updates}
In this subsection, we examine the impact of data updates. \cred{Without loss of generality, we assume that the distribution of underlying data does not change greatly over time.} Fig. \ref{fig:last}(a)(b) report the average query time and the number of page accesses for range queries after inserting 5K, 10K, 20K and 40K objects into 10M 8$d$ \textit{GaussMix} and \textit{Skewed} datasets, respectively. As expected, insertions cause query time increase since there are more points to query, and the index becomes less optimal. However, the rate of performance degradation is slow and steady.
For example, on \textit{Skewed}, after 40K insertions, LIMS still achieves 3.4X speedup compared to the second best index ($28.05ms$ v.s. $94.68ms$). \cred{Retraining can be triggered when the query performance deteriorates beyond a user-specified threshold.}
We also studied the impact of deletion operations, but omit those due to negligible influence on query performance and space constraints.

\vspace{-2ex}
\subsection{Ablation Study}
\begin{figure}[htbp]  
\begin{minipage}[r]{\linewidth}
		\centering
		\includegraphics[width=0.3\textwidth]{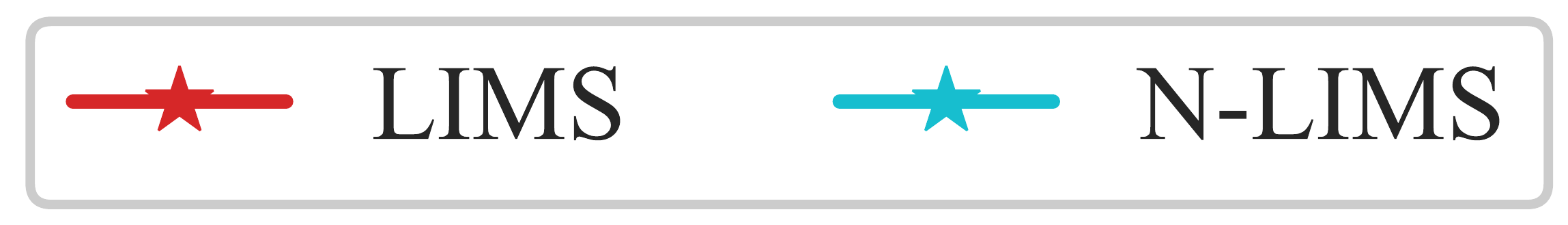}
 		\vspace{-0.5em}
	\end{minipage}
	\centering    
	\subfloat[CPU time]
	{
		\begin{minipage}[t]{0.48\linewidth}
			\centering          
			\includegraphics[width=\linewidth]{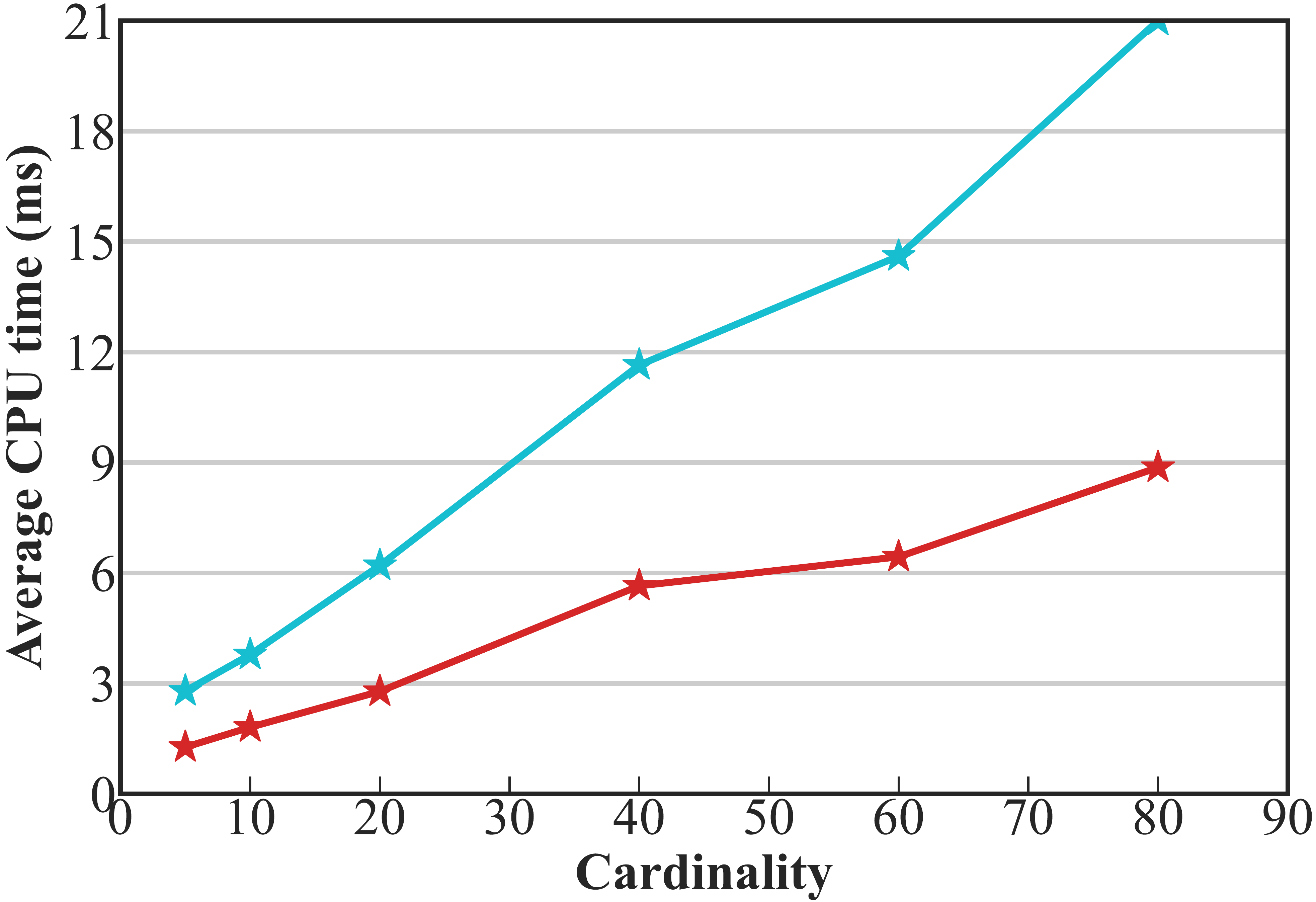}   
		\end{minipage}%
	}
	\subfloat [Query time]
	{
		\begin{minipage}[t]{0.48\linewidth}
			\centering      
			\includegraphics[width=\linewidth]{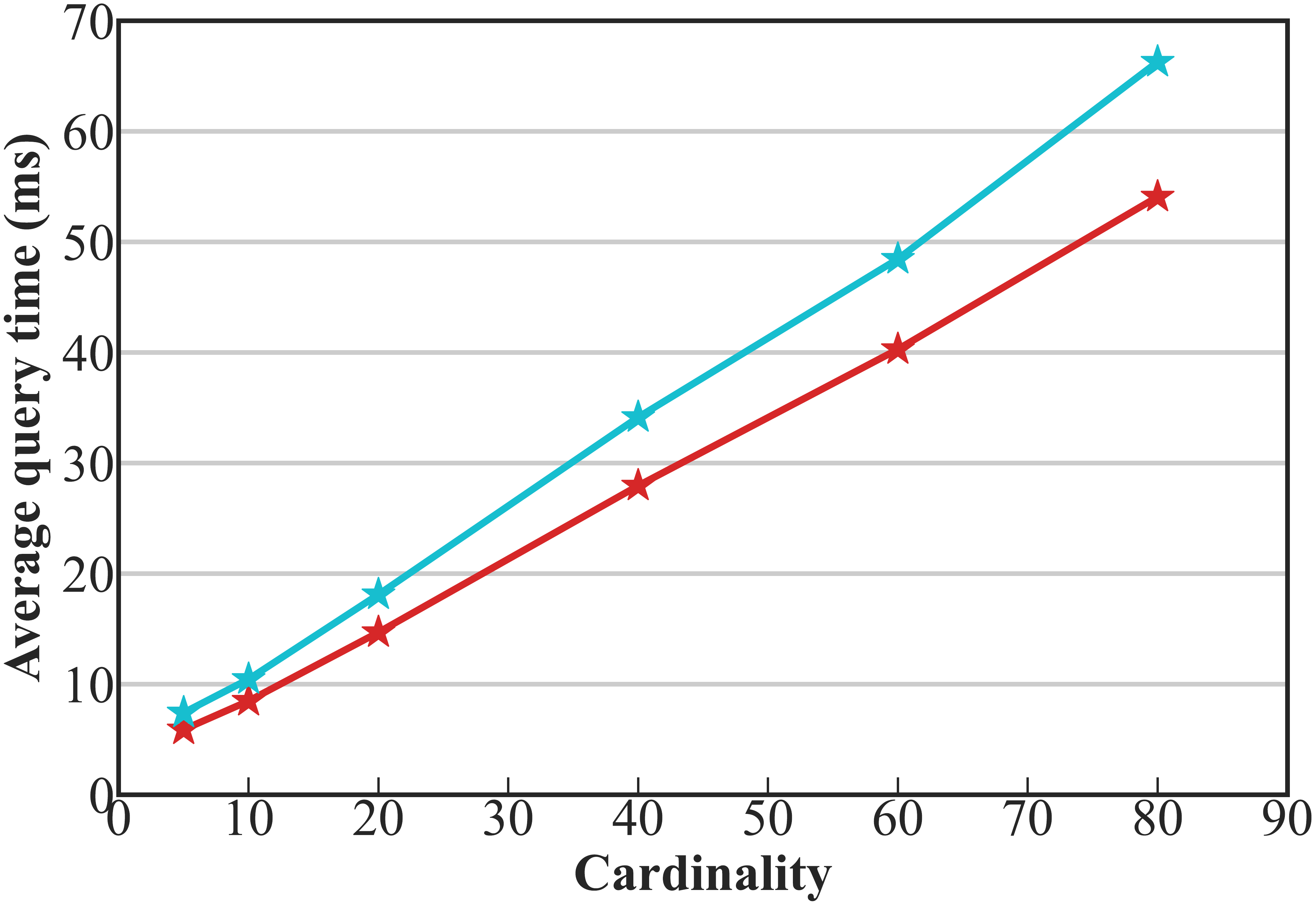}   
		\end{minipage}
	}%
	\vspace{-0.5ex}
	\caption{Comparison of LIMS and N-LIMS}
	\vspace{-2ex}
	\label{fig:ablation}
\end{figure}

Finally, we make a comparison of LIMS and N-LIMS to show the effectiveness of learning components in LIMS. Since the only difference between two methods is whether to use B$^+$-trees or the rank prediction models and exponential search to locate the start and end of a range query, both have the same number of page accesses (I/O cost).   
Fig. \ref{fig:ablation}(a) reports the average CPU time of range query by varying the cardinality of 8$d$ \textit{GaussMix} dataset. From the figure, we have the following observations: 1) LIMS outperforms N-LIMS on all data sizes. The reason is that rank prediction models in LIMS achieve a good grasp of \cbl{the} data distribution so that a query can be processed by a function invocation in $O(1)$ time instead of traversing B$^+$-trees in $O(\log_{}n)$ time, which implies the effectiveness of the machine learning model. 2) The advantage of LIMS becomes more obvious when the data size increases. This is because the query cost of LIMS does not directly depend on the cardinality, which allows LIMS scalable to very large data sets. To present the whole picture for a fair comparison, we also report the average query time (CPU time + I/O time)  of LIMS and N-LIMS in Fig. \ref{fig:ablation}(b). As expected, LIMS again offers the best query performance. Furthermore, N-LIMS is still faster than the second best competitor ($10.45 ms$ v.s. $12.21 ms$), which further confirms the superiority of LIMS index structure.
\vspace{-1ex}


\section{Conclusions} \label{sec:conclusion}
As a universal abstraction of various data types,  metric spaces and associated similarity search play an important role in many real-life applications. 
In this paper, we have developed LIMS, a novel learned index structure, for efficient exact similarity search query processing in metric spaces. LIMS takes advantage of compact-partitioning methods,  pivot-based techniques and the idea of \textit{learned index} to organize and access multi-dimensional data. Our extensive experimental results illustrate that  LIMS can respond significantly better to the problem of `curse of dimensionality' compared with other learned index structures. It also has a clear advantage over traditional indexes. In the future, we plan to take this work further by considering query workload information such that workload-aware optimization can be made.


\vspace{-1ex}
\ifCLASSOPTIONcompsoc
  \section*{Acknowledgments}
\else
  \section*{Acknowledgment}
\fi
This research is partially supported by Natural Science Foundation of China (Grant\# 62072125) and is conducted in the JC STEM Lab of Data Science Foundations funded by The Hong Kong Jockey Club Charities Trust.

\ifCLASSOPTIONcaptionsoff
  \newpage
\fi



%


\vspace{-2ex}
\bibliographystyle{ieeetr}
\bibliography{ref}

%
\vspace{-8ex}
\begin{IEEEbiography}[{\includegraphics[width=1in,height=1.2in,clip,keepaspectratio]{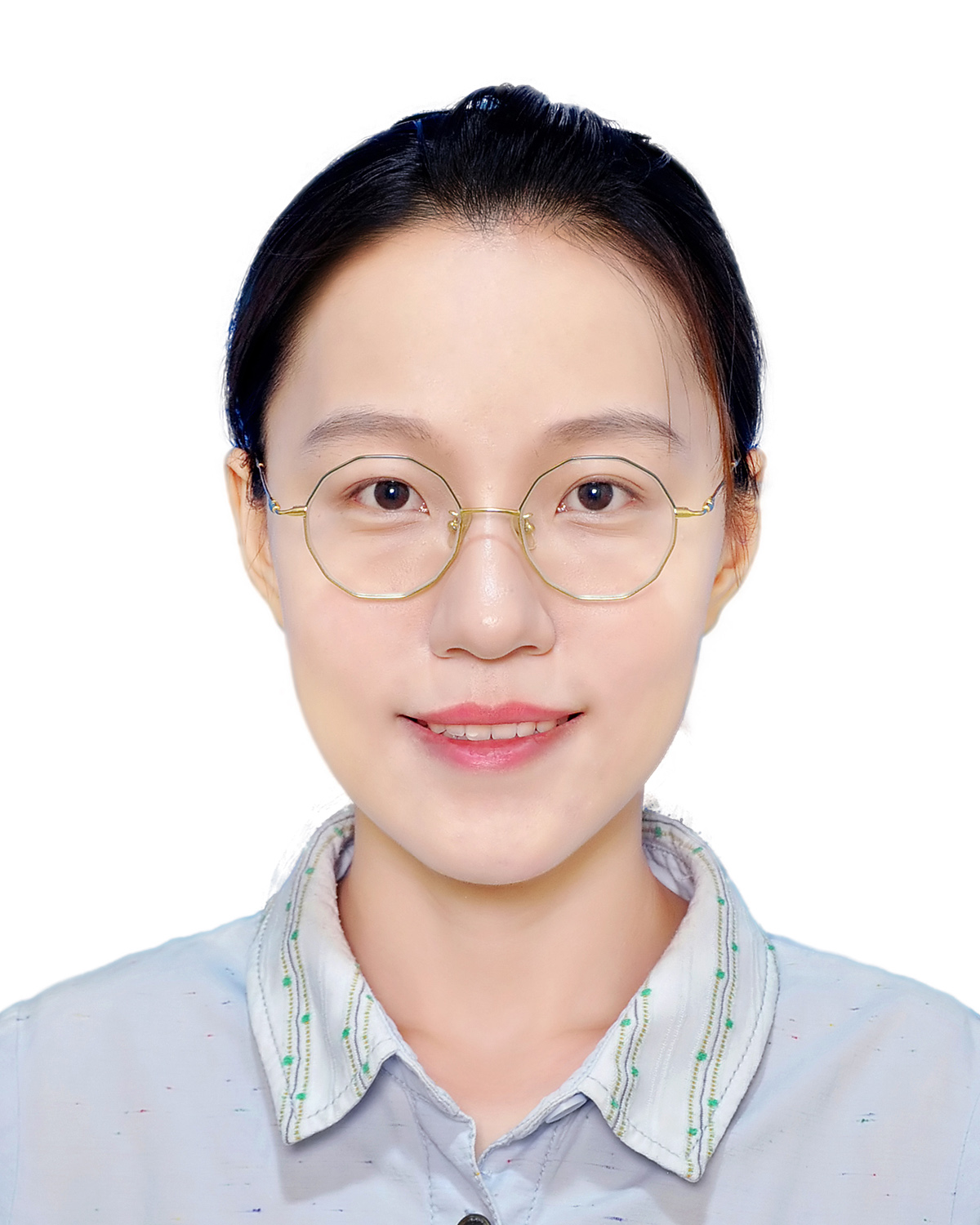}}]{Yao Tian} is currently a PhD student with the Department of Computer Science and Engineering, Hong Kong University of Science and Technology (HKUST), supervised by Prof. Xiaofang Zhou. She received her MSc degree in School of Mathematical Science from Zhejiang University in 2020. Her research interests include the learned index and approximate query processing in high dimensional spaces.
\end{IEEEbiography}
\vspace{-8ex}
\begin{IEEEbiography}[{\includegraphics[width=1in,height=1.2in,clip,keepaspectratio]{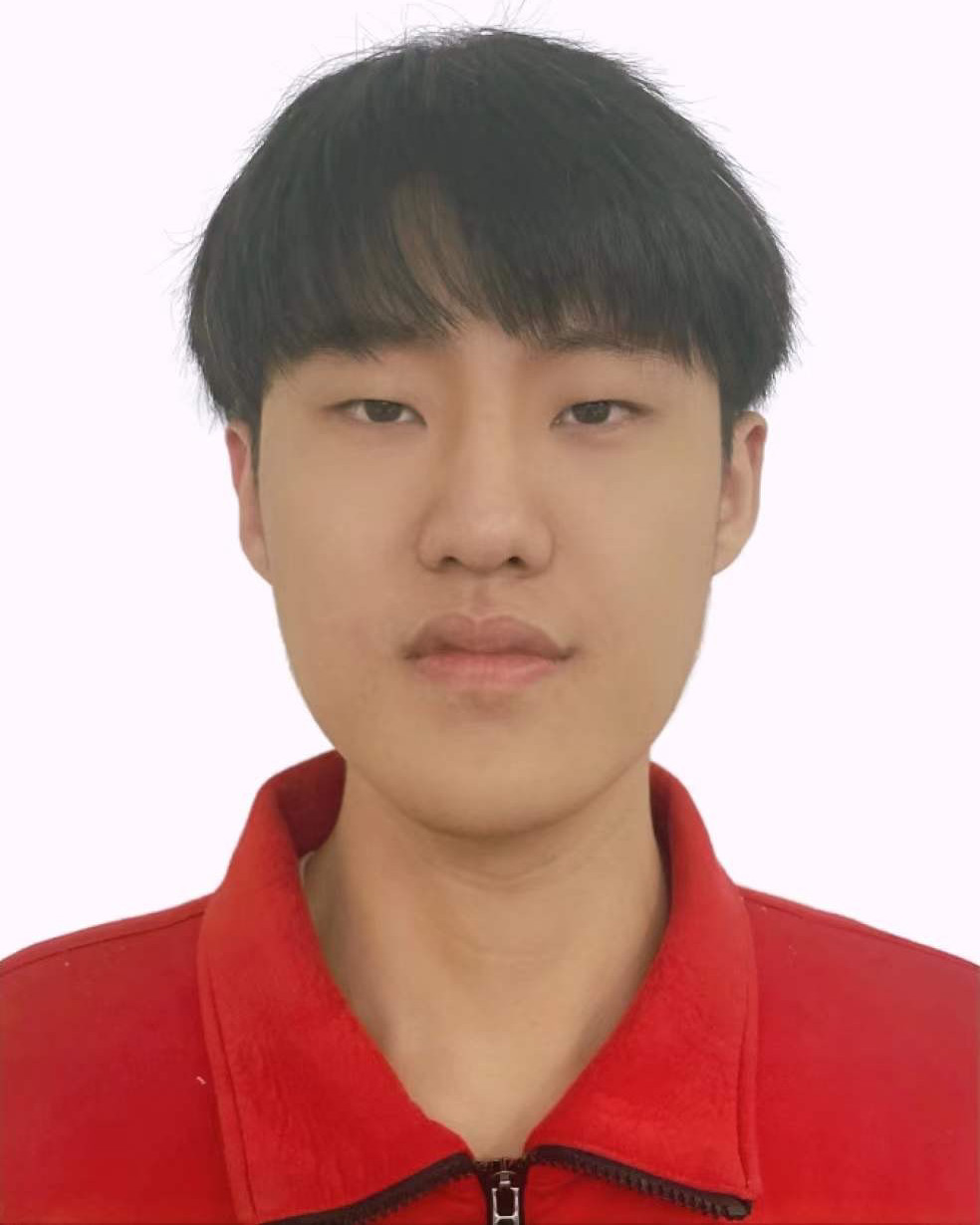}}]{Tingyun Yan} is currently a postgraduate student in Cyberspace Institute of Advanced Technology, Guangzhou University, supervised by Prof. Xiaofang Zhou. He received the Bachelor degree in Software College from Northeastern University in 2020. His research interests include spatial index and learned index.
\end{IEEEbiography}
\vspace{-8ex}
\begin{IEEEbiography}[{\includegraphics[width=1in,height=1.2in,clip,keepaspectratio]{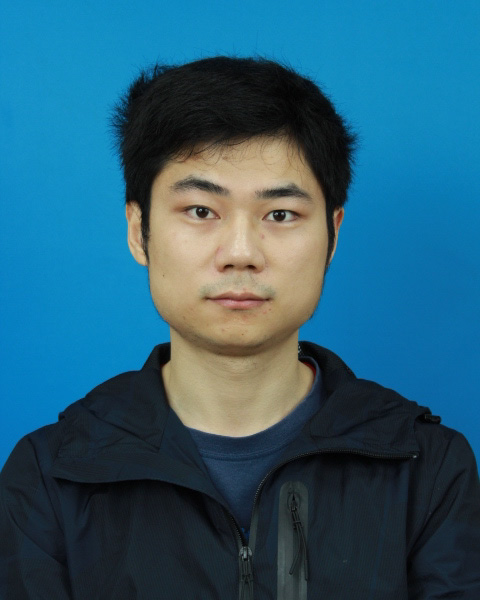}}]{Xi Zhao}
is currently a research assistant at HKUST, under the supervision of Prof. Xiaofang Zhou. He received the Master degree in Computer Science from Huazhong University of Science and Technology, China, in 2021. His research interests include the approximate query processing in high dimensional spaces, exact trajectory similarity search and exact textual similarity search. 
\end{IEEEbiography}
\vspace{-8ex}
\begin{IEEEbiography}[{\includegraphics[width=1in,height=1.2in,clip,keepaspectratio]{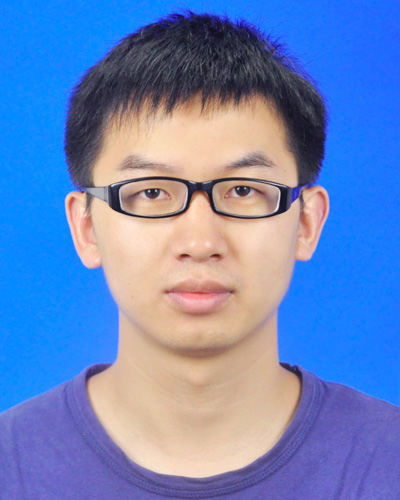}}]{Kai Huang} is a Postdoc at the Department of Computer Science and Engineering, HKUST, under the supervision of Prof. Xiaofang Zhou.  He received his PhD degree in School of Computer Science from Fudan University  in 2020, and  BEng degree in Software Engineering from East China Normal University
in 2014. His research interests include graph
database and privacy-aware data management.
\end{IEEEbiography}
\vspace{-8ex}
\begin{IEEEbiography}[{\includegraphics[width=1in,height=1.2in,clip,keepaspectratio]{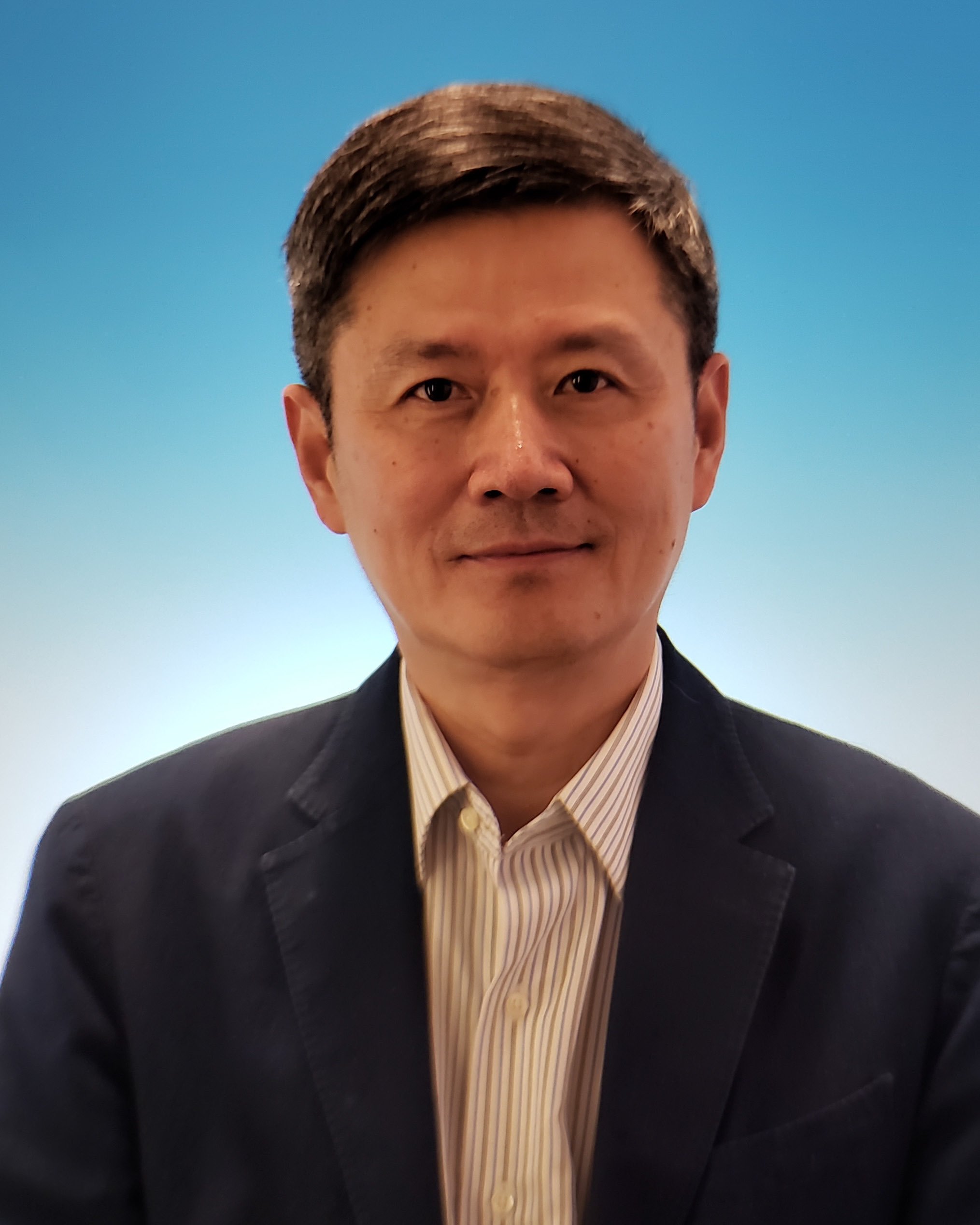}}]{Xiaofang Zhou}
 is Otto Poon Professor of Engineering and Chair Professor of Computer Science and Engineering at the Hong Kong University of Science and Technology. He received his Bachelor and Master degrees in Computer Science from Nanjing University, in 1984 and 1987 respectively, and his PhD degree in Computer Science from the University of Queensland in 1994. His research is focused on finding effective and efficient solutions for managing, integrating, and analysing very large amount of complex data for business, scientific and personal applications. His research interests include spatial and multimedia databases, high performance query processing, web information systems, data mining, data quality management, and machine learning. He is a Fellow of IEEE.
\end{IEEEbiography}





\end{document}